\documentclass[
superscriptaddress,
nofootinbib,
 amsmath,amssymb,
 aps,
 article,pre,
]{revtex4-2}

\usepackage{graphicx}
\usepackage{dcolumn}
\usepackage{cancel}
\usepackage{bm}
\usepackage{hyperref}

\usepackage{mathtools}
\usepackage{array}
\newcolumntype{P}[1]{>{\centering\arraybackslash}p{#1}}
\usepackage{color}
\usepackage{physics}

\newcommand{\nocontentsline}[3]{}
\newcommand{\tocless}[2]{\bgroup\let\addcontentsline=\nocontentsline#1{#2}\egroup}

\newcommand{\st}{\text{s}}
\newcommand{\rel}{\text{R}}
\newcommand{\irr}{\text{irr}}

\begin{document}
\title{Minimum time connection between non-equilibrium steady states: the Brownian gyrator}

\author{A. Patrón}%
\email{apatron@us.es}
\affiliation{Física Teórica, Universidad de Sevilla, Apartado de
  Correos 1065, E-41080 Sevilla, Spain}
\author{C. A. Plata}
\email{cplata1@us.es}
\affiliation{Física Teórica, Universidad de Sevilla, Apartado de
  Correos 1065, E-41080 Sevilla, Spain}
\author{A. Prados}
\email{prados@us.es}
\affiliation{Física Teórica, Universidad de Sevilla, Apartado de
  Correos 1065, E-41080 Sevilla, Spain}

\begin{abstract}
We study the problem of minimising the connection time between non-equilibrium steady states of the Brownian Gyrator. This is a paradigmatic model in non-equilibrium statistical mechanics, an overdamped Brownian particle trapped in a two-dimensional elliptical potential, with the two degrees of freedom $(x,y)$ coupled to two, in principle different, thermal baths with temperatures $T_x$ and $T_y$, respectively. Application of Pontryagin's Maximum Principle reveals that shortest protocols belong to the boundaries of the control set defined by the limiting values of the parameters $(k,u)$ characterising the elliptical potential. We identify two classes of optimal minimum time protocols, i.e. brachistochrones: (i) regular bang-bang protocols, for which $(k,u)$ alternatively take their minimum and maximum values allowed, and (ii) infinitely degenerate singular protocols. We thoroughly investigate the minimum connection time over the brachistochrones in the limit of having infinite capacity for compression. A plethora of striking phenomena emerge: sets of states attained at null connection times, discontinuities in the connection time along adjacent target states, and the fact that, starting from a state in which the oscillators are coupled, uncoupled states are impossible to reach in a finite time.
\end{abstract}

\pacs{Valid PACS appear here}
\maketitle

\section{\label{sec:intro} Introduction}

In the pursuit of enhancing the efficiency of physical processes, researchers have recently turned their attention towards the field of \textit{shortcuts to adiabaticity}~\cite{guery-odelin_shortcuts_2019} (STA). Originally born within the framework of quantum mechanics~\cite{chen_fast_2010,chen_shortcut_2010}, STA were developed with the aim of driving quantum systems towards a desired target state, while beating the natural relaxation timescale characterising the dynamics of the system under scrutiny. It was not long after when these ideas were extended to alternative contexts, such as those of classical mechanics~\cite{patra_shortcuts_2017} and non-equilibrium statistical mechanics~\cite{guery-odelin_nonequilibrium_2014,martinez_engineered_2016,li_shortcuts_2017,funo_shortcuts_2020}. For the latter, the term \textit{swift state-to-state transformations} (SST) have been recently proposed~\cite{guery-odelin_driving_2023} to include all the catalogue of different shortcut strategies, since other terms like engineered swift equilibration~\cite{martinez_engineered_2016} or shortcuts to isothermality~\cite{li_shortcuts_2017} are rather restrictive: in general, the shortcuts are neither an equilibration process nor an isothermal process.   

The field of STA and SST is deeply connected with control theory: for the physical system of interest, one aims at connecting given initial and final states by engineering a time-dependent driving---the control function---that gets the desired connection done.  Within the context of the control of Brownian particles,  one assumes most commonly that the controllable parameters characterise the external force or potential acting on the system---for instance, the amplitude and direction of an electric or magnetic field, or the stiffness or centre of an optical trap~\cite{schmiedl_optimal_2007,schmiedl_efficiency_2008,aurell_optimal_2011,aurell_boundary_2012,plata_optimal_2019}. Moreover, in some situations, the whole potential may be regarded as the control function~\cite{muratore-ginanneschi_extremals_2014,muratore-ginanneschi_application_2017,zhang_work_2020,zhang_optimization_2020,plata_taming_2021}. Also, instead of controlling the potential, one may engineer the thermal environment by effectively controlling the time dependence of the bath temperature~\cite{martinez_adiabatic_2015,martinez_colloidal_2017,chupeau_thermal_2018,plata_finite-time_2020,plata_building_2020,prados_optimizing_2021,patron_thermal_2022,ruiz-pino_optimal_2022}.

Once the connection is shown to be feasible, there appears the problem of optimising it in a certain sense~\cite{guery-odelin_driving_2023,blaber_optimal_2023}. Therefore, one enters the realm of optimal control theory~\cite{pontryagin_mathematical_1987,liberzon_calculus_2012}, in which the minimisation of a certain functional of the system's trajectory---entropy production~\cite{aurell_optimal_2011,aurell_boundary_2012,zhang_work_2020,plata_optimal_2019,muratore-ginanneschi_extremals_2014,muratore-ginanneschi_application_2017,plata_taming_2021}, connection time~\cite{prados_optimizing_2021,ruiz-pino_optimal_2022,patron_thermal_2022}, or other figures of merit~\cite{pires_optimal_2023}---is sought after. The optimisation problem is tackled by using the mathematical tools of optimal control theory, specifically Pontryagin's Maximum Principle when non-holonomic constraints are present~\cite{pontryagin_mathematical_1987,liberzon_calculus_2012}. With a few exceptions~\cite{prados_optimizing_2021,ruiz-pino_optimal_2022}, optimal SST protocols in non-equilibium statistical mechanics have only been analysed for connections between equilibrium states.

The Brownian Gyrator (BG) is a paradigmatic example of a system with a non-equilibrium steady state (NESS)~\cite{filliger_brownian_2007,dotsenko_two-temperature_2013,cerasoli_asymmetry_2018,baldassarri_engineered_2020,miangolarra_thermodynamic_2022}. An overdamped Brownian particle is confined in two dimensions  by an elliptic potential and, in addition, its two coordinates $(x,y)$ are submitted to the action of two white noise forces stemming from two thermal baths at different temperatures $T_x$ and $T_y$, respectively.  The two different temperatures make the system reach a NESS in the long-time limit, and the term ``gyrator'' refers precisely to this system presenting a non-vanishing average torque proportional to the temperature difference $|T_x-T_y|$ at the NESS.

From a theoretical perspective, the quadratic dependence of the potential on the system variables $(x,y)$ makes it possible to attack the problem analytically, since the evolution equations for the relevant moments are linear in them, and the probability distribution function remains Gaussian if it is so initially---moreover, the probability distribution function is Gaussian at the NESSs. Also, the linearity of the evolution equations makes it possible to introduce normal modes, which further simplify the mathematical description of this non-equilibrium system. From an experimental perspective, the BG is one of the simplest candidates for designing a non-equilibrium heat engine at the mesoscopic level. In fact, it has recently been realised in actual experiments with different techniques~\cite{ciliberto_heat_2013,chiang_electrical_2017,argun_experimental_2017,cerasoli_spectral_2022}.

SST connections between two NESSs of the BG corresponding to different values of the parameters of the elliptic potential have been shown to be possible~\cite{baldassarri_engineered_2020}, by employing inverse engineering techniques similar to the counterdiabatic method~\cite{deffner_classical_2014,patra_shortcuts_2017,li_shortcuts_2017,tu_stochastic_2014,li_geodesic_2022}. Here, we investigate the optimisation of the SST connection between two NESSs of the BG in the sense of minimising the connection time, i.e. we look into the brachistochrone problem for the BG. Moreover, we prove the existence of a speed-limit inequality for the connection between the two NESSs, which involves the connection time and the irreversible work. This inequality can be regarded as the extension to the connection between NESSs of similar inequalities for the connection between equilibrium states~\cite{sivak_thermodynamic_2012,aurell_refined_2012,dechant_minimum_2022,guery-odelin_driving_2023}. 

Specifically, we work with a BG in which the elliptical potential confining the particle is characterised by two controllable parameters $(k,u)$, which respectively account for the diagonal terms $k$, i.e. we have identical stiffnesses in the $(x,y)$ directions, and for the anti-diagonal terms $u$ coupling the two degrees of freedom. Throughout our whole work, the temperatures $(T_x,T_y)$ of the thermal baths are kept constant.
The problem we address in this paper is the following: we consider two NESSs of the BG, corresponding to different values of the parameters characterising the potential, $(k_i,u_i)$ and $(k_f,u_f)$ for the initial and final states, respectively, { and look for the time-dependent controls $(k(t),u(t))$ that make the connection time between these two {NESSs} minimum. Note that we must have $k_{i,f}>0$ and $|u_{i,f}|<k_{i,f}$ in order to have well-defined NESSs---the two eigenvalues of the quadratic potential are positive, as corresponds to a confining potential. It is physical motivation that encourages our studying the connection between an initial and a target NESSs, since any other state different from a NESS cannot persist in time---although more general SSTs might be considered from a purely mathematical point of view. } 

{In this work, we choose the control parameters of the BG to be bounded by} the non-holonomic constraints $0\leq k\leq k_{\max}$ and $k^2-u^2\geq 0$. {Following the usual terminology of control theory, the set of points in the $(k,u)$ plane verifying the inequalities $0\leq k\leq k_{\max}$ and $|u|\le k$ constitute our \textit{control set}, and we refer to controls belonging to the control set as admissible controls.} In many time-optimisation problems~\cite{chen_fast_2010,ding_smooth_2020,prados_optimizing_2021,ruiz-pino_optimal_2022,patron_thermal_2022}, the control parameters enter linearly in the dynamic evolution equations of the relevant physical observables, thus entailing that Pontryagin's Hamiltonian function is also linear in them. In that case, the optimal protocol leading to the brachistochrone between the initial and target states is usually of \textit{bang-bang} type: it comprises different time windows for which the control parameters are constant and attain their extremum---i.e. maximum or minimum---values {at the boundaries of the control set}. 

The constraints $k\geq 0$ and $k^2-u^2\geq 0$, or $k\geq 0$ and $|u|\leq k$, { mean that we are excluding the possibility of having controls $(k(t),u(t))$ corresponding to repulsive potentials. Points $(k,u)$ over the boundary belonging to the lines $|u|=k$ are \textit{marginally confining}, since at least one of the two eigenvalues of the quadratic potential vanishes. It must be stressed that the boundary lines must be included in the control set to have a mathematically well-defined control problem, since its solution may lie on the boundary---as is the case of bang-bang protocols. This is the reason why the inequalities defining the control set are not strict. From a physical point of view, the inclusion of marginally confining potentials is not problematic: for example, the point $(k,u)=(0,0)$ corresponds to shutting down the potential. } 

The constraint $k\le k_{\max}$ is of practical nature, since  there is typically a maximum value of the stiffness of the trap in actual experiments. {Therefore, we will relax this constraint at certain points to make analytical progress. In other words, we will consider the limit $k_{\max}\to\infty$, which physically corresponds to have a very large, formally infinite, compression power. This limit entails that certain processes, with a characteristic time proportional to $k_{\max}^{-1}$, can be regarded as instantaneous.}

{ The choice of control set plays a key role in posing, and solving, the control problem. It is on a physically motivated basis that our ``natural'' choice described above has been made, but it is not the only possibility. For example, if repulsive potentials do not constitute an experimental problem, one may allow $k(t)$ to become negative or $|u(t)|$ become larger than $k(t)$ during the connection, even if $k_{i,f}>0$ and $|u_{i,f}|<k_{i,f}$. For instance, relaxing these constraints by imposing $-k_{\max}\le (k,u) \le k_{\max}$ would lead to a different, mathematically well-posed, control problem for the same SSTs (connections between two NESSs). From a physical perspective, it is clear that the minimum time for such a less constrained SST would be shorter than (or equal to) the minimum connection time for the more constrained SST. } 

{ On the one hand, the work developed here} shows that there appear two different scenarios for the admissible controls over the brachistochrone: (i) regular bang-bang protocols, in which the controls $(k(t),u(t))$ switch among the three vertices of the triangular control set determined by the non-holonomic constraints $k^2-u^2\geq 0$ and $0\leq k\leq k_{\max}$: $(0,0)$, $(k_{\max},k_{\max})$ and $(k_{\max},-k_{\max})$, and (ii) infinitely degenerate \textit{singular} protocols, which belong to the boundaries of the control set but are more complex. {Note that two of the boundary segments, those joining the vertex $(0,0)$ with either $(k_{\max},k_{\max})$ or $(-k_{\max},-k_{\max})$ are marginally confining. As already stated above, the boundaries must be included in the control set---otherwise, the optimal values would not be attainable. 
On the other hand,} we also analyse the behaviour of the resulting minimum connection time as a function of the chosen initial and target NESS. A very rich phenomenology emerges, including discontinuities in the connection time for incremental changes of the target states, sets of target states that can be instantaneously reached---in the limit $k_{\max}\to\infty$, and sets of states that cannot be reached in a finite time.  

The paper is structured as follows. In Sec.~\ref{sec:model}, we introduce the dynamics of our BG model in detail, within the framework of non-equilibrium statistical mechanics. Section~\ref{sec:speed-limit} puts forward a derivation of a speed-limit inequality for the BG, which involves the connection time and the irreversible work done in the SST connection. In Sec.~\ref{sec:brachistochrone}, we pose the problem of minimising the connection time between NESS by making use of Pontryagin's Maximum Principle and present the main results of the paper. Section~\ref{sec:constructing-bang-bang} is devoted to building the optimal-time bang-bang protocols stemming from Pontryagin's Maximum Principle. The minimum connection time as a function of the initial and target states is analysed in Sec.~\ref{sec:connection-time}.
A brief recap of the main results, together with the main conclusions of this work, is presented in Sec.~\ref{sec:conclusions}. Finally, we relegate most of the rigorous proofs and  technical details to the appendices.

\section{\label{sec:model} Model}

We consider a BG, which can be thought as an overdamped Brownian particle submitted to a two-dimensional confining potential $U = U(\bm{r})$ and in contact with two thermal baths of temperatures $T_x$ and $T_y$, which affect separately each degree of freedom. Hence, the system is described by the Fokker-Planck equation (FPE)
\begin{equation}
    \label{fpe}
    \gamma \,\partial_t P = \nabla_{\bm{r}}^{\sf{T}} \left[P \, \nabla_{\bm{r}} U + k_B \mathbb{T}\, 
 \nabla_{\bm{r}} P\right], \quad \mathbb{T} \equiv \left(\begin{array}{cc}
    T_x & 0 \\
    0 & T_y 
    \end{array}\right),
\end{equation}
where $P=P(\bm{r},t)$ is the probability distribution function for the position of the particle, $\gamma$ the friction coefficient, $k_B$ the Boltzmann constant, and we have introduced the notation
\begin{equation}
    \bm{r}=\begin{pmatrix} x\\ y \end{pmatrix}, \quad \bm{r}^{\sf{T}} = (x,y), \quad \nabla_{\bm{r}}=\begin{pmatrix}\partial_x \\ \partial_y \end{pmatrix}, \quad \nabla_{\bm{r}}^{\sf{T}}=(\partial_x,\partial_y),
\end{equation}
i.e. the superindex $\sf{T}$ in a matrix denotes its transpose. The confining potential is given by
\begin{equation}
\label{potential}
    U(\bm{r}) = \frac{1}{2}\bm{r}^{\sf{T}} \mathbb{K} \bm{r}, \quad \mathbb{K} \equiv \left(\begin{array}{cc} 
    k & u \\
    u & k
    \end{array}\right).
\end{equation}
{ In order to be confining, $(k,u)$ must verify the conditions $k>0$ and $k^2-u^2>0$, which ensure} the existence of a NESS for $T_x\ne T_y$. If $T_x=T_y=T$, the NESS reduces to the canonical equilibrium distribution at temperature $T$. 

The parameters $(k,u)$ characterise the harmonic trap: the parameter $k$ corresponds to the stiffness of the potential, which for the sake of simplicity we assume that is identical in both directions, while $u$ accounts for the coupling between the two degrees of freedom. We assume that both parameters can be externally controlled, so they play the role of control functions in our problem. On experimental grounds, the BG model describes the behaviour of a colloidal particle trapped in an elliptical trap, and simultaneously in contact with two thermal baths at different temperatures in the two orthogonal directions. Such thermal baths have been realised in the laboratory with different experimental setups~\cite{ciliberto_heat_2013,chiang_electrical_2017,argun_experimental_2017,cerasoli_spectral_2022}.

Now let us introduce the normal modes $\bm{q}^{\sf{T}}=(q_1,q_2)$ via the linear transformation
\begin{equation}
    \bm{q} = \mathbb{M}\bm{r}, \quad \mathbb{M} \equiv \frac{1}{\sqrt{2}}\left(\begin{array}{cc}
    1 & 1 \\
    1 & -1
    \end{array}\right).
\end{equation}
Note that $\mathbb{M}$ is a symmetric and orthogonal matrix, its orthogonality entails that the Jacobian of the coordinate transformation equals unity, and thus $P(\bm{q},t) = P(\bm{r},t)$. Therefore, the FPE can now be easily rewritten in terms of the normal modes as
\begin{equation}
    \label{fpe-2}
    \gamma \, \partial_t P = \nabla_{\bm{q}}^{\sf{T}} \left[P \, \nabla_{\bm{q}}U + k_B \tilde{\mathbb{T}}\, \nabla_{\bm{q}} P \right], \quad \tilde{\mathbb{T}} \equiv \mathbb{M} \mathbb{T} \mathbb{M} = \frac{1}{2} \left(\begin{array}{cc}
    T_x + T_y & T_x - T_y \\
    T_x - T_y & T_x + T_y
    \end{array}\right),
\end{equation}
with 
\begin{equation}
    U(\bm{q}) = \frac{1}{2}\bm{q}^{\sf{T}}\tilde{\mathbb{K}}\bm{q}, \quad \tilde{\mathbb{K}} \equiv \mathbb{M} \mathbb{K} \mathbb{M} = \left(\begin{array}{cc}
    k + u & 0 \\
    0 & k - u 
    \end{array} \right).
\end{equation}
{We have natural boundary conditions at infinity, i.e. $P(\bm{q},t)$ vanishes fast enough for $|\bm{q}|\to\infty$.} Although the confining potential may be expressed as the sum of two independent potentials for each normal mode, the new FPE presents a cross-derivative term, which is related to the fact that the noises associated with the normal modes are now coupled. This entails that the long-time state for $T_x\neq T_y$ is a NESS, with a non-zero torque proportional to the temperature difference $|T_x-T_y|$~\cite{filliger_brownian_2007,dotsenko_two-temperature_2013}.

Rather than working directly with the FPE, it is convenient to work with the dynamical equations for the moments of the probability distribution. At the NESS, the probability distribution is Gaussian~\cite{dotsenko_two-temperature_2013,baldassarri_engineered_2020}, and the linearity of the potential entails that the probability distribution remains Gaussian if it is so initially. Since we are interested in the connection between NESSs corresponding to different values of the potential parameters $(k,u)$, the probability distribution is Gaussian for all times and thus it is completely characterised by its first  and second moments, i.e. 
\begin{equation}\label{eq:av-q-Q2-defs}
 \expval{\bm{q}}=\begin{pmatrix}
     \expval{q_1} \\ \expval{q_2}
 \end{pmatrix},
 \qquad
 \expval{\mathbb{Q}_2}\equiv \expval{\bm{q}\bm{q}^{\sf{T}}}=\begin{pmatrix}
     \expval{q_1^2} & \expval{q_1 q_2} \\ \expval{q_2 q_1} & \expval{q_2^2}
 \end{pmatrix},
\end{equation}
respectively. The evolution equations for the moments are
\begin{subequations}\label{eq:BG-evol-eqs-moments}
  \begin{align}
  \label{eq:BG-evol-eqs-momentsA}
    \gamma\, \frac{d}{dt}\expval{\bm{q}} &= - \tilde{\mathbb{K}}\, \expval{\bm{q}}, \\
    \label{eq:BG-evol-eqs-momentsB}
    \gamma\, \frac{d}{dt}\expval{\mathbb{Q}_2} & = -\left\{   \tilde{\mathbb{K}},\expval{\mathbb{Q}_2} \right\} + 2 k_B\tilde{\mathbb{T}},
\end{align}  
\end{subequations}
with $\left\{A,B \right\} \equiv AB + BA$ being the anticommutator between two matrices. { More explicitly, we have that
\begin{subequations}\label{eq:BG-evol-eqs-correl-explicit}
    \begin{align}
  \gamma\frac{d\expval{q_1^2}}{dt}  & = -2(k+u)\expval{q_1^2}+k_B (T_x+T_y),\\
  \gamma\frac{d\expval{q_2^2}}{dt}  & = -2(k-u)\expval{q_2^2}+k_B (T_x+T_y),\\
  \gamma\frac{d\expval{q_1 q_2}}{dt} & = -2 k \expval{q_1 q_2}+k_B (T_x-T_y).
    \end{align}
\end{subequations}
Here, it is where we can appreciate the main advantage of employing the normal modes: had we chosen to work with our original variables $\bm{r}^{\sf{T}} = (x,y)$, the dynamical equations for their respective relevant moments would be coupled. However, the above equations for the second moments $\expval{q_j q_k}$ are not completely uncoupled, it has to be taken into account that the controls $(k,u)$ appearing inside them are the same for all of them. For example, if we looked for an inverse engineering solution,  i.e. for the controls $(k(t),u(t))$ stemming from a given time evolution for the second moments, the three functions $\{\expval{q_1^2}(t),\expval{q_2^2}(t),\expval{q_1 q_2}(t)\}$  would not be independent{---see also next section.}
} 

For time-independent values of $(k,u)$, one readily obtains the values of the moments at the NESS from Eq.~\eqref{eq:BG-evol-eqs-correl-explicit}: 
\begin{equation}\label{eq:stationary-moments}
    \expval{\bm{q}}_{\st} = \bm{0}, \quad \expval{q_1^2}_{\st} = \frac{k_B(T_x + T_y)}{2(k+u)}, \quad \expval{ q_2^2}_{\st} = \frac{k_B(T_x+T_y)}{2(k-u)}, \quad \expval{q_1q_2}_{\st} = \frac{k_B(T_x - T_y)}{2k}.
\end{equation}
As already stated, we would like to connect two NESSs, corresponding to different values of the potential parameters. i.e. we start from the NESS for $(k_i,u_i)$ and would like to end up in the NESS for $(k_f,u_f)$. Therefore, $\expval{\bm{q}}_i \equiv \expval{\bm{q}}(t=0)=\bm{0}$ and then Eq.~\eqref{eq:BG-evol-eqs-momentsA} implies that  $\left<\bm{q}\right>(t)=0$ at all times.  As a consequence, the time evolution of the system is entirely determined by the dynamics of the  covariance matrix $\expval{\mathbb{Q}_2}$ in Eq.~\eqref{eq:BG-evol-eqs-momentsB}, since the probability distribution reads
\begin{equation}
    P(\bm{q},t) = \frac{1}{2\pi \sqrt{\det\expval{\mathbb{Q}_2(t)}}}\text{exp}\left(-\frac{1}{2}\bm{q}^{\sf{T}}\expval{\mathbb{Q}_2(t)}^{-1}\bm{q}\right),
\end{equation}
with $\expval{\mathbb{Q}_2}^{-1}$ being the inverse of the matrix $\expval{\mathbb{Q}_2}$ defined in Eq.~\eqref{eq:av-q-Q2-defs}.

{ \section{Speed limit for the Brownian Gyrator}\label{sec:speed-limit}

Let us consider the energetic balance, in average, for our system. In terms of the normal modes just introduced above, the average energy reads
\begin{equation}
    \expval{U}=\frac{1}{2}(k+u)\expval{q_1^2}+\frac{1}{2}(k-u)\expval{q_2^2}.
\end{equation}
The average energy changes in time because both the controls $(k,u)$  and the variances $\expval{q_j^2}$ vary in time. The first contribution corresponds to the average work done on the system, whereas the second corresponds to the average heat exchanged with the thermal baths:
\begin{equation}
    \frac{d}{dt}\expval{U}=\underbrace{\frac{1}{2}(\dot{k}+\dot{u})\expval{q_1^2}+\frac{1}{2}(\dot{k}-\dot{u})\expval{q_2^2}}_{\expval{\dot{W}}}+\underbrace{\frac{1}{2}(k+u)\frac{d\expval{q_1^2}}{dt}+\frac{1}{2}(k-u)\frac{d\expval{q_2^2}}{dt}}_{\expval{\dot{Q}}}.
\end{equation}

We focus now on the work done on the system in a process where $(k,u)$ are controlled during the time interval $(0,t_f)$:
\begin{equation}\label{eq:evol-av-W}
    \expval{W}=\int_{0}^{t_f} dt\, \left[ \frac{1}{2}(\dot{k}+\dot{u})\expval{q_1^2}+\frac{1}{2}(\dot{k}-\dot{u})\expval{q_2^2}\right]=\Delta\expval{U}-\int_{0}^{t_f} dt\, \left[ \frac{1}{2}(k+u)\frac{d\expval{q_1^2}}{dt}+\frac{1}{2}(k-u)\frac{d\expval{q_2^2}}{dt} \right]
\end{equation}
where $\Delta\expval{U}\equiv \expval{U}(t=t_f)-\expval{U}(t=0)$. Bringing to bear Eqs.~\eqref{eq:BG-evol-eqs-correl-explicit}, we have
\begin{align}\label{eq:k+u-and-k-u}
    k+u &= \frac{k_B(T_x+T_y)}{2\expval{q_1^2}}-\frac{\gamma}{2}\frac{1}{\expval{q_1^2}}\frac{d\expval{q_1^2}}{dt}, &
    k-u &= \frac{k_B(T_x+T_y)}{2\expval{q_2^2}}-\frac{\gamma}{2}\frac{1}{\expval{q_2^2}}\frac{d\expval{q_2^2}}{dt},
\end{align}
which, when inserted into Eq.~\eqref{eq:evol-av-W}, gives
\begin{equation}\label{eq:evol-av-W-v2}
    \expval{W}=\Delta \left[ \expval{U}-\frac{k_B(T_x+T_y)}{4}\ln \left(\expval{q_1^2}\expval{q_2^2}\right) \right]+\frac{\gamma}{4}\int_0^{t_f} dt\, \left[\frac{1}{\expval{q_1^2}}\left(\frac{d\expval{q_1^2}}{dt}\right)^2+\frac{1}{\expval{q_2^2}}\left(\frac{d\expval{q_2^2}}{dt}\right)^2\right].   
\end{equation}

Equation~\eqref{eq:evol-av-W-v2} suggests the definitions
\begin{align}
    \mathcal{F}&\equiv \expval{U}-\frac{k_B(T_x+T_y)}{2}\ln \left(\sigma_1\sigma_2\right), \\
    \expval{W_{\irr}}& \equiv \gamma\int_0^{t_f} dt\, \left[\left(\frac{d\sigma_1}{dt}\right)^2+\left(\frac{d\sigma_2}{dt}\right)^2\right], \label{eq:irr-work}
\end{align}
where
\begin{equation}
    \sigma_j\equiv \sqrt{\expval{q_j^2}}.
\end{equation}
With these definitions,
\begin{equation}
    \expval{W}=\Delta\mathcal{F}+\expval{W_{\irr}}\ge \Delta\mathcal{F}.
\end{equation}
The first contribution $\Delta\mathcal{F}$ to $\expval{W}$ is the variation of the function of state $\mathcal{F}$, which only depends on the initial and final values of the variances $\expval{q_j^2}$. For $T_x=T_y$, $\mathcal{F}$ equals Hemlhotz's free energy for a system of two oscillators; therefore, $\mathcal{F}$ can be considered as its generalisation to a non-equilibrium situation for the BG. The second contribution $\expval{W_{\irr}}$  depends on the protocol employed to connect the initial and final states. In other words, it is a functional of the protocol, which is non-negative and only vanishes when $d{\sigma_j}/dt=0$ for all times---i.e. for an infinitely slow protocol in which the variances have their instantaneous equilibrium values at all times. Therefore, we physically interpret $\expval{W_{\irr}}$ as the \textit{irreversible} contribution to the average work.

Interestingly, the above extension of the irreversible work to the connection between NESSs entails the emergence of a speed-limit for it. By using the Cauchy-Schwarz inequality, we have that
\begin{equation}\label{eq:Cauchy-Schwartz}
    \left| \int_0^{t_f} dt \, \frac{d\sigma_j}{dt} \right|^2 =\left| \sigma_{j,f}-\sigma_{j,i}\right|^2 \leq t_f \int_0^{t_f} dt \, \left(\frac{d\sigma_j}{dt}\right)^2.
\end{equation}
Therefore, we conclude that
\begin{equation}
    t_f \expval{W_{\irr}}\geq \gamma \sum_{j=1}^2 \left| \sigma_{j,f}-\sigma_{j,i}\right|^2 .
\end{equation}
For the connection between NESSs in which we are interested in, the above inequality becomes
\begin{equation}\label{eq:speed-limit}
    t_f \expval{W_{\irr}}\geq \gamma \,\frac{k_B(T_x+T_y)}{2} \left[ \left(\frac{1}{\sqrt{k_f+u_f}}-\frac{1}{\sqrt{k_i+u_i}}\right)^2+\left(\frac{1}{\sqrt{k_f-u_f}}-\frac{1}{\sqrt{k_i-u_i}}\right)^2\right] .
\end{equation}
Equation~\eqref{eq:speed-limit} is a speed limit inequality for the desired connection between two NESSs of our system. Since the right hand side only vanishes for $(k_f=k_i,u_f=u_i)$, this speed limit hints at the existence of a minimum time for the connection.  This is the problem we will address in the following. 

{ Note that the inequality derived above gives a bound for the product $t_f \expval{W_{\irr}}$, so it also hints at the existence of a minimum value of the irreversible work for fixed connection time. Equation~\eqref{eq:speed-limit} resembles recently derived speed limit inequalities for the connection between equilibrium states~\cite{sivak_thermodynamic_2012,aurell_refined_2012,dechant_minimum_2022,guery-odelin_driving_2023}; in fact, it can be considered as their generalisation to non-equilibrium connections. Then, one could expect Eq.~\eqref{eq:speed-limit} to be saturated for the protocol that minimises $\expval{W_{\irr}}$, but this is not the case---even for the completely unconstrained problem where $-\infty\le k,u\le +\infty$. The Cauchy-Schwartz inequality \eqref{eq:Cauchy-Schwartz} is saturated if and only if $\sigma_j$ varies linearly between their fixed initial and final values, which completely determines their shapes. Inserting these $\sigma_j(t)$ into Eq.~\eqref{eq:k+u-and-k-u} gives the corresponding controls $k(t)$ and $u(t)$, so it seems that everything is fine. However, the issue comes from the  non-vanishing cross-correlation $\expval{q_1 q_2}$ for the non-equilibrium case $T_x\ne T_y$: the so obtained controls $k(t)$ and $u(t)$ do not connect the initial and target values of $\expval{q_1 q_2}$. Therefore, the problem of minimising $\expval{W_{\irr}}$ for fixed $t_f$ is far from trivial even for the completely unconstrained problem and, in particular, it is not straightforward that $\expval{W_{\irr}}_{\min}\propto t_f^{-1}$. The consideration of non-holonomic constraints, like the ones introduced in this paper, would further complicate the optimisation problem.

In the brachistochrone problem we address below, the bound provided by Eq.~\eqref{eq:speed-limit} is not tight either. More specifically, our analytical approach will show that $\expval{W_{\irr}}$ increases linearly for large $k_{\max}$, which implies that the bound for the connection time goes to zero---despite the brachistochrone lasting a finite time.}
}

\section{\label{sec:brachistochrone} Control problem for the brachistochrone. Main results }

Let us now pose the control problem we have described in the introduction. Given an initial NESS of the BG, characterised by the control parameters $(k_i,u_i)$, and a final NESS, characterised by the parameters $(k_f,u_f)$, we aim at finding the optimal protocol $(k^*(t),u^*(t))$ that provides the fastest connection between them. In other words, we are interested in obtaining the protocol that connects two arbitrary NESSs in the shortest time $t_f$. 

\subsection{Dimensionless variables. Control set. Natural relaxation time}

{
For our analytical calculations, it is handy to introduce dimensionless variables as follows:
\begin{equation}\label{eq:dimensionless-variables}
    z_{1,2} \equiv \frac{k_i\expval{q_{1,2}^2}}{k_B(T_x + T_y)}, \quad z_3 \equiv \frac{k_i\expval{q_1 q_2}}{k_B(T_x - T_y)}, \quad k^* \equiv \frac{k}{k_i}, \quad u^* \equiv \frac{u}{k_i}, \quad t^* \equiv \frac{k_i}{\gamma}t.
\end{equation}
We are assuming that $T_x\ne T_y$, which is the relevant situation for our purposes: if $T_x=T_y=T$,  the initial and final states would be equilibrium states at the common temperature $T$. From now on, asterisks are dropped not to clutter our formulas.  The $z$ variables in Eq.~\eqref{eq:dimensionless-variables} correspond to the dimensionless relevant moments for the normal modes. 

In dimensionless variables, the evolution equations corresponding to Eq.~\eqref{eq:BG-evol-eqs-correl-explicit} are
\begin{equation}
\label{dynamic-equations}
    \frac{dz_j}{dt} = f_j(\bm{z},\bm{\omega}) \equiv -2\omega_j z_j + 1, 
\end{equation}
where we have defined
\begin{equation}\label{eq:omega-defs}
    \omega_1 = k + u, \quad \omega_2 = k - u, \quad \omega_3 = k.
\end{equation}
We have introduced the compact notation $\bm{z}$ for the moments of the normal modes, such that $\bm{z}^{\sf{T}}\equiv (z_1,z_2,z_3)$. Equation \eqref{dynamic-equations} is analogous to the dynamical equation for the variance of a one-dimensional Brownian particle confined in a harmonic trap, with each component $\omega_j$ of $\bm{\omega}$ (with $\bm{\omega}^{\sf{T}}\equiv (\omega_1,\omega_2,\omega_3)$) playing the role of an effective stiffness of the trap for each relevant moment $z_j$; in fact, equilibrium states fulfill $z_{j,\st}=(2\omega_j)^{-1}$. We must highlight here an interesting feature: due to the linearity of the dynamical equations, regardless of the chosen protocol $(k(t),u(t))$ for the control variables, the time evolution of the system does not depend at all on the values of the bath temperatures $T_x$ and $T_y$, when working with dimensionless variables.

}

\begin{figure}
  \centering\includegraphics[width=0.5\textwidth]{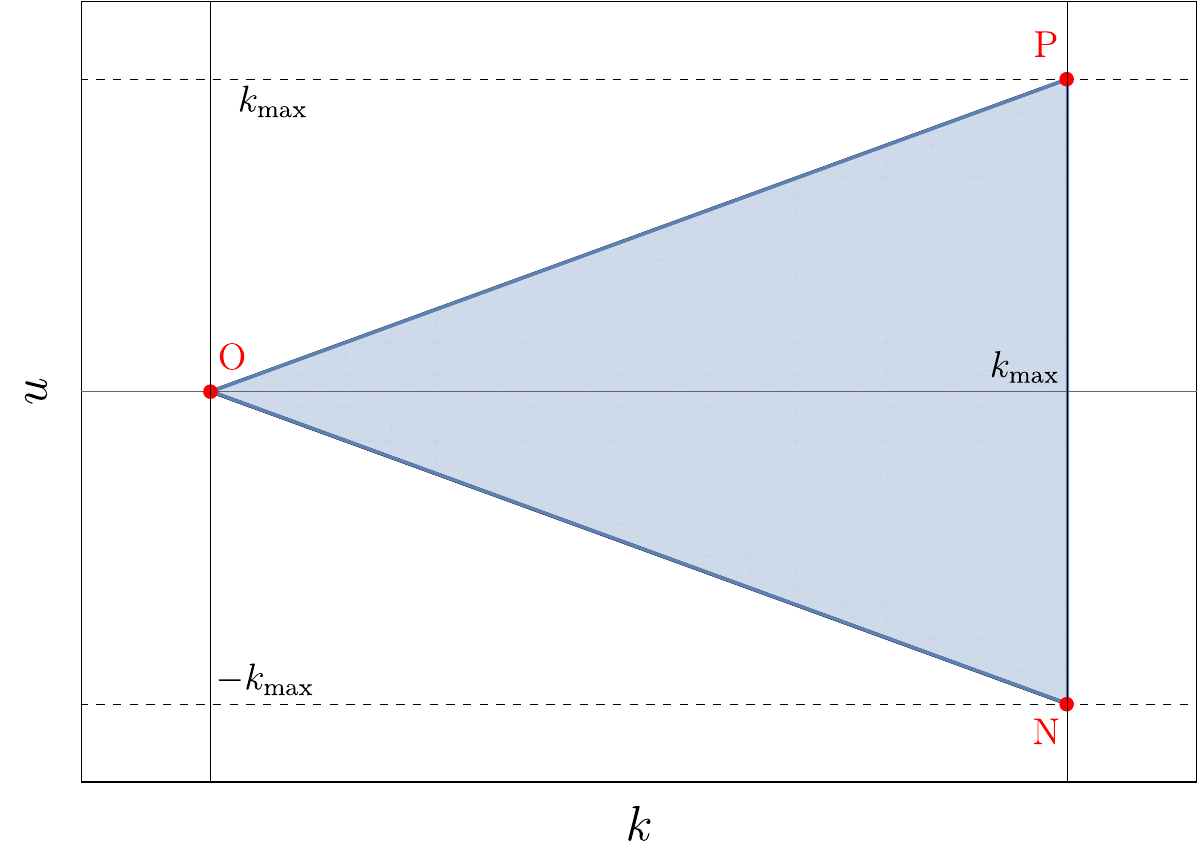}
  \caption{Sketch of the control set in the $(k,u)$ plane. The control set is defined by the inequalities in Eq.~\eqref{eq:control-set}: the blue, triangle-shaped, area constitutes the region where admissible protocols $(k(t),u(t))$ lie. In red, the vertices of the triangle-shaped set have been depicted, which correspond to the points O$ = (0,0)$, P$=(k_{\max},k_{\max})$ and N$=(k_{\max},-k_{\max})$.
  }
  \label{fig:control-set}
\end{figure}

In terms of the dynamical variables $z_j$, we want to drive our dynamical system \eqref{dynamic-equations} with initial condition $z_j(0) = (2\omega_{j,i})^{-1}$ towards a final state with $z_j(t_f) = (2\omega_{j,f})^{-1}$, with $\omega_{j,i} \equiv \omega_j(0)$ and  $\omega_{j,f} \equiv \omega_j(t_f)$, in the fastest way. The connection is made by controlling the time dependence of $(k,u)$. As discussed in the introduction, we assume that (i) the potential is non-repulsive and (ii) the stiffness has an upper bound, for all times. Therefore, we have the non-holonomic constraints
\begin{equation}
\label{eq:control-set}
    -k \leq u \leq k, \quad 0 \leq k \leq k_{\max}.
\end{equation}
{ In the usual terminology of optimal control theory~\cite{pontryagin_mathematical_1987,liberzon_calculus_2012}, Eq.~\eqref{eq:control-set} defines the control set, i.e. the region in parameter space where all admissible controls $(k(t),u(t))$ must lie. In our case, we have the triangular control set depicted in Fig.~\ref{fig:control-set}. We stress that the constraint $-k \leq u \leq k$ ensures that the potential is always confining except for the points over the segments $u=\pm k$, where the potential is marginally confining---with the terminology employed in the introduction. The additional constraint $k\leq k_{\max}$ accounts for the range of possible values of the strength of the confinement, stemming from experimental limitations on the amplitude of an externally induced electric or magnetic field, or the intensity of optical tweezers.}

On general grounds, we expect that the fastest SST protocols achieve the desired connection in a finite time. This entails a huge improvement with respect to the direct STEP process, which consists in switching the control parameters from $(1,u_i)$ to $(k_f,u_f)$ instantaneously at $t=0^+$ and letting them constant afterwards. The relaxation to the final NESS corresponding to $(k_f,u_f)$ is exponential in the relevant moments, Eq.~\eqref{dynamic-equations} tells us that the moment $z_j$ has a characteristic relaxation time $\tau_j\equiv (2\omega_{j,f})^{-1}$.
Therefore,  we can define the characteristic relaxation time $t_{\rel}$ for the STEP process of the BG as
\begin{equation}\label{eq:characteristic-time-trel}
    t_{\rel} = t_{\rel}(k_f, u_f) = \max\left[\frac{1}{2(k_f-u_f)},\frac{1}{2(k_f+u_f)}\right] = \frac{1}{2(k_f-|u_f|)},
\end{equation}
which gives the relaxation timescale for the slowest mode. From an experimental point of view, it is attractive to implement protocols that beat the natural timescale for relaxation, which one may estimate as $3\, t_{\rel}$\footnote{An exponential relaxation of the form $e^{-t/t_{\rel}}$ is completed up to 95\% after $t = 3t_{\rel}$.}. {Still, it must be stressed that the system only reaches the target NESS after an infinite time in the STEP process. Thus any finite-time process represents an improvement from a theoretical point of view.}

\subsection{Pontryagin's Maximum Principle for the time-optimal problem}\label{sec:PMP-time-opt}

Pontryagin's Maximum Principle (PMP) provides necessary conditions for the optimal protocols $(k^*(t),u^*(t))$ that minimise a certain functional of the system variables and the controls. For the brachistochrone, also known as the time-optimal control problem~\cite{pontryagin_mathematical_1987,liberzon_calculus_2012}, the functional to be minimised is 
\begin{equation}
    \label{eq:time-functional}
    J[k,u] \equiv \int_0^{t_f}dt \ \underbrace{f_0(\bm{z},\bm{\omega})}_{=1} ,
\end{equation}
which corresponds to the total time for the protocol. The functional dependence on $(k,u)$ takes place through the effective stiffnesses $\bm{\omega}$ defined in Eq.~\eqref{eq:omega-defs}. For the SST connection, we have the following boundary conditions for the control and the variables
\begin{subequations}\label{eq:bc-control}
\begin{align}
    &(k(0),u(0)) = (1,u_i), & &(k(t_f),u(t_f)) = (k_f,u_f), \label{eq:bc-control-k-u}\\
    &z_j(0) = \frac{1}{2\omega_{j,i}}, &  &z_j(t_f) = \frac{1}{2\omega_{j,f}}, \qquad j = 1, 2, 3. \label{eq:bc-control-zs}
\end{align}   
\end{subequations}
This minimisation has to be done while keeping $(k(t),u(t))$ in the control set, i.e. $(k,u)$ must fullfill the non-holonomic constraints~\eqref{eq:control-set} for all times. Now, we define an additional variable $z_0$, such that  $z_0(0) = 0$ and 
\begin{equation}
    \dot{z}_0 = f_0(\bm{z},\bm{\omega}) \implies  z_0(t_f) = \int_0^{t_f}dt \ f_0(\bm{z},\bm{\omega}) = t_f.
\end{equation}
Next, we introduce the conjugate momenta $(\psi_0,\bm{\psi}^{\sf{T}}) \equiv (\psi_0,\psi_1,\psi_2,\psi_3)$ and Pontryagin's Hamiltonian as
\begin{align}
    \Pi(\bm{z},\psi_0,\bm{\psi},k,u) & = \psi_0 f_0(\bm{z},\bm{\omega}) + \bm{\psi}^{\sf{T}}\bm{f}(\bm{z},\bm{\omega})=\psi_0+\sum_{j=1}^3 \psi_j f_j(\bm{z},\bm{\omega}) \nonumber \\
    &{= \psi_0+\psi_1+\psi_2+\psi_3- 2 k \left(\psi_1 z_1+\psi_2 z_2+\psi_3 z_3\right) - 2 u \left(\psi_1 z_1 -\psi_2 z_2 \right)},
    \label{eq:Pontryagin-Hamiltonian-def}
\end{align}
where $\bm{f}^{\sf{T}}(\bm{z},\bm{\omega}) \equiv (f_1(\bm{z},\bm{\omega}),f_2(\bm{z},\bm{\omega}),f_3(\bm{z},\bm{\omega}))$ is the right hand side of the dynamic equations~\eqref{dynamic-equations}. Note that the Hamiltonian does not depend on $z_0$ by construction.

PMP for our time-optimal problem states the following: let $(k^*(t),u^*(t))$ be the optimal control and $\bm{z}^*(t)$ the corresponding trajectory of the system variables. There exist a time-dependent vector $\bm{\psi}^*(t)$ and a constant $\psi_0^*\le 0$ satisfying $(\psi_0^*,\bm{\psi}^{\sf{T}})\ne (0,0,0,0)$ for all $t\in(0,t_f)$ and having the following properties:
\begin{enumerate}
    \item The variables $\bm{z}$ and their conjugate momenta $\bm{\psi}$ obey the canonical system
    \begin{subequations}\label{eq:canonical}
    \begin{align}
    \dot{z}_j &= \frac{\partial \Pi}{\partial \psi_j} = f_j(\bm{z},\bm{\omega})=-2\omega_j z_j+1, 
    \\
    \dot{\psi}_j &= - \frac{\partial \Pi}{\partial z_j} = 2\omega_j \psi_j,
    \end{align}
    \end{subequations}
    for $j=1,2,3$, with the boundary conditions~\eqref{eq:bc-control-zs}.
    \item For each fixed $t\in(0,t_f)$, the Hamiltonian attains a global maximum at $(k,u)=(k^*,u^*)$ as a function of the controls $(k,u)$, i.e.
    \begin{equation}
        \Pi(\bm{z}^*(t),\psi_0^*,\bm{\psi}^*(t),k^*(t),u^*(t))\ge \Pi(\bm{z}^*(t),\psi_0^*,\bm{\psi}^*(t),k,u),
    \end{equation}
    for all $(k,u)$ in the triangular control set defined by Eq.~\eqref{eq:control-set}.
    \item $\Pi(\bm{z}^*(t),\psi_0^*,\bm{\psi}^*(t),k^*(t),u^*(t))=0$ for all $t\in(0,t_f)$.
\end{enumerate}

Some comments are in order: for the time-optimal problem, it is not necessary to write the evolution equations for $z_0$ and $\psi_0$, which also follow the canonical form, since
\begin{equation}
    \dot{z}_0=\frac{\partial \Pi}{\partial \psi_0}=1, \quad \dot{\psi}_0=- \frac{\partial \Pi}{\partial z_0}=0
\end{equation}
do not provide additional information---the statement of PMP already states that $\psi_0$ is constant, and the equation for $z_0$ simply tells us that $z_0=t$. For {twice} differentiable protocols, PMP is equivalent to solving the control problem using the ordinary tools of variational calculus~\cite{pontryagin_mathematical_1987,liberzon_calculus_2012}. In such scenario,  Pontryagin's Hamiltonian function can be mapped to a Lagrangian function that includes one Lagrange multiplier (the momenta $\psi_j$) for each of the time-dependent constraints given by the dynamical equations \eqref{dynamic-equations} for the variables $z_j$. This variational framework gives rise to the corresponding Euler-Lagrange equations, which are equivalent to the canonical system of PMP. The advantage of PMP relies on the fact that it extends the variational framework by including the possibility of (i) introducing non-holonomic constraints, in the form of inequalities such as those we have in Eq.~\eqref{eq:control-set}, and (ii) considering less regular control functions, which in PMP's approach can have finite-jump discontinuities~\cite{pontryagin_mathematical_1987,liberzon_calculus_2012}. Neither (i) nor (ii) can be tackled with the simpler approach of variational calculus.

PMP entails that we can classify the time-optimal protocols into three distinct categories:
\begin{enumerate}
    \item \textbf{Euler-Lagrange protocols}: They  correspond to situations for which the maximum of $\Pi$ is attained inside the control set, so it is necessary that
    \begin{equation}\label{eq:E-L-eq}
        \frac{\partial \Pi}{\partial k} = 0, \qquad \frac{\partial \Pi}{\partial u} = 0.
    \end{equation}
    These correspond to twice-differentiable functions that might also be determined via the variational calculus approach described above in the open interval $(0,t_f)$. Still, within the PMP framework, sudden changes of the control parameters at both the initial and final times for the protocol, i.e. at $t = 0^+$ and $t=t_f^-$, respectively, are allowed.
    \item \textbf{Bang-bang protocols}: They correspond to situations in which the maximum of $\Pi$ is attained over the boundary of the control set, so that Eq.~\eqref{eq:E-L-eq} no longer holds. Specifically, we refer to a  protocol when the maximum is reached at the vertices of the control set as a bang-bang protocol, as explained below.  
    
    For the case of our concern, the boundaries of the control set correspond to the sides of the triangle depicted in Fig.~\ref{fig:control-set}. The specific point at which $\Pi$ reaches its maximum value depends on the sign of the \textit{switching functions}
    \begin{subequations}\label{eq:switching-functions}
    \begin{align}
        \phi_{\text{OP}}(\bm{z},\bm{\psi})& \equiv \frac{\partial \Pi}{\partial k} + \frac{\partial \Pi}{\partial u}=-2\left(2\psi_1 z_1+\psi_3 z_3\right), 
        \label{eq:switching-functions-OP}   \\
        \phi_{\text{ON}}(\bm{z},\bm{\psi}) &\equiv \frac{\partial \Pi}{\partial k} - \frac{\partial \Pi}{\partial u}=-2\left(2\psi_2 z_2+\psi_3 z_3\right), 
        \label{eq:switching-functions-ON}   \\
        \phi_{\text{NP}}(\bm{z},\bm{\psi}) &\equiv  \frac{\partial \Pi}{\partial u}=2\left(\psi_2 z_2-\psi_1 z_1\right),
        \label{eq:switching-functions-NP}
    \end{align}   
    \end{subequations}
    which give the component of $\nabla\Pi$ along the lines OP, ON, and NP, respectively, where O$\equiv (0,0)$, P$\equiv (k_{\max},k_{\max})$ and N$\equiv (k_{\max},-k_{\max})$ are the vertices of the triangle. The sign of $\phi_{\text{AB}}$ gives the direction in which Pontryagin's Hamiltonian $\Pi$ increases, $\phi_{AB}>0$ means that $\Pi$ increases from A to B. Note that (i) the switching functions only depend on $\bm{z}$ and $\bm{\psi}$, and not on $(k,u)$, because the control functions enter linearly in Pontryagin's Hamiltonian, and (ii) the possible signs of the switching functions are restricted by the relation $\phi_{\text{OP}}-\phi_{\text{ON}}=2\phi_{\text{NP}}$.
    
    At any given time, the signs of the switching functions determine which of the three vertices of the control set gives the maximum value of $\Pi$. Specifically, for $\phi_{\text{OP}}<0$ and $\phi_{\text{ON}}<0$, the maximum is attained at  O, for $\phi_{\text{OP}}>0$ and $\phi_{\text{NP}}>0$,  the maximum is attained at  P, and for $\phi_{\text{ON}}>0$ and $\phi_{\text{NP}}<0$, the maximum is attained at  N. 
    Let us note that the sign of two switching functions determine which vertex of the triangle is chosen: for visualisation purposes, this is schematically represented in Fig.~\ref{fig:switching-funcs-regular}. The times at which these switching functions change their sign determine the switching times between the different vertices during a bang-bang protocol.  
    \begin{figure}
    \centering\includegraphics[width=0.8\textwidth]{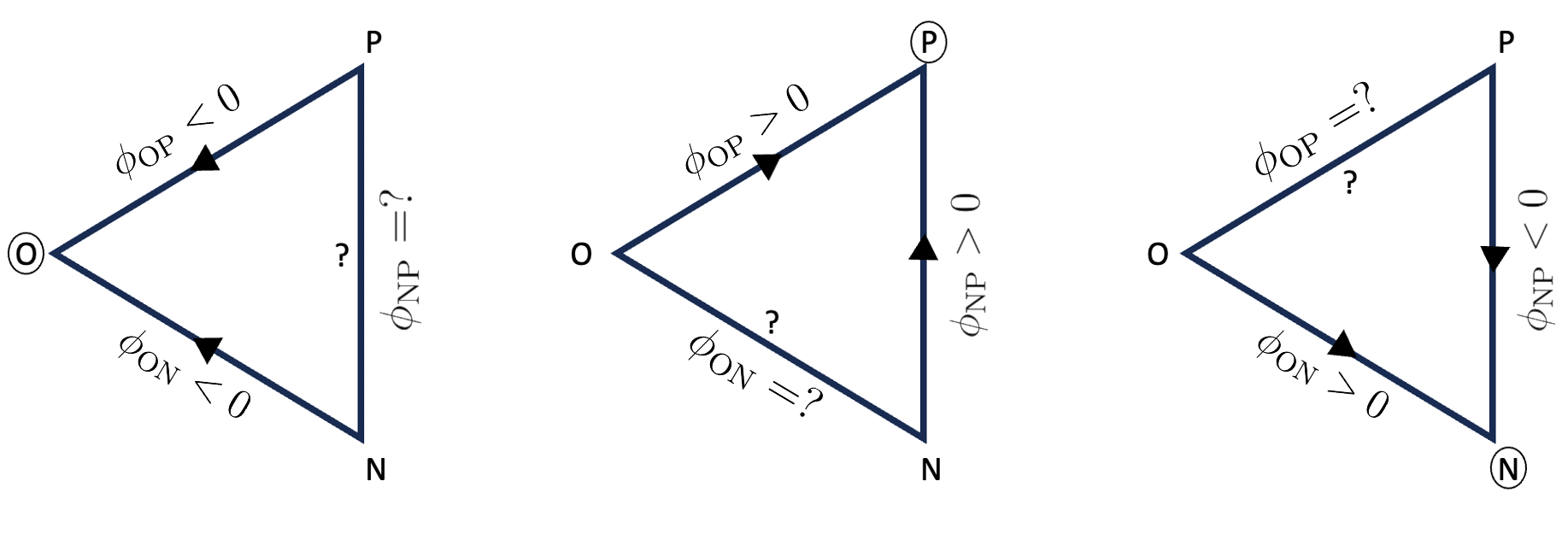}
    \caption{Schematic representation of the different situations for a bang-bang protocol. The arrows mark the direction of the gradient of Pontryagin's Hamiltonian along each of the edges of the triangular set from Fig. \ref{fig:control-set}---i.e. the signs of the switching functions from Eq.~\eqref{eq:switching-functions}. From left to right, the maximum of Pontryagin's Hamiltonian over the boundary is attained at the vertex O, P, and N.
    }
    \label{fig:switching-funcs-regular}
    \end{figure}  
    \item \textbf{Singular protocols}: The above picture for bang-bang protocols breaks down if (at least) one of the switching functions in Eq.~\eqref{eq:switching-functions} vanishes during a finite time interval $[t_1,t_2]$ with $0\leq t_1 < t_2 \leq t_f$. We will refer to these protocols as \textit{singular} protocols: any point along the singular branch---i.e.~the branch of the control set over which the corresponding switching function identically vanishes in a finite interval---is a candidate for the optimal protocol, since Pontryagin's Hamiltonian is constant over the singular branch. Similarly to the bang-bang case, Fig.~\ref{fig:switching-funcs-singular} depicts a schematic representation of the three possible singular situations. 
    \begin{figure}
    \centering\includegraphics[width=0.8\textwidth]{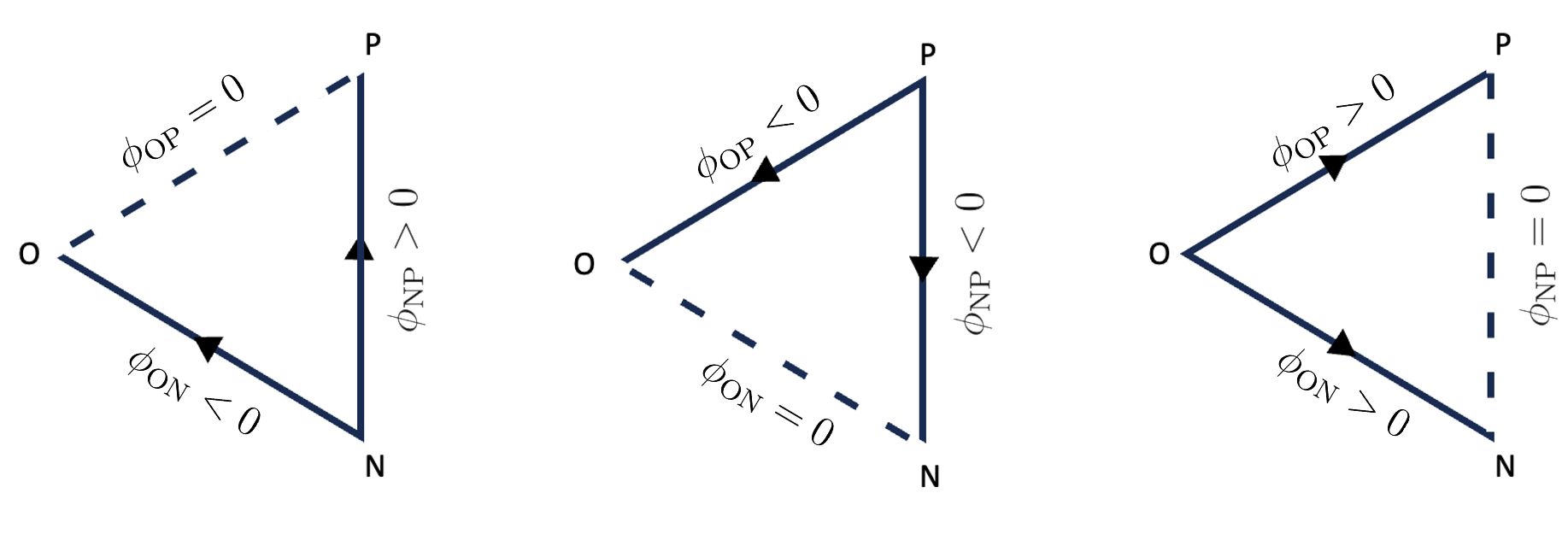}
    \caption{Schematic representation for the different situations for a singular protocol. The dashed lines mark the edge over which Pontryagin's Hamiltonian is constant, due to the vanishing of the corresponding switching function, whereas the arrows have the same meaning as in Fig.~\ref{fig:switching-funcs-regular}, they mark the gradient of Pontryagin's Hamiltonian  along each of the edges of control set. The relation $\phi_{\text{OP}}-\phi_{\text{ON}}=2\phi_{\text{NP}}$ restricts the possible arrow configurations.
    }
    \label{fig:switching-funcs-singular}
    \end{figure}
\end{enumerate}

{ We must emphasise that, although our dynamical system is linear in the control variables, and thus so it is Pontryagin's Hamiltonian function, our system does not belong to the class of what is called a ``linear system'' in optimal control theory. Therein, for dynamical variables $\bm{z}$ and control $\bm{\chi}$,  a system is called linear if the dynamic equations have the form
\begin{equation}\label{eq:linear-opt-control}
    \dot{\bm{z}} = \mathbb{A} \bm{z} + \mathbb{B} \bm{\chi},
\end{equation}
where $\mathbb{A}$ and $\mathbb{B}$ are two matrices  of adequate dimensions with constant elements. It is for this kind of ``linear systems'' that there are theorems ensuring that the optimal-time control problem has a solution of bang-bang type, with as many bangs as dynamical variables $z_j$~\cite{liberzon_calculus_2012,pontryagin_mathematical_1987}~\footnote{This was the case for the thermal brachistochrone of a system of oscillators in dimension $d$, in contact with a unique heat bath with controllable temperature $T(t)$~\cite{patron_thermal_2022}.}.

In the BG, we do not have this structure, due to the products $k z_j$ and $u z_j$ in the dynamical equations~\eqref{dynamic-equations}---which are sometimes called bilinear~\cite{liberzon_calculus_2012}. Therefore, protocols belonging to any of the three categories above, Euler-Lagrange, bang-bang, and singular, are candidates for the brachistochrone. Also, optimal-time protocols could also combine different categories, concatenating time intervals corresponding to different classes of protocols. For example, we could have an optimal protocol starting as an Euler-Lagrange solution, then becoming bang-bang at some time $t_1 \in (0,t_f)$, and then  becoming singular at some other time $t_2 \in (t_1,t_f)$. In principle, PMP does not exclude these combinations for a general problem, so we need to thoroughly study the behaviour of the dynamical system and their conjugate variables in order to check which category combinations are possible in the BG's brachistochrone. 
}

\subsection{\label{subsec:main-results} Main results}

We now put forward the key findings stemming from the application of PMP to the time-optimal problem for the BG. In such presentation, we relegate the mathematical details to the appendices, since here we focus on emphasising the main underlying physical ideas. 

To start with, there is no brachistochrone of Euler-Lagrange type nor mixed protocols involving Euler-Lagrange time intervals.  Therefore, time-optimal protocols must always lie at the boundaries of the control set. As already stated, this is not a direct consequence of the linearity of  Pontryagin's Hamiltonian in the control functions $(k,u)$. A detailed analysis of the possible solutions from Eq.~\eqref{eq:E-L-eq} is required in order to prove this statement, since the BG does not belong to the class of ``linear-Pontryagin'' systems, for which there are theorems ensuring the absence of Euler-Lagrange time-intervals~\cite{liberzon_calculus_2012,pontryagin_mathematical_1987}. The corresponding proof for the inexistence of Euler-Lagrange brachistochrones in the BG can be found in Appendix~\ref{app:Euler-Lagrange}.

Optimal protocols achieving the fastest connection between NESSs can thus be split into two classes:
\begin{enumerate}
    \item  The first class of time-optimal protocols correspond to ``pure'', regular, bang-bang protocols. The term ``pure'' reflects that these bang-bang protocols do not mix with Euler-Lagrange protocols nor with singular protocols. In this case, the bang-bang protocol extends to the whole interval $(0,t_f)$. 
    
    Bang-bang protocols emerge as a combination of different time windows where the control parameters $(k,u)$ switch between the values corresponding to the points O, P, and N. In principle, there are infinitely many ways of combining these points in order to construct a bang-bang protocol. However, as we show later in Sec.~\ref{sec:constructing-bang-bang}, bang-bang protocols involving three time windows suffice for connecting any two NESSs.

    \item The second class of time-optimal protocols correspond to ``pure'' singular protocols, which, again,  extend to the entire time interval $(0,t_f)$. The vanishing of one of the switching functions in the whole interval entails that there are infinitely many optimal protocols $(k^*(t),u^*(t))$ along such a branch, which satisfy the necessary conditions provided by PMP---see Appendix~\ref{app:singular-protocols} for details. 
\end{enumerate}

In the remainder of the paper, we mainly work with regular bang-bang protocols, which can be completely determined using the tools provided by PMP. We will analyse the behaviour of the minimum connection time along these candidates to brachistochrone, as a function of the initial and target states---i.e. $(1,u_i)$ and $(k_f,u_f)$. { Similarly to what we did before with the dimensionless variables, in the following, we drop the asterisks for the optimal control functions and associated optimal variables and conjugate momenta to simplify our notation, since from now on all results refer to optimal protocols.}

\section{\label{sec:constructing-bang-bang} Constructing optimal bang-bang protocols}

We devote this section to building optimal bang-bang protocols with different numbers of bangs---i.e. different number of switchings between the vertices O, P, N of the control set. Here, we use the general notation V to refer to one of these vertices. { We will show that (i) one-bang protocols only allow for connecting the initial point $\textit{I}\equiv (1,u_i)$ with some isolated points $(k_f,u_f)$ within the control set, (ii) two-bang protocols allow for connecting I with some specific curves in the $(k_f,u_f)$ plane, (iii) three-bang protocols allow for connecting I with all the points of $(k_f,u_f)$ within the control set. 
}

As the controls $(k,u)$ take constant values at any of the vertices, the evolution equations~\eqref{dynamic-equations} can be analytically solved. Let us assume that the switching to a certain vertex V occurs at a certain time $t_0$, for which $z_j(t_0)=z_{j,0}$, and denote by $\omega_{j,\! V}$ the value of the $j$-th effective stiffness at the considered vertex; the time evolution of $z_j$ is given by 
\begin{equation}\label{eq:operator-k-finite}
    \mathcal{E}^{(\tau)}_{\omega_{j,\text{V}}}(z_{j,0}) \equiv z_j(t) = \frac{1}{2\omega_{j,\! V}} + \left(z_{j,0}-\frac{1}{2\omega_{j,\! V}}\right)e^{-2\tau\omega_{j,\! V}},
\end{equation}
with $\tau \equiv t - t_0\ge 0$. We have defined the operator $\mathcal{E}^{(\tau)}_{\omega_{j,\text{V}}}(z_{j,0})$, which generates the time evolution of the considered moment $z_j$ during a time interval $\tau$, with constant driving $\omega_{j,\! V}$. This operator has a well-defined  limit for $\omega_{j,\! V} \to 0$,
\begin{equation}\label{eq:operator-k-0}
    \mathcal{E}_0^{(\tau)}(z_{j,0}) = \lim_{\omega_{j,\! V} \to 0}\mathcal{E}^{(\tau)}_{\omega_{j,\text{V}}}(z_{j,0}) = z_{j,0} + \tau,
\end{equation}
$z_j$ linearly increases with time. This will be useful for our analysis, since at each of the vertices (O,P,N) at least one $\omega_{j,\! V}$ vanishes and then the corresponding moment linearly increases. { In fact, at the origin O we have  $\omega_{j,O}=0$ and thus $\mathcal{E}^{(\tau)}_{\omega_{j,\text{O}}}=\mathcal{E}_0^{(\tau)}$, $\forall j$.}

A  bang-bang protocol $\text{V}_1 \text{V}_2 \text{V}_3$\ldots, where $\text{V}_1 \text{V}_2 \text{V}_3$\ldots is any permutation of the vertices of the control set, such that $\text{V}_{\alpha+1}\ne \text{V}_{\alpha}$, is obtained by sequentially applying the time evolution operators corresponding to each of the vertices $\text{V}_{\alpha}$ during a certain time window $\tau_{\alpha}$. The total duration of the bang-bang protocol is thus given by $t_f=\sum_{\alpha} \tau_{\alpha}$, and the final value of the variables $z_j$ is 
\begin{equation}\label{eq:general-bang-bang}
    z_j(t_f) = \underbrace{\left( \cdots \circ \mathcal{E}_{\omega_{j,\text{V}_3}}^{\tau_3} \circ \mathcal{E}_{\omega_{j,\text{V}_2}}^{\tau_2} \circ \mathcal{E}_{\omega_{j,\text{V}_1}}^{\tau_1} \right)}_{{\scriptsize \textnormal{composition of }} M {\scriptsize \textnormal{ operators }}}
\left(z_{j,i}\right), \quad j=1,2,3.
\end{equation}
The mathematical task ahead is looking for a consistent choice of the sequence $\text{V}_1 \text{V}_2 \text{V}_3$\ldots with the right values of the corresponding time intervals $\tau_1$, $\tau_2$, $\tau_3$, \ldots to satisfy the boundary conditions for the moments \eqref{eq:bc-control-zs}, i.e. to drive the system from the initial NESS with $z_j(0)= (2\omega_{j,i})^{-1}$ to the target NESS with $z_j(t_f)= (2\omega_{j,f})^{-1}$.

In order to simplify our analysis, we  consider the limit $k_{\max} \rightarrow +\infty$, i.e. the limit of having infinite capacity for compression. Infinite capacity for compression entails vanishing times $\tau \rightarrow 0^+$ for the windows corresponding to the points P and N, since they involve exponential decays for the dynamic variables with characteristic timescales of the order of $1/k_{\max}$. { See Sec.~\ref{sec:connection-time} for further discussion on the approach of $\tau$ to zero in the limit $k_{\max}\to\infty$, for finite but large $k_{\max}$. As $k_{\max}$ increases, we have that $k_{\max} \tau$ remains finite and this suggests our introducing a parameter $\xi$ that substitutes $\tau$ for determining the optimal control:}
\begin{equation}\label{eq:quenching-factor}
    \xi \equiv  e^{-2k_{\max} \tau} \implies k_{\max} \tau = -\frac{1}{2}\ln \xi \ ,
\end{equation}
with $\xi \in [0,1]$ being the quenching factor corresponding to the relaxing dynamic variable. 
Let us consider the vertex P, thereat we have $\omega_{1,\text{P}}=2 k_{\max}$, $\omega_{2,\text{P}}=0$, $\omega_{3,P}=k_{\max}$. Therefore, for $k_{\max}\to\infty$, while keeping $\xi$ constant, one has the following instantaneous evolutions, from $t_0$ to $t_0^+=t_0+\tau$:
\begin{equation}
    z_1(t_0^+)= z_{1,0}\,\xi^2 , \quad z_2(t_0^+)= z_{2,0}, \quad z_3(t_0^+)=z_{3,0}\,\xi.
\end{equation}
If we define the operator
\begin{equation}
    \tilde{\mathcal{E}}_{\xi}(z_{j,0}) \equiv z_{j,0}\xi,
\end{equation}
the above instantaneous evolutions can be rewritten as
\begin{equation}\label{eq:operator-kmax-inf}
    z_1(t_0^+)= \tilde{\mathcal{E}}_{\xi^2}(z_{1,0}) , \quad z_2(t_0^+)= \tilde{\mathcal{E}}_{1}(z_{2,0}), \quad z_3(t_0^+)=\tilde{\mathcal{E}}_{\xi}(z_{3,0}).
\end{equation}
For the vertex N, the situation is completely analogous, but the evolutions for $z_1$ and $z_2$ are exchanged because $\omega_{1,\text{N}}=0=\omega_{2,\text{P}}$ and $\omega_{2,\text{N}}=2k_{\max}=\omega_{1,\text{P}}$. 

Bang-bang controls in the limit $k_{\max}\to\infty$ are constructed analogously to Eq.~\eqref{eq:general-bang-bang}, but with the operators $\mathcal{E}_{\omega_{j,\text{V}}}^{\tau}$, with V being either P or N, being substituted with the corresponding operator from Eq.~\eqref{eq:operator-kmax-inf}. Our discussion above entails that we thus have a concatenation of operators $\mathcal{E}_0^{(\tau)}$, corresponding to the vertex O, and operators $\tilde{\mathcal{E}}_{\xi^2}$, $\tilde{\mathcal{E}}_{\xi}$, and $\tilde{\mathcal{E}}_{1}$, corresponding to the vertices P or N, in the time evolution of the moments.

\subsubsection{\label{subsubsec:0-switching} One-bang protocols}

This is the most basic case, for which the optimal protocol remains at one of the vertices, O, P or N, during the entire time interval $(0,t_f)$. In the following, we analyse the evolution corresponding to each of the three vertices separately. 

We start our analysis with the vertex O, for which $\omega_{j,\text{O}}=0$ for all $j$. Therefore, we have 
\begin{equation}\label{eq:bang-bang-O}
    z_j(t_f) = \frac{1}{2\omega_{j,f}} =\mathcal{E}_0^{(t_f)}\left(\frac{1}{2\omega_{j,i}}\right) = \frac{1}{2\omega_{j,i}} + t_f, \quad j=1,2,3.
\end{equation}
Note that the above corresponds to an algebraic system of three equations with just one variable; the connection time. Thus, it will only be consistent for specific choices of both the initial and final conditions. Recalling the definition of the $\omega_j$'s, Eq.~\eqref{eq:omega-defs}, the above system presents the following solutions: (i) $t_f = 0$ for $(k_f,u_f) = (1,u_i)$---i.e. the initial and final states are the same, ad thus the system does not evolve, (ii) $t_f \rightarrow + \infty$ for $(k_f,u_f) = (0,0)$---i.e. the target stationary state becomes the point O, {in which case the optimal protocol becomes a direct STEP protocol that involves an infinite time,} and (iii) $t_f = k_f^{-1} - 1$ for $u_f = u_i = 0$ and $k_f < 1${---i.e. connecting uncorrelated states of the BG, for which the dynamic variables degenerate into a unique one.} 

Now, me move onto the analysis of a one-bang protocol at the vertex P. In this case, we have the instantaneous evolution found in Eq.~\eqref{eq:operator-kmax-inf},
\begin{subequations}\label{eq:bang-bang-P}
    \begin{align}
        z_1(0^+) &= \frac{1}{2(k_f+u_f)}=\tilde{\mathcal{E}}_{\xi^2}\left(\frac{1}{2(1+u_i)}\right) = \frac{\xi^2}{2(1+u_i)}, \label{subeq:bang-bang-P-z1}
        \\
        z_2(0^+) &= \frac{1}{2(k_f-u_f)} = \tilde{\mathcal{E}}_{1}\left(\frac{1}{2(1-u_i)}\right) = \frac{1}{2(1-u_i)},
        \\
        z_3(0^+) &= \frac{1}{2k_f} = \tilde{\mathcal{E}}_{\xi}\left(\frac{1}{2} \right) = \frac{\xi}{2}.
    \end{align}
\end{subequations}
Note that $t_f=0$ for this protocol, in the limit $k_{\max}\to\infty$ we are considering. In the above, we have taken into account the explicit expressions for the initial and final values of the moments, as given by Eq.~\eqref{eq:bc-control-zs} with the stiffnesses $\omega_j$ of Eq.~\eqref{eq:omega-defs}. Equation~\eqref{eq:bang-bang-P} presents the following solutions: (i) $\xi = 1$ for $(k_f, u_f) = (1, u_i)$---i.e., again, the initial and final states are the same, and (ii) $\xi = (1+u_i)/(1-u_i)$ for $u_i<0$, which corresponds to a final point P$^*$ with coordinates
\begin{equation}\label{eq:P*}
    k_{\text{P}^*}=\frac{1-u_i}{1+u_i}, \qquad u_{\text{P}^*}=-u_i\, k_{\text{P}^*}, \quad u_i<0.
\end{equation}
The point $\text{P}^*$ is the only non-trivial NESS, i.e. different from the initial state, reachable with a one-bang protocol at vertex P. We recall that $|u_i|\le k_i=1$. 

Lastly, the one-bang N protocol is completely analogous to the P one; we have the same algebraic equations as in Eqs.~\eqref{eq:bang-bang-P}, but upon the variable changes $z_1 \leftrightarrow z_2$, $u_{i,f}\leftrightarrow -u_{i,f}$. Thus, the corresponding solutions in this case are (i) the trivial solution $\xi = 1$, i.e. again $(k_f, u_f) = (1, u_i)$, and (ii) $\xi = (1-u_i)/(1+u_i)$ for $u_i > 0$,  which corresponds to a final point $\text{N}^*$ such that
\begin{equation}\label{eq:N*}
    k_{\text{N}^*}=\frac{1+u_i}{1-u_i}, \qquad u_{\text{N}^*}=-u_i\, k_{\text{N}^*}, \quad u_i>0.
\end{equation}
Note that both $\text{P}^*$ and $\text{N}^*$ tend to the initial point in the limit $u_i\to 0$, i.e. for an uncorrelated initial state.

The symmetry between the P and N protocols will be further exploited when studying higher-order bang-bang protocols in the forthcoming sections.

\subsubsection{\label{subsubsec:1-switching} Two-bang protocols}

Optimal two-bang protocols involve a switching time ${t_1\in (0,t_f)}$ that splits the whole time interval into two different time windows, in which the control parameters correspond to one of the vertices, i.e. O, P, or N. Permutations of two points avoiding consecutive repetition give rise to six possible protocols: OP, ON, PO, NO, PN and NP. These protocols involve the action of two operators of the type \eqref{eq:operator-k-0} or \eqref{eq:operator-kmax-inf} on each of the
dynamic variables. 

The two-bang protocols involve the action of two successive evolution operators, which in turn introduce two unknowns---$(t_f,\xi)$ for the protocols OP, ON, PO and NO, and $(\xi_1,\xi_2)$ for the protocols PN and NP. Since the boundary conditions for the target state give us three algebraic equations, one for each $z_{j,f}$, there will be solutions for the unknowns only for specific choices of the initial and target NESSs. However, for a given fixed initial condition, there will be curves of the form $f(k_f,u_f)=0$ consistent with each of the algebraic equations, instead of the isolated points found in the one-bang case. The analytical expressions for these curves are rather cumbersome and not especially illuminating, thus we relegate them to Appendix~\ref{app:explicit-bang-bang}.

\begin{figure} 
{\centering \includegraphics[width=3.2in]{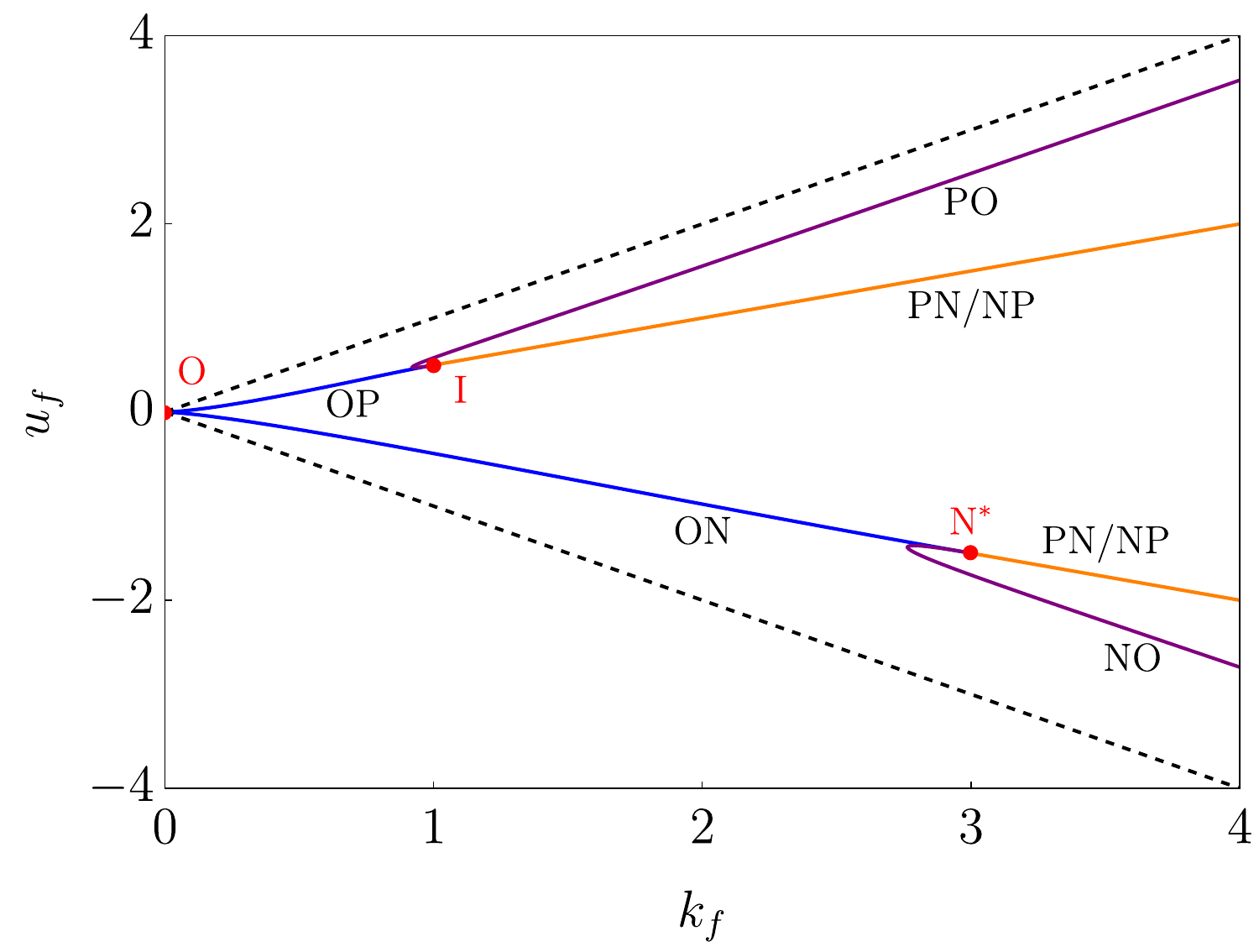} \includegraphics[width=3.2in]{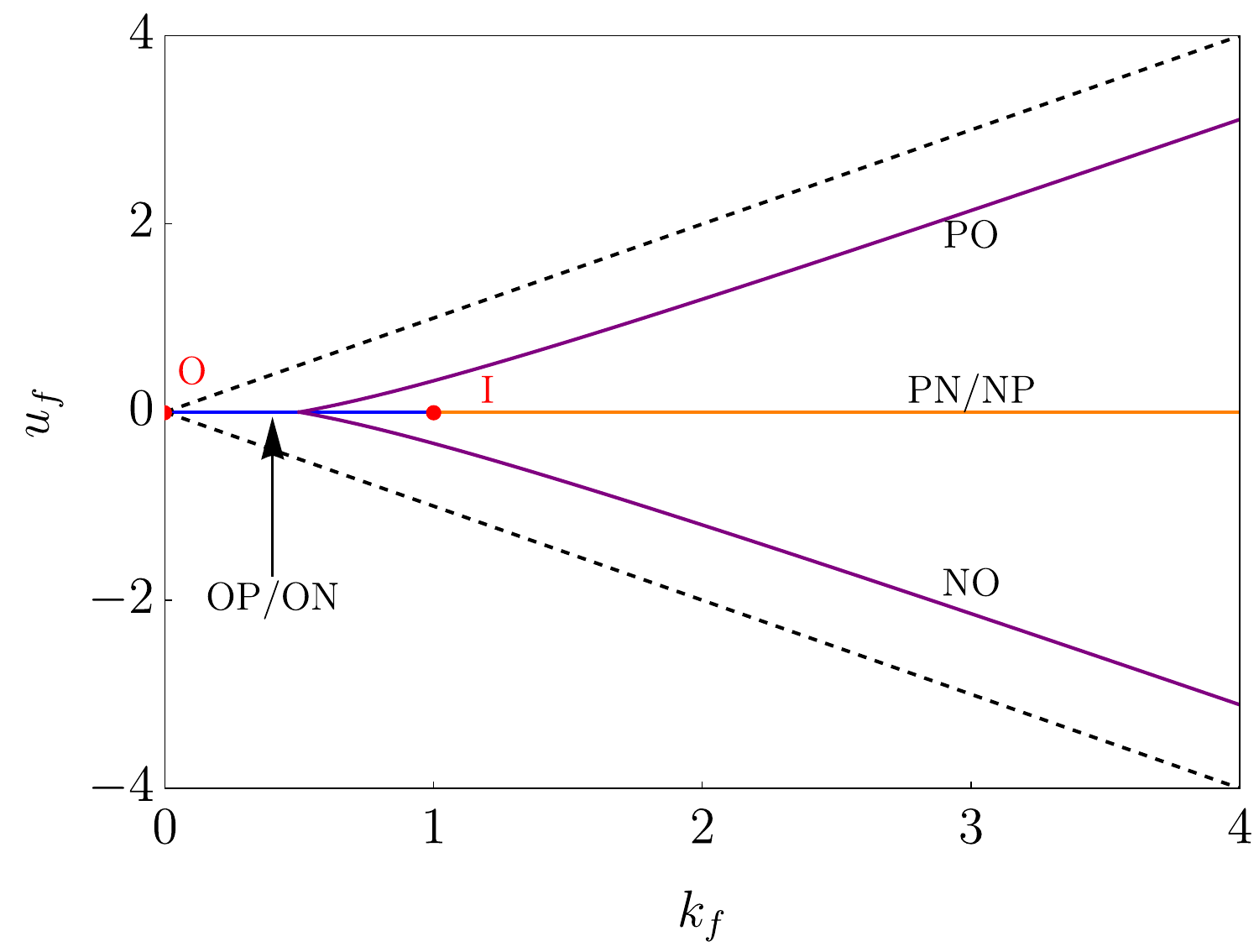} }
\caption{Reachable curves in the $(k_f,u_f)$ plane with two-bang protocols. The curves have been obtained in the infinite capacity of compression, $k_{\max}\to\infty$. Specifically, the left panel corresponds to $u_i=0.5$ and the right panel to $u_i=0$---recall that $k_i=1$ in our dimensionless variables. Dashed lines represent the boundaries of the control set, given by the constraints $k_f>0$, $k_f^2-u_f^2 > 0$. The red points mark the reachable points with one-bang protocols. On the left panel, we have the origin O, the initial state $I=(1,u_i)$, and the point $\text{N}^*$ defined in the text. On the right panel, I and $\text{N}^*$ merge into a unique point, { and the plot is symmetric with respect to the $u_f=0$ axis upon exchanging N$\leftrightarrow$P.}
}
    \label{fig:two-bang-protocols}
\end{figure}

Figure~\ref{fig:two-bang-protocols} shows the curves in the $(k_f,u_f)$-plane that are reachable with two-bang protocols, for two different choices of the initial condition $u_i$. First of all, it is worth highlighting that we have fixed the initial condition $u_i$ to be non-negative, without loss of generality. Due to the symmetry between the P and N vertices, the figure for $u_i<0$ would be obtained by reflecting the figure for $u_i>0$ {(left panel)} through the $u_f=0$ axis, with the exchange N$\leftrightarrow$P in all protocols. This symmetry is particularly evident in the $u_i = 0$ case, whose figure {(right panel)} is symmetric with respect to the $u_f = 0$ axis. We have also chosen the color code of each curve for the same reason: for example, the blue curves, corresponding to the protocols OP and ON, are exchanged upon the flipping of the figure, and they merge for the case $u_i = 0$. In fact, they become a one-bang protocol at vertex O, since for $u_i=u_f=0$, the operation of P or N during the two-bang protocol becomes the identity. The same colour code applies for the purple curves, corresponding to the protocols PO/NO, and the orange ones, accounting for the PN/NP protocols. Moreover, we have also marked with red dots the points reachable with a one-bang protocol, described in Sec.~\ref{subsubsec:0-switching}. On the left panel, we have the origin O, the initial state $(1, u_i)$, and the point $\text{N}^*$, while on the right panel we also have the origin O, but the initial state and the point $\text{N}^*$ have merged into a unique point. We note that these points constitute limiting values for all the two-bang protocols: OP starts from the origin O and ends at the initial state I, ON starts at O and ends at the point $\text{N}^*$, both PO and PN/NP start at I, and both NO and PN/NP start at the point $\text{N}^*$. These are not coincidences: the one-bang solutions must belong to a subgroup of the two-bang ones, they are only attained when (at least) one of the operators corresponding to one of the time windows of the considered two-bang protocol becomes the identity operator. For instance, let us consider the two-bang protocol ON. When the operator corresponding to the N time window becomes the identity, the ON protocol reduces to the O protocol that makes it possible to reach the origin; when the operator corresponding to the O time window becomes the identity, the ON protocol reduces to the N protocol that makes it possible to reach the point N$^*$.

\subsubsection{\label{subsubsec:2-switching} Three-bang protocols}

The last case we look into is three-bang optimal protocols, which split the time interval $(0,t_f)$ into three time windows with two switching times. In each time window, the controls $(k,u)$ attain the values corresponding to one of the three vertices: O, P, and N. Permutations of these points---eliminating those having two identical consecutive vertices---give rise to twelve three-bang protocols: OPO, POP, ONO, NON, OPN, ONP, PON, NOP, PNO, NPO, PNP, NPN. For the sake of brevity, as we did in the previous section, the specific algebraic systems of equations associated to these candidates to optimal-time protocol are not given here---see Appendix~\ref{app:explicit-bang-bang} for details. As clearly observed there, the symmetry between P and N entails that many of the aforementioned three-bang protocols have the same algebraic systems; specifically the pairs (OPN,ONP), (PNO,NPO), and (PNP,NPN). Thus, they allow to reach the same final states with the same connection time, so in practice we only need to deal with nine different three-bang protocols. 

{  All the above mentioned three-bang protocols are regular---i.e. their switching times are completely determined via the switching functions from Eq.~\eqref{eq:switching-functions}, except for the protocols POP and NON. In Appendix~\ref{app:singular-bang-bang}, we prove that the POP and NON protocols are singular: in order to have such a protocol, the corresponding switching function ($\Phi_{\text{OP}}(t)$ for POP, $\Phi_{\text{ON}}(t)$ for NON) must vanish over the whole time interval $[0,t_f]$. Recalling our definition of singular protocols in Sec.~\ref{sec:PMP-time-opt}, Pontryagin's Hamiltonian is constant over the corresponding edge (OP or ON). All sequences of points on the edge connecting the initial and target states give the same connection time. Thus, the optimal protocol becomes degenerate and there is a continuous family of singular protocols---as discussed in Appendix~\ref{app:singular-protocols}. In this section, we stick to the POP and NON protocols as representative members of these singular families, because they have the same structure of the remainder of regular three-bang protocols.}

As was the case for one- and two-bang protocols, the evolution equations provide three algebraic equations from the three boundary conditions at the final time for the moments $z_{j,f}$ in the target NESS. Since the three-bang protocols introduce three unknown parameters, this algebraic system provides a solution for them and thus univocally characterises the three-bang protocols. However, the fact that the ``quenching'' parameters for the P and N vertices must verify $0\le \xi \le 1$ and that the lengths $\tau$ of the time windows for the O vertex must verify $\tau \ge 0$ impose restrictions that make each V$_1$V$_2$V$_3$ protocol reach a certain region of the $(k_f,u_f)$ plane---given an initial condition $u_i$. 

\begin{figure}
\centering\includegraphics[width=0.7\textwidth]{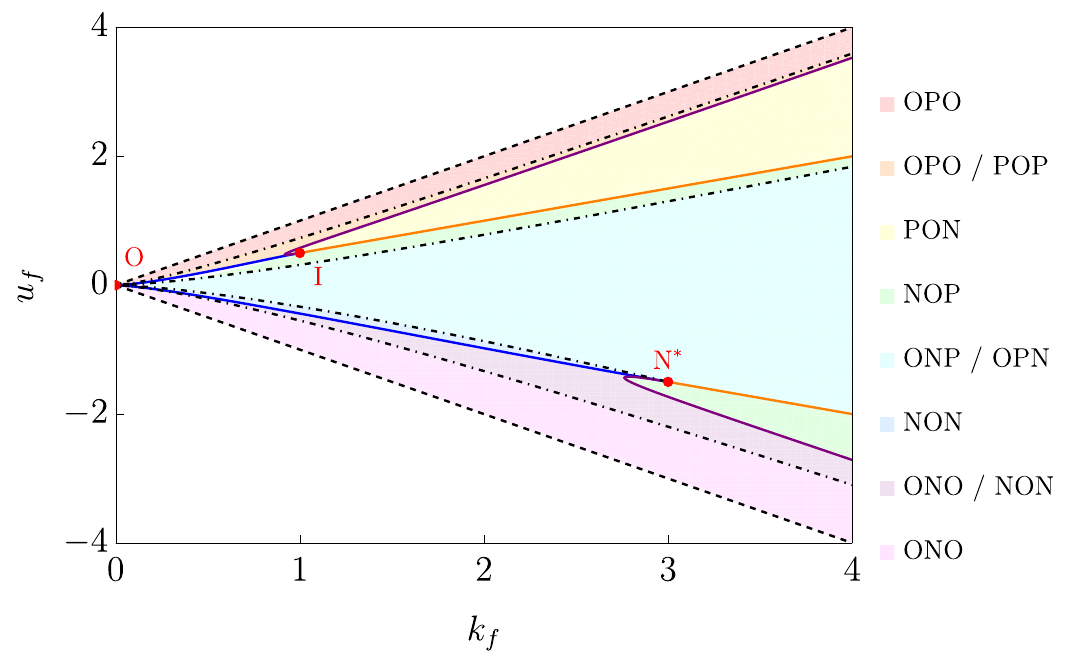}
  \caption{{
  Reachable regions in the $(k_f, u_f)$-plane by means of  optimal three-bang protocols, from an initial NESS with $u_i = 0.5$. The regions are labelled in the legend, ordered from top to bottom. The red points and the solid curves are the same ones as in the left panel of Fig.~\ref{fig:two-bang-protocols}. Specifically, the solid curves correspond to the lines reachable by two-bang protocols. The additional dot-dashed curves also mark the boundary of certain regions reachable by three-bang protocols, over these curves the three-bang protocols cease to exist but they do not reduce to two-bang protocols.}
  }
  \label{fig:three-bangs-uipos}
\end{figure}

\begin{figure}
\centering\includegraphics[width=0.7\textwidth]{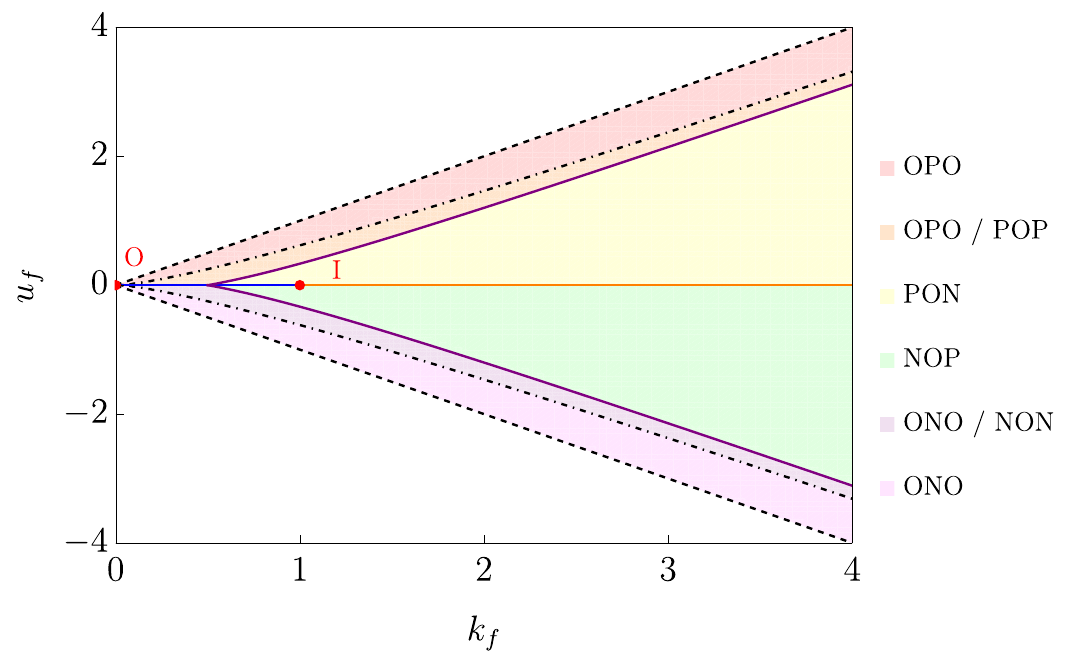}
  \caption{Same as in Fig.~\ref{fig:three-bangs-uipos}, but for a different initial state, now corresponding to $u_i=0$. { Again, note the symmetry with respect to the $u_f=0$ axis, upon exchanging N$\leftrightarrow$P.}
  }
  \label{fig:three-bangs-ui0}
\end{figure}

Figures~\ref{fig:three-bangs-uipos} and \ref{fig:three-bangs-ui0} corroborate the insights above for the same values of the initial condition considered in Fig.~\ref{fig:two-bang-protocols}: $u_i = 0.5$ for Fig.~\ref{fig:three-bangs-uipos}, and $u_i = 0$ for Fig.~\ref{fig:three-bangs-ui0}. In both figures, we plot the $(k_f, u_f)$ regions attained via three-bang protocols where the connection time is minimum. {This is an important clarification, since there are several regions corresponding to different three-bang protocols that intersect, as it is the case 
for the NOP and ONP/OPN protocols in Fig. \ref{fig:three-bangs-uipos}, where the NOP region (green) also belongs to the ONP/OPN one (cyan), but NOP becomes the optimal protocol in that common region because its connection time is shorter.} That is the main reason why the PNO/NPO protocols neither appear in Fig. \ref{fig:three-bangs-uipos} nor in Fig. \ref{fig:three-bangs-ui0}: they are always suboptimal, i.e. give longer connection times, with respect to other {coexisting} bang-bang protocols.

The curves corresponding to two-bang protocols from Fig.~\ref{fig:two-bang-protocols} play for three-bang protocols a role similar to that played by the points corresponding to one-bang protocols for two-bang ones: the curves from  Fig.~\ref{fig:two-bang-protocols} delimit the boundaries of the regions reachable by the different three-bang protocols. This is again logical: two-bang protocols are a limiting case for three-bang protocols~\footnote{In terms of the algebraic systems of equations displayed in Appendix~\ref{app:explicit-bang-bang} for the three-bang protocols, which are characterised by three variables, it is equivalent to the case in which at least one of such variables attains its lower or upper bound---if such variable corresponds to the length of a time window $\tau$, its lower bound would be $\tau = 0$, while if it corresponds to a quenching parameter $\xi$, its lower and upper bounds correspond to $\xi = 0$ and $\xi = 1$, respectively.}. However, there are additional dot-dashed curves that also delimit regions reached by three-bang protocols and do not correspond to two-bang protocols. At these dot-dashed curves, the algebraic systems for the three-bang protocols cease to have physical solutions because one (or more) of the unknown parameters become complex.

It is worth emphasising that three-bang protocols already make it possible to reach any point in the $(k_f, u_f)$-plane---thus any target NESS of the BG--- within the boundaries of the control set from Eq.~\eqref{eq:control-set}. In principle, we could keep constructing higher-order-bang protocols, but on a physical basis one expects the three-bang protocols to give the minimum connection time, since they comprise the minimum number of time windows. In this sense, it must be stressed that we have not been able to found shorter connections with a higher number of bangs~\footnote{Still, from a strict mathematical viewpoint, we cannot assure that higher-order-bang protocols are suboptimal against our three-bang solutions.}

We devote the forthcoming section to the detailed study of the connection time for the optimal three-bang protocols presented in this section. Interestingly, these ``simple'' protocols already beat the characteristic relaxation timescale of the dynamics for certain sets of initial and final states.


\section{\label{sec:connection-time} Minimum connection time as a function of the target state}

In the previous section we have shown that, given a fixed initial state $\text{I}=(1,u_i)$ in our dimensionless variables, we are able to reach any final NESS of the BG, characterised by one point in the $(k_f,u_f)$-plane within the control set, by means of a three-bang protocol. Now, we turn our attention into the original matter of concern, which is the minimum connection time between two NESSs. Therefore, we study here the behaviour of the connection time over the aforementioned three-bang protocols, candidates to be the brachistochrone, as a function of the target state.
\begin{figure} 
{\centering \includegraphics[width=3.5in]{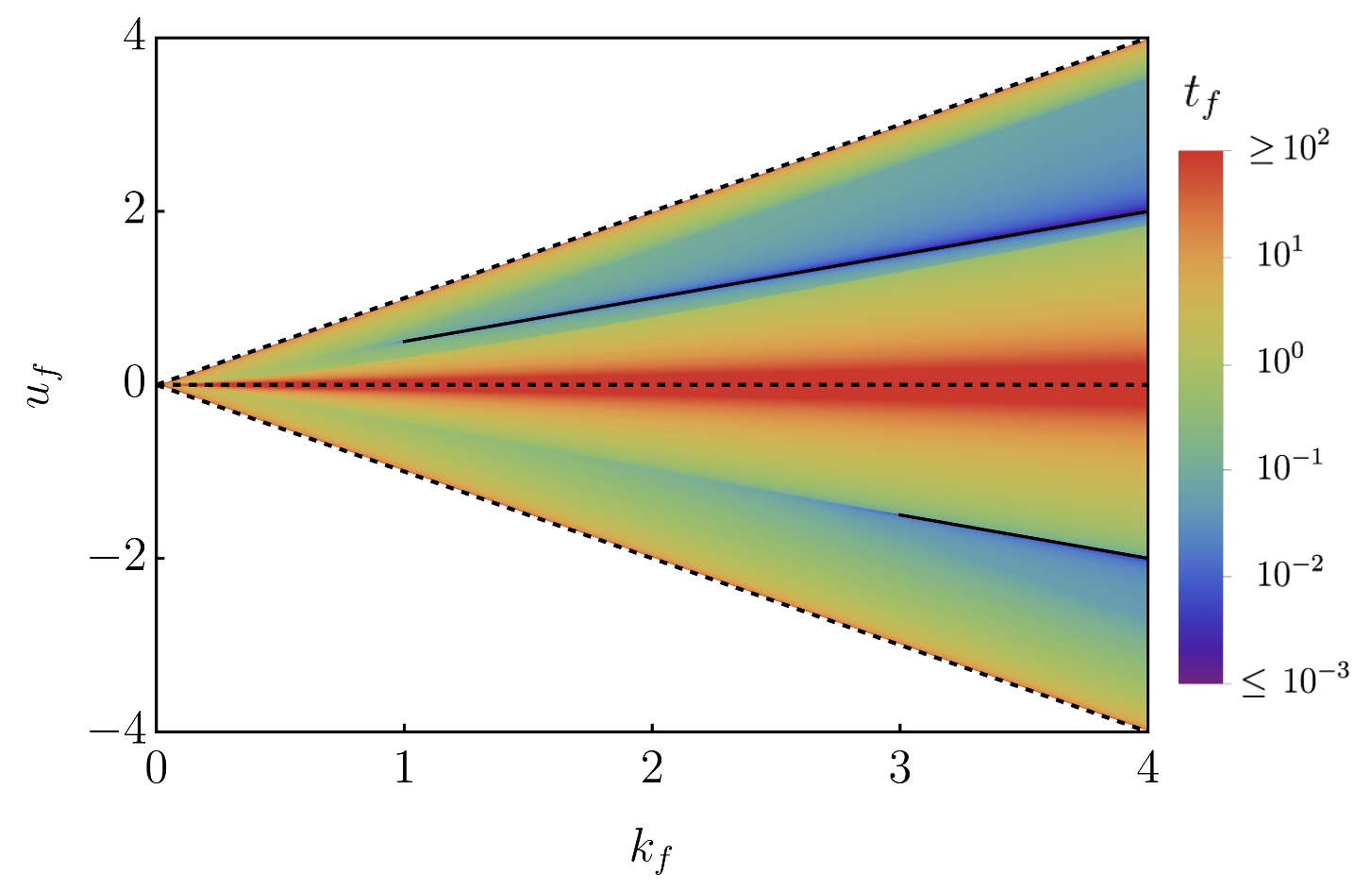} \includegraphics[width=3.5in]{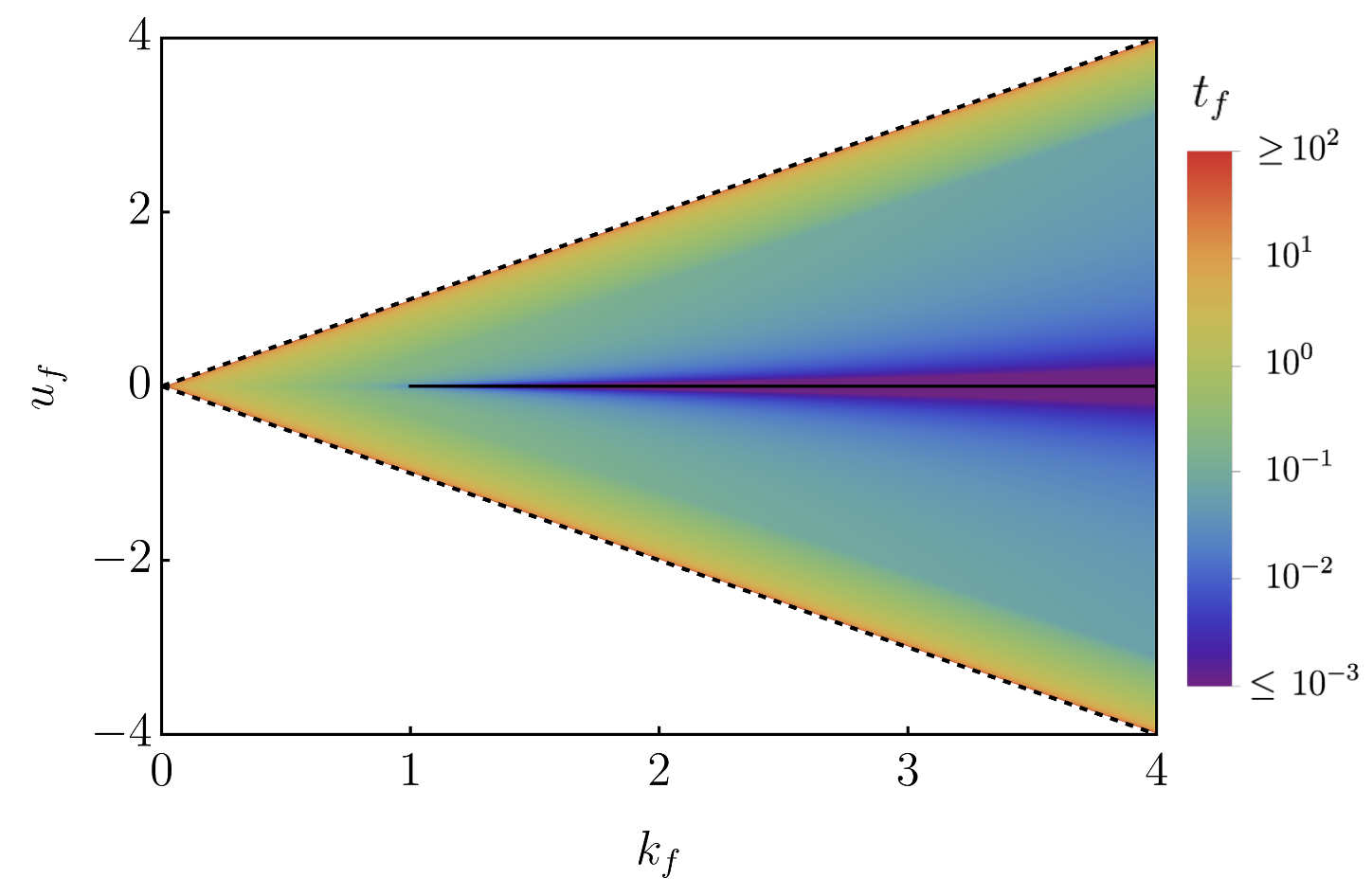} }
\caption{Density maps of the minimum connection time $t_f$ on the phase-plane $(k_f,u_f)$. Note the logarithmic scale for $t_f$. Similarly to Fig.~\ref{fig:two-bang-protocols}, we have considered different initial states in the left and right panels: $u_i = 0.5$ (left) and $u_i = 0$ (right); recall that $k_i=1$ in our dimensionless units. For the left panel, the corresponding optimal three-bang protocols providing the plotted minimum connection time $t_f$ are shown in Fig.~\ref{fig:three-bangs-uipos}; for the right panel, in Fig.~\ref{fig:three-bangs-ui0}. { Dashed lines represent the sets of points at which the connection time diverges, whereas the black solid ones account for the points with null connection time in the $k_{\max}\to +\infty$ limit. Note that the time scale has been chosen to be equal in both panels.} 
}
    \label{fig:heat-maps-nonull-ui}
\end{figure}

Figure~\ref{fig:heat-maps-nonull-ui} shows density maps for the connection time $t_f$ over each of the optimal three-bang protocols built in Sec.~\ref{subsubsec:2-switching}. The behaviour of the minimum connection time $t_f$ radically changes from the $u_i \neq 0$ case to the $u_i = 0$ case. The most remarkable change refers to the final states lying on the line $u_f = 0$. On the one hand, for $u_i = 0$, points lying on the $u_f = 0$ line are reached in (i) a finite time for $k_f < 1$ and (ii) in a null connection time for $k_f > 1$~\footnote{If $k_{\max}$ were finite, the corresponding minimum connection time would be of the order of $1/k_{\max}$.}. { The case $u_i=u_f=0$ is equivalent to connecting equilibrium states of two uncoupled oscillators with different bath temperatures $T_x\ne T_y$, for which the minimum connection time is explicitly derived in Appendix~\ref{app:uncoupled-oscillators}.} On the other hand, for $u_i \neq 0$, states lying in the line $u_f = 0$ can only be reached in an infinite time. In Appendix~\ref{app:explicit-bang-bang}, we show that
\begin{equation}\label{eq:diverging-1}
    t_f \propto u_f^{-2}, \qquad |u_f| \ll 1, \quad u_i \neq 0,
\end{equation}
which correspond to the relevant OPN/ONP three-bang protocols for $|u_f| \ll 1$.

The above has striking implications. From a physical standpoint, if we initially start from a state in which the two coordinates of the BG are uncoupled, $u_i = 0$, which is equivalent to having two independent oscillators at equilibrium with their respective baths at temperatures $T_x$ and $T_y$, any NESS of the BG, with $u_f\ne 0$ can be reached in a finite time. However, had we started from an initial NESS with $u_i\ne 0$, it would take an infinite amount of time to reach any uncoupled final state with $u_i=0$. Hence, the candidates OPN/ONP to optimal-time SST protocol  would present no advantage with respect to the direct STEP protocol in this case, {which also connects the initial and target states in an infinite time. The inability to extinguish correlations in a finite time is a, unexpected a priori, physically interesting result. It is relevant to keep in mind that, as stated in the introduction, the control set plays a crucial role in the solution of the control problem, so it may be possible to find protocols able to extinguish correlations in a finite time by considering a larger, less constrained, control set.}

Apart from the $u_f = 0$ line, the connection times also diverge along the $u_f = \pm k_f$ lines, which bound the control set of admissible controls, both for $u_i=0$ and $u_i\ne 0$. Similarly to the previous analysis carried out for the $u_f = 0$ line, asymptotic expressions for the minimum connection time along the lines $u_f = \pm k_f$ can be derived from the algebraic systems that we explicitly write in Appendix~\ref{app:explicit-bang-bang}. Specifically, along the $u_f = k_f$ line, the OPO protocol gives
\begin{equation}\label{divergence-2}
    t_f(k_f,u_f) = \frac{1}{2(k_f - u_f)} - \frac{1}{2(1-u_i)} \sim \frac{1}{2(k_f - u_f)}, \qquad k_f-u_f \ll 1-u_i
\end{equation}
while for the $u_f = -k_f$ line, the ONO protocol gives
\begin{equation}\label{divergence-3}
    t_f(k_f,u_f) = \frac{1}{2(k_f + u_f)} - \frac{1}{2(1+u_i)} \sim \frac{1}{2(k_f + u_f)}, \quad k_f+u_f \ll 1+u_i.
\end{equation}
Let us note that the minimum times in Eqs.~\eqref{divergence-2} and \eqref{divergence-3} have the same divergence of the characteristic relaxation time for the STEP process in Eq.~\eqref{eq:characteristic-time-trel}, which entails this property to be a consequence of the physical bounds of our model: the fact that $k_f^2 - u_f^2 > 0$ for the system to attain a NESS.
\begin{figure}[h!]
{\centering \includegraphics[width=3.3in]{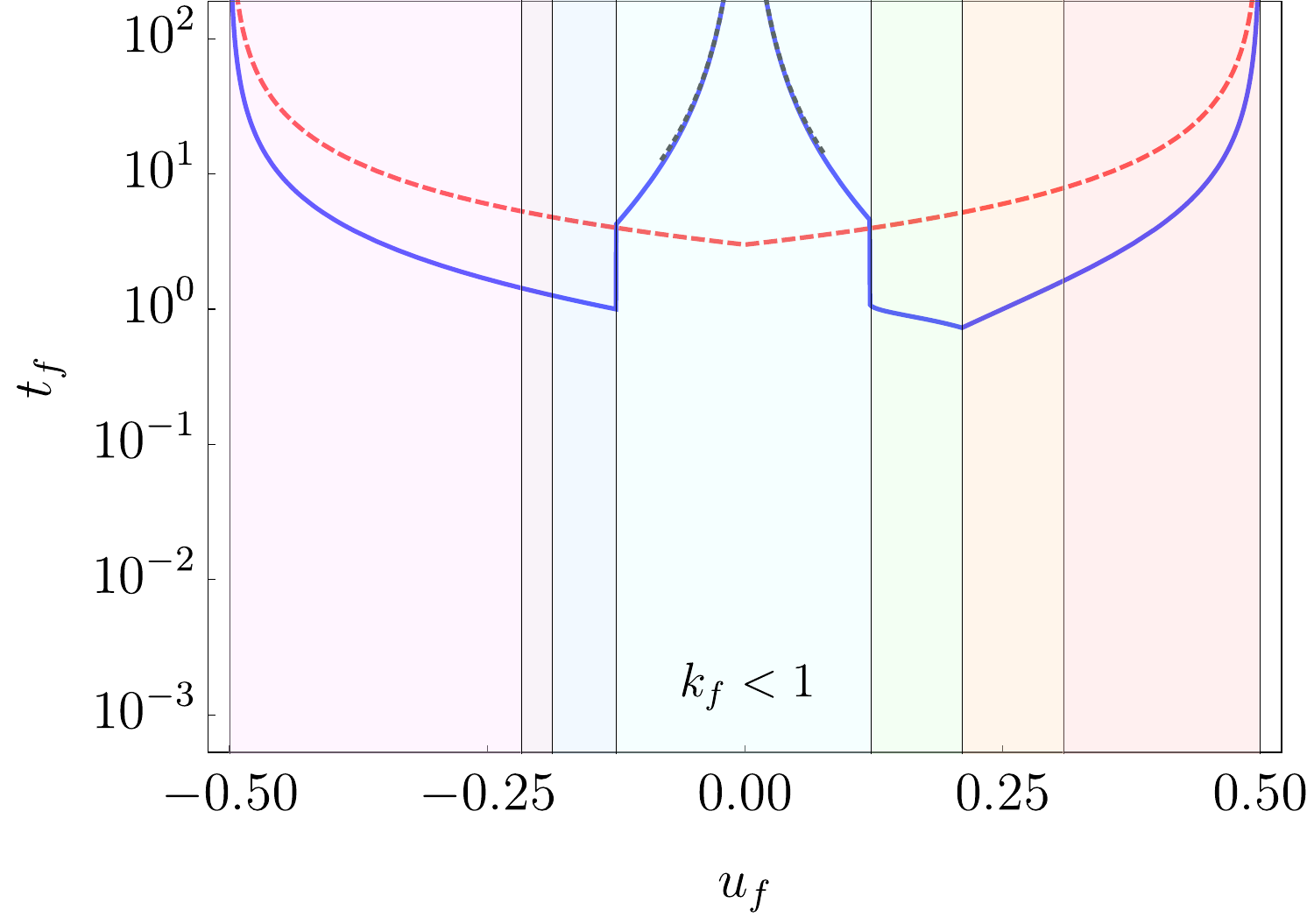} \includegraphics[width=3.3in]{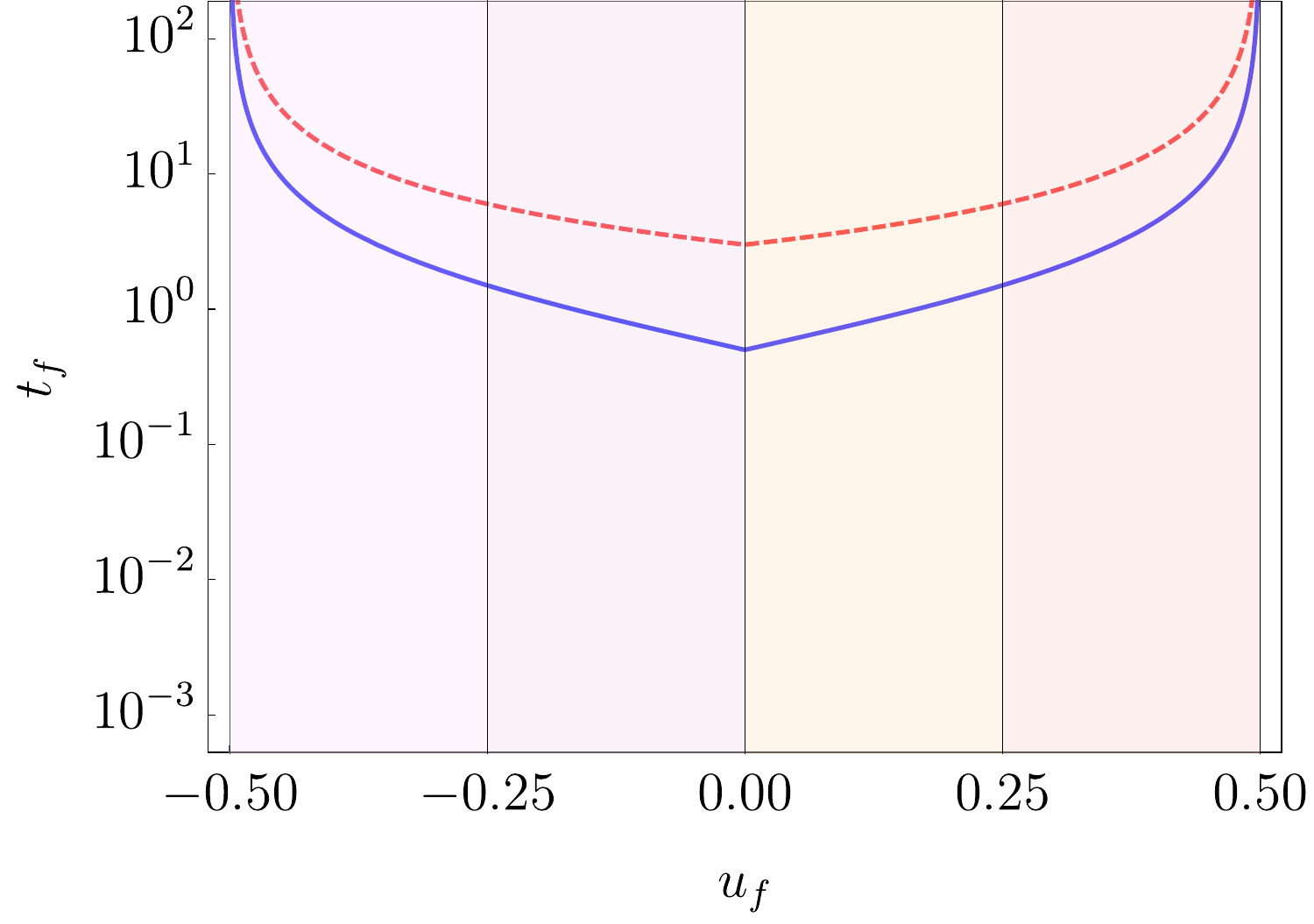}
\\
\includegraphics[width=3.3in]{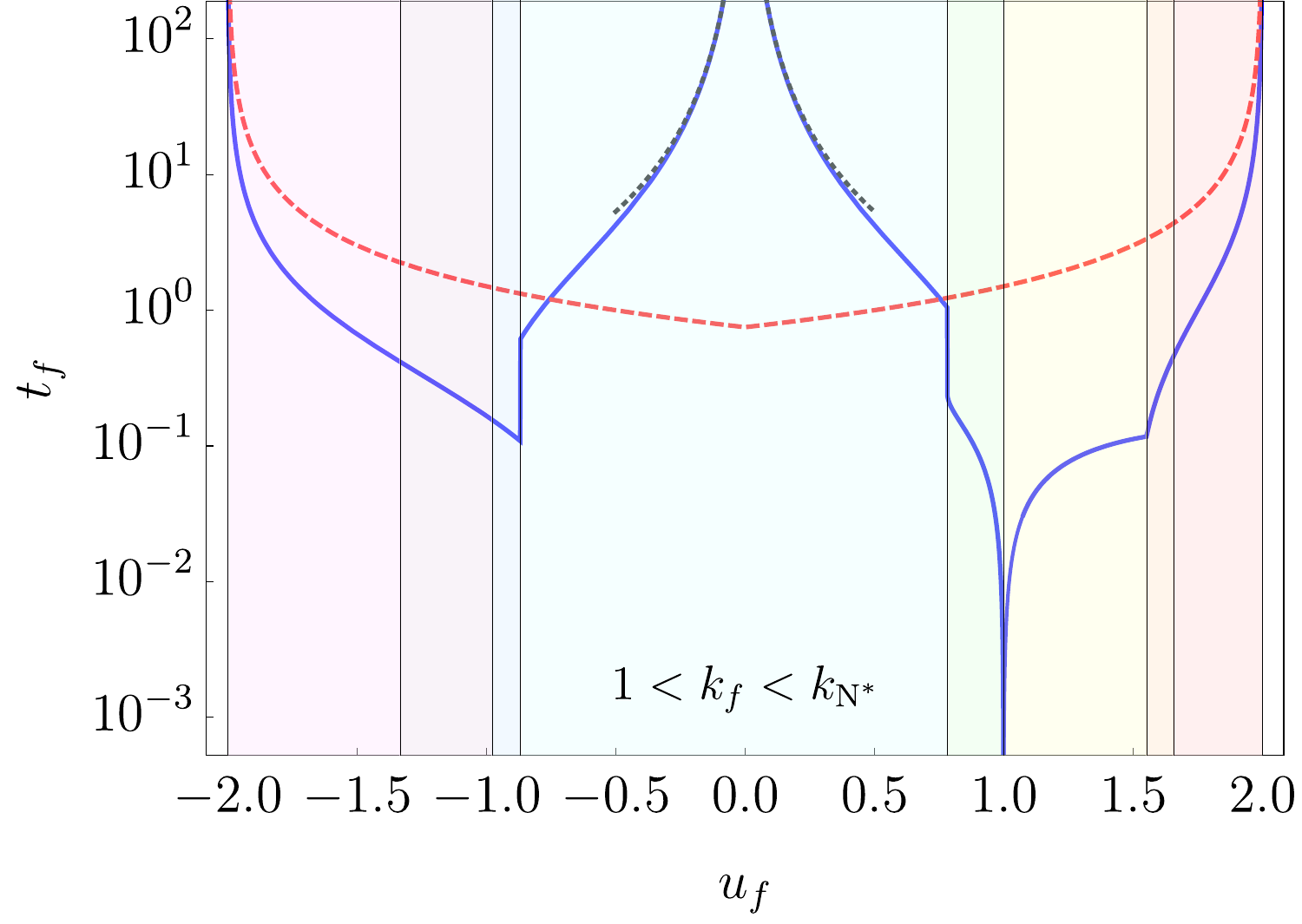}
\includegraphics[width=3.3in]{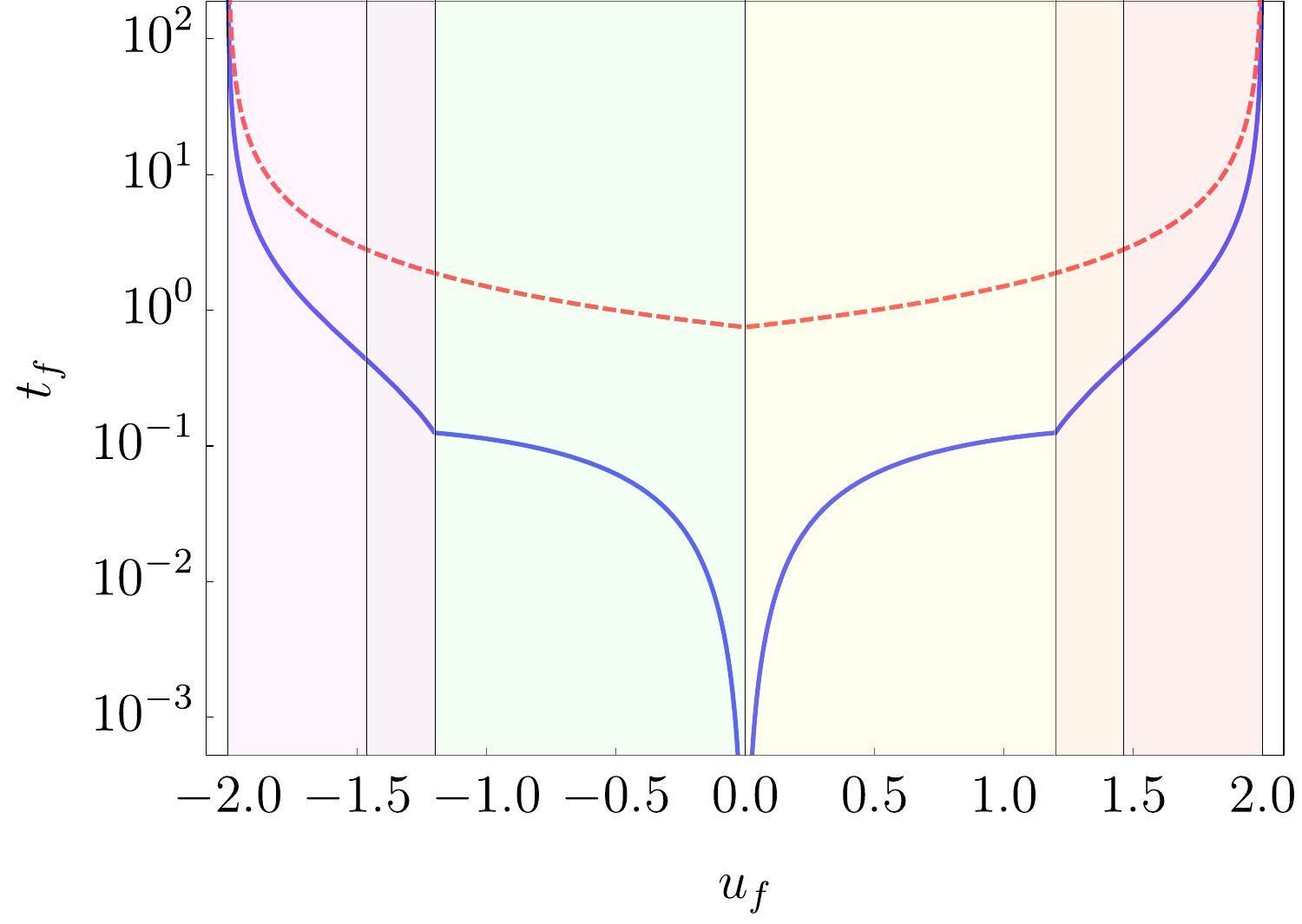}
\\
\includegraphics[width=3.3in]{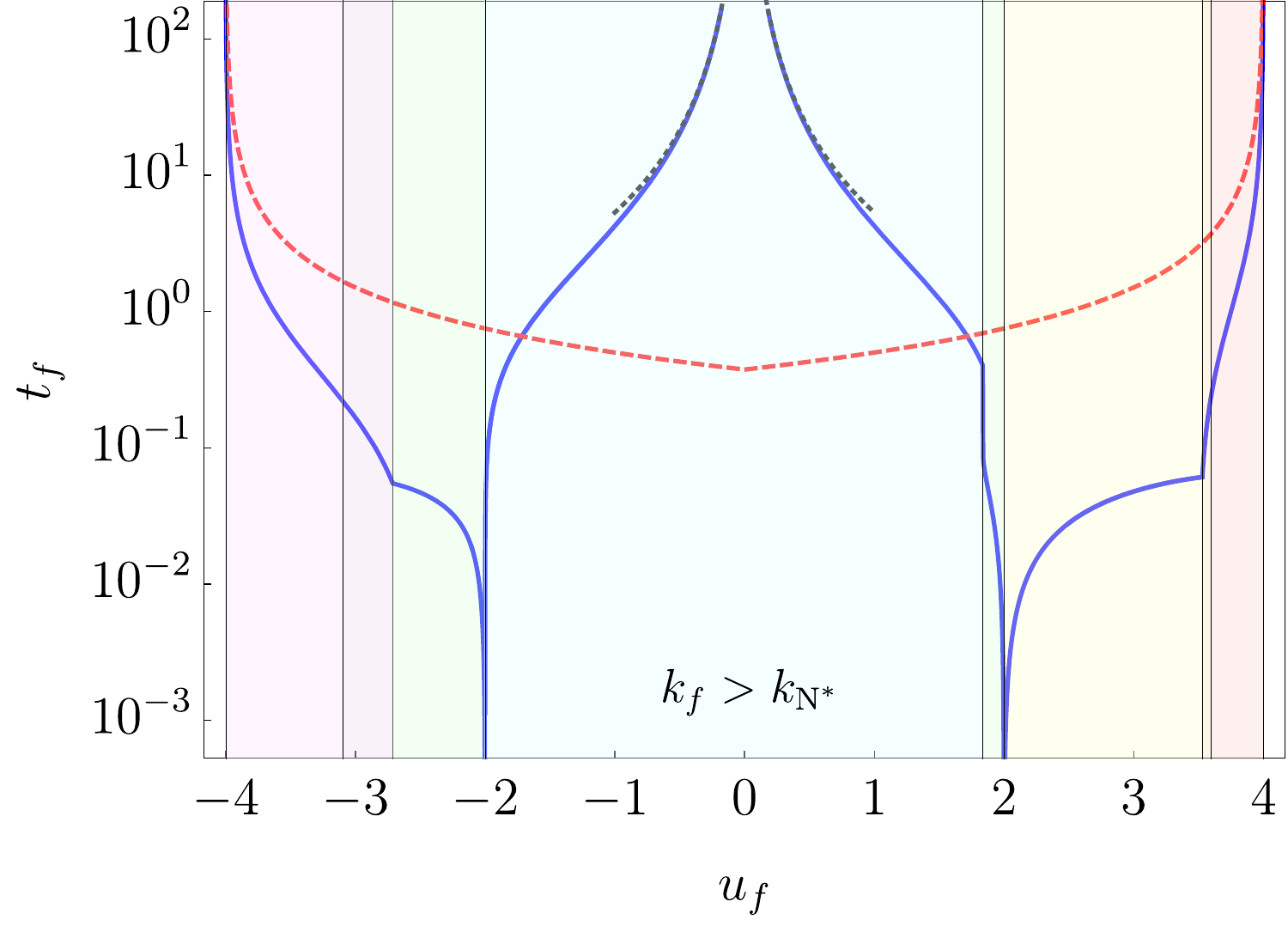} \includegraphics[width=3.3in]{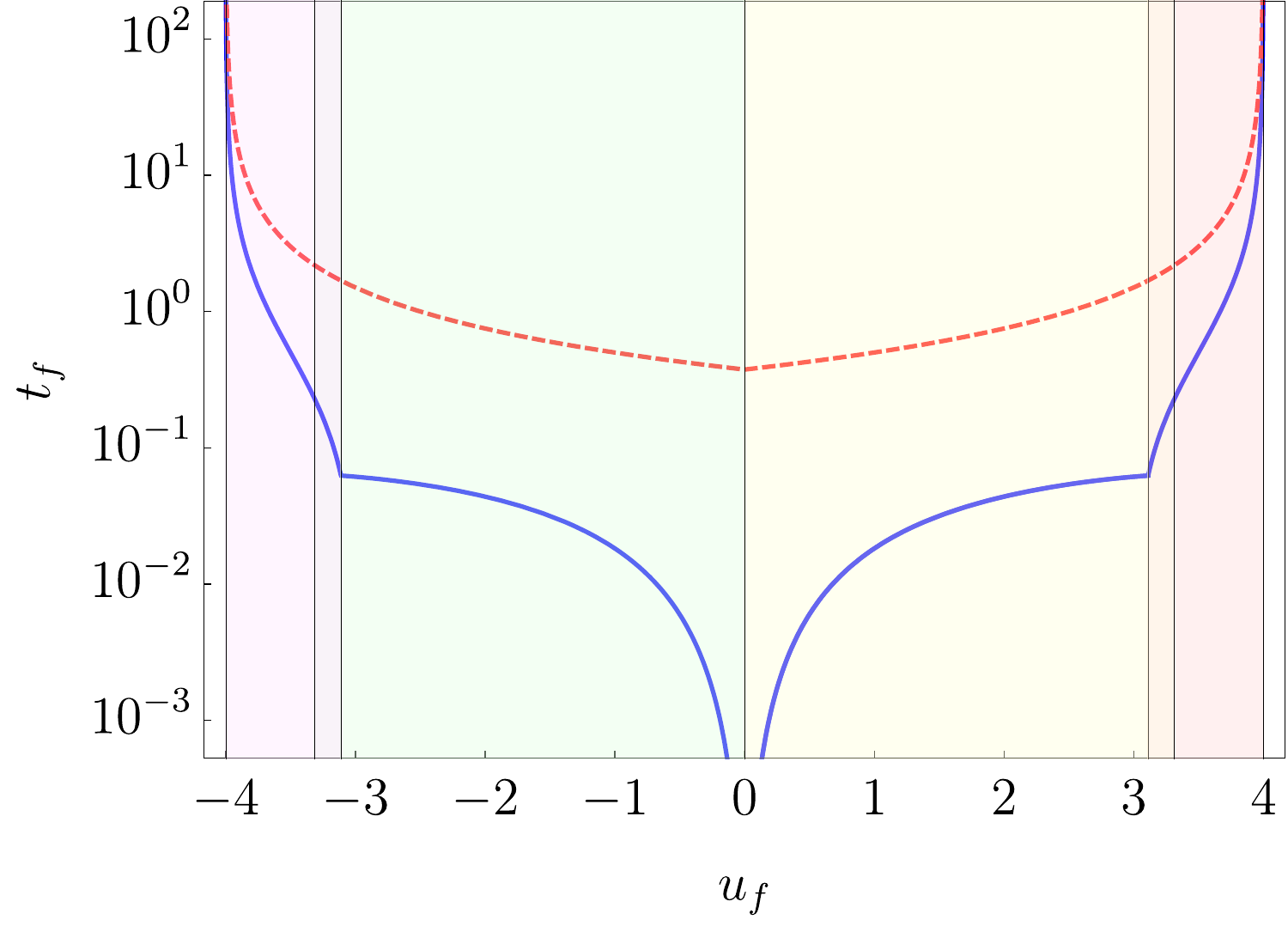}}
\caption{Minimum connection time $t_f$ as a function of $u_f$, for different fixed values of $k_f$. In other words, we are representing vertical slices of the two panels of Fig.~\ref{fig:heat-maps-nonull-ui}---left, $u_i=0.5$ with $k_{\text{N}^*}=3$; right, $u_i=0$---at several values of $k_f$: from top to bottom, we have $k_f=0.5$, $2$, and $4$, as indicated in the legend. The solid line corressponds to $t_f$, whereas the dashed curve corresponds to three times the characteristic relaxation time, $3 t_{\rel}$, for the STEP protocol. The {dotted} curves on the left panels correspond to the asymptotic behaviour from Eq.~\eqref{eq:diverging-1} for $|u_f|\ll 1$. The coloured regions on the background follow the colour code from Figs.~\ref{fig:three-bangs-uipos} (left) and  \ref{fig:three-bangs-ui0} (right). Note that the timescale has been chosen to be  equal in all the panels.
}
    \label{fig:time-slices}
\end{figure}
In order to improve our understanding on the global landscape of minimum connection times, several sections of Figs.~\ref{fig:heat-maps-nonull-ui} for  different fixed values of $k_f$ are plotted in Fig.~\ref{fig:time-slices}. Again, this is done for the two choices of initial NESSs we have repeatedly employed: $u_i = 0.5$ and $u_i = 0$. Recall that, due to the symmetry between the P and N vertices, the results corresponding to the $u_i = 0.5$ would be equivalent to those for $u_i = -0.5$, upon reflection through the $u_f = 0$ axis. In Fig.~\ref{fig:time-slices}, the solid lines correspond to the the minimum connection times already shown in Figs.~\ref{fig:heat-maps-nonull-ui}, for the considered value of $k_f$. The background colour follows the same colour code as in Figs.~\ref{fig:three-bangs-uipos} and \ref{fig:three-bangs-ui0}, thus showing which three-bang protocol the minimal time connection belongs to. The red dashed lines correspond to the characteristic relaxation time $3 t_{\rel}$ for the STEP process as defined in Eq.~\eqref{eq:characteristic-time-trel}, thus allowing us to quantify the efficiency of the minimum-time protocols in beating the STEP one. { Still, it must be stressed that the bang-bang protocols exactly reach the target state in the plotted time $t_f$, whereas we recall that the STEP process needs an infinite time---at $t=3 t_{\rel}$ its exponential relation is completed at 95\%.}

As a general trend, the minimum connection time decreases with $k_f$. This is reasonable from a physical point of view: high values of $k_f$ can only be reached by employing bang-bang protocols in which the P and N time windows dominate the dynamics over the O one. Since the former involve vanishing ---{ in an actual experiment very short,} of the order of $1/k_{\max}$---time windows, the resulting minimum connection time is small. { In order to corroborate the latter,  we explicitly study the PON three-bang protocol for a finite value of $k_{\max}$ in Appendix~\ref{app:finite-kmax}. Specifically, we numerically analyse the behaviour of the time windows corresponding to the P, O and N vertices, and we develop a perturbative theory for large $k_{\max}$. This analytical approach predicts the $1/k_{\max}$ leading behaviour for large $k_{\max}$ and accurately reproduces the numerical results.}

The choice for the three values of $k_f$ displayed in Fig.~\ref{fig:time-slices} has not been arbitrary. For $u_i \neq 0$ (left panels), three qualitatively different behaviors are observed for the connection time. For $k_f < k_i=1$ (top left), minimum connection times are non-zero and finite---with the exception of the asymptotic divergences when approaching $u_f = 0$ and $u_f = \pm k_f$. For $1<k_f<k_{\text{N}^*}$ (left middle), where N$^*$ is the only point reached by a one-bang N protocol---see Fig.~\ref{fig:two-bang-protocols},  the upper PN/NP line in the left panel of Fig.~\ref{fig:two-bang-protocols} comes into play with a zero connection time. Lastly, for $k_f>k_{\text{N}^*}$ (left bottom), the lower branch of the PN/NP line comes into play, which also involve a vanishing $t_f$---providing a second instantaneous brachistochrone. Recall that the PN/NP line of Fig.~\ref{fig:two-bang-protocols} describes the set of final NESS attainable with a PN/NP two-bang protocol,

We may understand the difference between the phenomenology above for the cases $k_f<k_i=1$ and $k_f>k_i=1$ on a physical basis. Let us start by considering $k_f < 1$: this requires a decompression to attain a NESS with a target value of the stifness that is smaller than the initial one. Then, regardless of the specific target value of $k_f$, the corresponding bang-bang optimal protocol must include a time window in the O vertex, and therefore the connection time is non-vanishing. Next, let us move onto the case $1\leq k_f < k_{\text{N}^*}$: there appear states with $u_f>0$ reachable by instantaneous---in the limit $k_{\max}\to\infty$---two-bang PN/NP protocols, with the P window being the dominating one{, because we are considering $u_i>0$ and $u_P=k_{\max}>0$. These are the states over the positive branch, with $u_f>0$, of the PN/NP line. Finally, for $k>k_{\text{N}^*}$, there appear states with $u_f<0$ reachable by instantaneous PN/NP protocols, with the N window being dominating now. This is reasonable: N$^*$ is the only point that can be reached from the initial one with a one-bang N protocol, so states close to it over the negative branch, with $u_f<0$, of the PN/NP line are expected to be dominated by the N time window.}

The central OPN/ONP region, which includes $u_f=0$, in the left panels of Fig.~\ref{fig:time-slices} deserves further attention, as it presents interesting features. First, the minimum connection time $t_f$ diverges for $u_f\to 0$, which implies that, from a practical perspective, optimal bang-bang protocols become inefficient with respect to the direct STEP protocol, because the $3t_{\rel}$ curve lies below $t_f$ over most of the region OPN/ONP. Second,  the minimum connection time $t_f$ presents discontinuities at the boundaries of the region OPN/NOP: on the left, for $u_f<0$, when the region NON emerges; on the right, for $u_f>0$, when the region NOP emerges. Such discontinuities stem from the emergence of the NON and NOP regions, since the OPN/ONP protocols continue to exist in such regions but give longer connection times than the NON and NOP protocols. Therefore, the latter become the optimal ones.

The right panels of Fig.~\ref{fig:time-slices}, for which $u_i = 0$, are simpler to understand. There are fewer three-bang regions stemming from the scheme in Fig.~\ref{fig:three-bangs-ui0}, since the existing OPN/ONP regions for $u_i\ne 0$ disappear in the case $u_i=0$: as we have already commented, this vanishing stems from the degeneration of the PN/NP lines into a unique line $u_f=0$, $k_f\ge 1$, as seen in Fig.~\ref{fig:two-bang-protocols}. Therefore, there are only two cases of interest: (i) $k_f < 1$, for which the minimum connection time is always non-zero, due to the presence of time window corresponding to the vertex O in the optimal protocols, and (ii) $k_f > 1$, for which the minimum connection time vanishes along the $u_f = 0$ line---i.e. the PN/NP line, with a vanishing time in the limit $k_{\max}\to\infty$. { For $u_i=0$,} the minimum connection time from optimal control systematically beats the characteristic relaxation time of the STEP protocol.

In Appendix~\ref{app:speed-limit-bound}, we briefly discuss the bound for the minimum connection time stemming from the speed limit inequality~\eqref{eq:speed-limit}. Therein, it is shown that $\expval{W_{\irr}}$ linearly increases with $k_{\max}$ in the limit $k_{\max}\to\infty$ we have considered to simplify our discussion. This entails that the bound given by Eq.~\eqref{eq:speed-limit} becomes very loose, since it vanishes as $k_{\max}^{-1}$. 
This behaviour is similar to the one found in other systems for thermal control, when the upper bound of bath temperature goes to infinity and the corresponding bang become instantaneous~\cite{prados_optimizing_2021,patron_thermal_2022}.


\section{Summary and discussion}\label{sec:conclusions}

A relevant question within the field of shortcuts, or SST, concerns the brachistochrone---i.e. the protocol involving the minimum connection time---between two states of a physical system. For systems at the mesoscopic level of description, the latter is significant from a theoretical point of view, since it brings together concepts from the field of non-equilibrium statistical mechanics, stochastic thermodynamics~\cite{sekimoto_stochastic_2010,peliti_stochastic_2021}, and optimal control theory~\cite{liberzon_calculus_2012,pontryagin_mathematical_1987}. From a practical perspective, it is relevant for the design of nanodevices, for which fluctuations are non-negligible and may affect their performance.

In this work, we have tackled, both numerically and analytically, the above question for a paradigmatic model in non-equilibrium statistical mechanics: the Brownian Gyrator (BG)~\cite{filliger_brownian_2007,dotsenko_two-temperature_2013,cerasoli_asymmetry_2018,baldassarri_engineered_2020,miangolarra_thermodynamic_2022}. The SST connection we would like to build aims at connecting an initial NESS, corresponding to potential parameters $(k_i,u_i)$, with a target NESS, with parameters $(k_f,u_f)$ in the minimum possible time, while keeping the potential { non-repulsive} for all times. Therefore, the controls $(k,u)$ belong to the triangular control set illustrated in Fig.~\ref{fig:control-set}, which has three vertices O$=(0,0)$, P$=(k_{\max},k_{\max})$, N$=(k_{\max},-k_{\max})$, where $k_{\max}$ is the maximum value of the stiffness---typically, in experiments one cannot have a harmonic trap with an arbitrarily large value of the elastic constant. 

A first main result of our paper is that the protocols providing the minimum time connection between two NESSs  are of bang-bang type, i.e. the controls $(k,u)$ lie on the boundaries of the triangular control set of Fig.~\ref{fig:control-set}. In other words, the optimal time protocols do not have time windows with Euler-Lagrange solutions, for which $(k,u)$ would belong in the interior of the control set. Bang-bang protocols are divided into two classes: (i) regular protocols, in which the maximum of Pontryagin's Hamiltonian is reached at the vertices of control set, and thus the controls switch between the values corresponding to the three vertices O, N, and P of Fig.~\ref{fig:control-set}, and (ii) singular protocols,  Pontryagin's Hamiltonian $\Pi$ is constant along one of the edges of the control set and the maximum of $\Pi$ is reached precisely over that edge, being thus degenerate. To make analytical progress, we have considered mainly the limit $k_{\max}\to\infty$ throughout the paper. { On the one hand, it simplifies the mathematical description; on the other hand, it corresponds to the physical situation in which one has a very large capacity for compression with a very short relaxation time.}

We have focused on three-bang regular protocols, i.e. protocols with three time windows, with each time window corresponding to the controls being given by one of the vertices of the control set. We have denoted them by the sequence of vertices involved: e.g. the protocol ONP means that the first time window corresponds to the vertex O, the second to the vertex N, and the third to the vertex P. These three-bang protocols (i) can be univocally determined as a function of the target state $(k_f,u_f)$, for an arbitrary given initial state $(k_i,u_i)$, and (ii) connect all possible initial states with all possible final states. Simpler one-bang and two-bang protocols only make the connection to specific target points and lines, respectively, of the control set. Over the optimal three-bang protocols, we have studied the behaviour of the final connection time $t_f$ as a function of the target state $(k_f,u_f)$, for a given initial state $(k_i = 1,u_i)$. The connection time landscape radically changes from the $u_i \neq 0$ to the $u_i = 0$ cases, with the former being much more complicated than the latter.

There are some connections that imply an infinite time: in particular, an infinite time is needed to reach the boundary lines of the control set $k_f=\pm u_f$. This was expected on a physical basis, since over these lines one of the characteristic relaxation times of the normal modes diverge. More surprising is the divergence of the connection time for reaching uncoupled target states, with $u_f=0$, from initial states that are coupled, with $u_i\ne 0$, since all the characteristic relaxation times of the normal modes remain finite over the line $u_f=0$. From a physical point of view, this is related to needing an infinite time at vertex O to ``kill'' the correlations between the original variables $(x,y)$ at the final state.

There are some connections that are instantaneous---in the limit $k_{\max}\to\infty$ we are considering, in which the time windows corresponding to the P and N vertices involve no time. Therefore, any bang-bang protocol containing only the vertices P and N give a null connection time. On the one hand, we have the one-bang P and N protocols, which only reach a specific point P$^*$ and N$^*$ of the control set, for $u_i<0$ and $u_i>0$, respectively. On the other hand, we have the two-bang PN/NP protocols, which reach two specific lines of the control set, as illustrated in Fig.~\ref{fig:two-bang-protocols}. For the case $u_i=0$, the PN and NP lines merge into the $\{u_f=0,k_f>1\}$ line~\footnote{Three-bang PNP/NPN protocols reach the same lines as the PN/NP lines, see Appendix~\ref{app:explicit-bang-bang} for details.}. Therefore, at variance with the $u_i\ne 0$ case, in which the connection time at the $u_f=0$ line diverges, for $u_i=0$ not only are states along the $u_f = 0$ line reachable, but instantaneously reachable.

It must also be remarked that there appear discontinuities in the minimum connection time as a function of $(k_f,u_f)$, for an initial state with $u_i\ne 0$---i.e. an actual NESS in which the two degrees of freedom are coupled. We have illustrated these discontinuities in Fig.~\ref{fig:time-slices}, in which we have plotted the minimum connection time as a function of $u_f$, for several fixed values of $k_f$. For $k_f>1$, a finite-jump discontinuity of $t_f$ emerges when the optimal protocol changes from the central OPN/ONP region, which includes the $u_f=0$ line, to the NOP or NON regions~\footnote{{For the shown case $u_i>0$, for $u_i<0$ it would be the PON and POP regions.}. In the NOP and NON regions, the OPN and ONP protocols continue to exist---and would give a continuous connection time---but the connection time for the NOP and NON protocols are shorter in their respective domains.}.

This work also opens the door for future research. First, the speed limit inequality derived here for the connection between NESSs of the BG is analogous to others for connections between equilibrium states, which also involve the connection time and the irreversible work~\cite{sivak_thermodynamic_2012,aurell_refined_2012,dechant_minimum_2022,guery-odelin_driving_2023}. { Yet, it is worth looking for a tighter bound, since the inequality derived here  for the connection between two NESSs cannot be saturated. On another note,} speed limit inequalities based on thermodynamic geometry quantities like the Fisher information have been obtained for the connection between arbitrary states~\cite{shiraishi_speed_2018,shanahan_quantum_2018,funo_speed_2019,shiraishi_speed_2020,ito_stochastic_2020,van_vu_geometrical_2021,lee_speed_2022}, and applied to the connection between two NESSs of a uniformly heated granular system~\cite{prados_optimizing_2021,ruiz-pino_optimal_2022}. Still, it must be remarked that the thermodynamic geometry quantities in the latter speed limit inequalities have a less transparent physical meaning than the irreversible work. The possible generalisation of the speed limit inequality for the BG to a more general out-of-equilibrium system  thus deserves further research. 

Second, the current state of the art in optical trapping allows for the experimental implementation of the optimal protocols developed here for a BG, following the techniques employed in its recent practical realisations~\cite{ciliberto_heat_2013,argun_experimental_2017,chiang_electrical_2017,cerasoli_spectral_2022}. The fact that bang-bang protocols involve instantaneous switchings of the control variables may be implemented by engineering the potential to vary over a timescale much shorter than that corresponding to the timescale of the dynamics of the BG. 

{ Third, our work paves the way for the consideration of other physically relevant control problems for the BG. On the one hand, it would be interesting to analyse other physically relevant control sets, with the possible inclusion of repulsive potentials during the connection, for the brachistochrone. Enlarging the set of admissible controls would lead, in general, to shorter connection times. In this regard, it would be interesting to find SST capable of extinguishing cross-correlations in a finite time. On the other hand, the minimisation of the irreversible work  for a given connection time seems to be a highly non-trivial problem, much more difficult than in the equilibrium case. This is so even for the completely unconstrained case where $(k,u)$ are both unbounded, due to the impossibility of saturating the right hand side of Eq.~\eqref{eq:speed-limit}. Including non-holonomic constraints, like the ones we consider in our paper, would further increase the complexity of the control problem.} 

{Finally,} due to the infinite degeneracy of singular protocols, we may consider the  optimisation of additional figures of merit, such as the irreversible work or the entropy production~\cite{muratore-ginanneschi_extremals_2014,muratore-ginanneschi_application_2017,landi_irreversible_2021,aurell_optimal_2011,aurell_boundary_2012,aurell_refined_2012,zhang_work_2020,plata_optimal_2019,plata_taming_2021}, the information cost or thermodynamic length~\cite{crooks_measuring_2007,sivak_thermodynamic_2012,ito_stochastic_2018},  in order to further constrain the subspace of optimal protocols. This seems particularly relevant, due to the competition between the connection time and irreversible work expressed by our speed limit inequality. These perspectives would become significant for the design of optimal heat engines with Brownian particles~\cite{schmiedl_efficiency_2008,esposito_efficiency_2010,blickle_realization_2012,deng_boosting_2013,tu_stochastic_2014,muratore-ginanneschi_efficient_2015,martinez_brownian_2016,abah_shortcut--adiabaticity_2019,albay_thermodynamic_2019,albay_realization_2020,plata_building_2020,nakamura_fast_2020,zhang_optimization_2020,tu_abstract_2021,frim_optimal_2022,li_geodesic_2022,pedram_quantum_2023,deng_exploring_2024}, which constitutes an active field of research. 

\section{Acknowledgements}

We thank Marco Baldovin for fruitful discussions. We acknowledge financial support from Grants PID2021-122588NB-I00, funded by MCIN/AEI/10.13039/501100011033/ and by ``ERDF A way of making Europe'', and ProyExcel\_00796, funded by Junta de Andalucía's PAIDI 2020 programme. CAP acknowledges the funding received from European Union’s Horizon Europe–Marie Skłodowska-Curie 2021 program through the Postdoctoral Fellowship
with Reference 101065902 (ORION). A. Patr\'on also acknowledges support from the FPU programme through Grant FPU2019-4110.

\section*{Data Availability Statement}

The Mathematica notebooks employed for generating the data and figures that support the findings of this study are openly available in the~\href{https://github.com/fine-group-us/Brownian-gyrator}{GitHub page} of University of Sevilla's FINE research group.


\appendix


\section{\label{app:Euler-Lagrange} On the inexistence of Euler-Lagrange optimal-time protocols}

In this section, we explicitly show that Euler-Lagrange solutions---i.e those for which the optimal control $(k^*(t),u^*(t))$ lies in the interior of the control set---do not exist for the time-optimisation problem of concern. We recall that our dynamical system does not belong to the class of ``linear systems'' of optimal control theory, since the evolution equations do not have the form~\eqref{eq:linear-opt-control}. Therefore, the question of whether we may rule out or not Euler-Lagrange solutions is non-trivial.

\subsection{Pure Euler-Lagrange solutions}
We start by considering ``pure'' Euler-Lagrange solutions, i.e. those for which the control variables lie in the interior of the control set for all $t \in (0,t_f)$. Therefore,  Eq.~\eqref{eq:E-L-eq} must be satisfied at all times: specifically, we have that
    \begin{align}\label{pontry-der}
    \frac{\partial \Pi}{\partial k}  &= -2(\psi_1z_1 + \psi_2z_2 + \psi_3z_3) = 0,
    &
    \frac{\partial \Pi}{\partial u} & = -2(\psi_1z_1 - \psi_2z_2) = 0.
    \end{align}
Since these two equalities hold $\forall t \in (0,t_f)$, their time derivatives must also vanish. On the one hand, the first time derivative gives
\begin{subequations}
    \begin{align}
        &\frac{d}{dt}\left(\psi_1z_1 + \psi_2z_2 + \psi_3z_3 \right) = \psi_1 + \psi_2 + \psi_3 = 0,\label{subec:derk}
        \\
        &\frac{d}{dt}\left(\psi_1z_1 - \psi_2z_2\right) = \psi_1 - \psi_2 = 0.\label{subec:deru}
    \end{align}
\end{subequations}
Let us start with Eq.~\eqref{subec:deru}. Together with Eq.~\eqref{pontry-der} and the fact that $\Pi=0$ over the optimal solution, they imply that $\psi_1=\psi_2 \ne 0$ and thus $z_1 = z_2$---we cannot have $\psi_1=\psi_2=0$, because then Eq.~\eqref{subec:derk} implies that $\psi_3=0$, and $\Pi=0$ in turn implies that $\psi_0=0$~\footnote{One of the conditions in PMP is that $(\psi_0,\bm{\psi}^{\sf{T}})\ne (0,0,0,0)$, the so-called non-triviality condition~\cite{liberzon_calculus_2012}.}. Moreover, their initial and final values must coincide as well,
\begin{subequations}
\begin{align}
    z_1(0) &= \frac{1}{2(1+u_i)} = \frac{1}{2(1-u_i)} = z_2(0) & \Leftrightarrow  u_i &= 0,
    \\
    z_1(t_f) &= \frac{1}{2(k_f+u_f)} = \frac{1}{2(k_f-u_f)} = z_2(t_f) & \Leftrightarrow   u_f &= 0.
\end{align}
\end{subequations}
Thus, regardless of the form of the optimal control $(k,u)$, Euler-Lagrange protocols only allow to connect uncorrelated states. Now, Eq.~\eqref{subec:derk} provides further insights. Recalling again that $\Pi(t) = 0$ over the optimal solution, we have
\begin{align}\label{eq:second-condition-pmp}
    0 = \Pi &= \psi_0 + \psi_1\left[-2(k+u)z_1 + 1\right] + \psi_2\left[-2(k-u)z_2 + 1\right] + \psi_3\left[-2kz_3 + 1\right] \nonumber
    \\
    &= \psi_0 + \cancelto{0}{(\psi_1 + \psi_2 + \psi_3)} -2k\cancelto{0}{\left(\psi_1z_1 + \psi_2z_2 + \psi_3z_3 \right)} -2u\cancelto{0}{\left(\psi_1z_1 - \psi_2z_2\right)} .
\end{align}
Therefore, $\psi_0 = 0$ for Euler-Lagrange protocols. Although this does not contradict PMP, it will become useful later in this section to disregard ``mixed'' protocols. 

The second time derivatives of Eq.~\eqref{pontry-der} must vanish too:
\begin{subequations}
    \begin{align}
        &\frac{d}{dt}\left(\psi_1 + \psi_2 + \psi_3 \right) = 2k\cancelto{0}{(\psi_1 + \psi_2 + \psi_3)} + 2u\cancelto{0}{(\psi_1-\psi_2)} = 0, \label{subeq:derrk}
        \\
        &\frac{d}{dt}\left(\psi_1 - \psi_2\right) = 2u(\psi_1 + \psi_2) = 0. \label{subeq:derru}
    \end{align}
\end{subequations}
Equation~\eqref{subeq:derrk} identically vanishes due to the relations \eqref{subec:derk} and \eqref{subec:deru} and, thus, does not provide any further information. On the other hand, Eq.~\eqref{subeq:derru} leaves us with two different cases:
\begin{enumerate}
    \item $\psi_1 + \psi_2=0$. This again leads to  $\psi_0=\psi_1=\psi_2=\psi_3=0$, which makes us ignore this possibility.
    \item $u=0$, for all $t\in(0,t_f)$. Recalling the dynamical system from Eq.~\eqref{dynamic-equations}, together with the fact that $u_i = u_f = 0$, we have that all the dynamical variables $z_j$ degenerate onto the same one, and the structure of our optimisation problem radically changes, as we only need to study the behaviour of one dynamical variable now. In such scenario, it can be shown that optimal-time protocols belong to the subclass of one-bang protocols, for which the control parameter $k$ takes either the value $0$ or $k_{\max}$---depending on the choice of the target state---during the entire time window. See Appendix~\ref{app:uncoupled-oscillators} for further details.
\end{enumerate}


\subsection{Mixed Euler-Lagrange solutions}

Once we have disregarded ``pure'' Euler-Lagrange solutions, we consider here ``mixed'' solutions involving at least one time interval of Euler-Lagrange type. In the following, we investigate whether such Euler-Lagrange intervals can be mixed with regular bang-bang ones---we do not consider mixing with singular intervals, the impossibility of which is shown in Appendix~\ref{app:singular-protocols}.

Let us consider a bang-bang window starting from O at some time $t_0$. Then, the switching functions must verify $\phi_{\text{OP}}(t_0)<0$ and $\phi_{\text{ON}}(t_0)<0$---as shown in Fig.~\ref{fig:switching-funcs-regular}. Making use of Eq.~\eqref{eq:switching-functions}, we have the conditions
\begin{equation}\label{eq:inequalities-1}
    2\psi_1(t_0)z_1(t_0) + \psi_3(t_0)z_3(t_0) >0, \quad 2\psi_2(t_0)z_2(t_0) + \psi_3(t_0)z_3(t_0) >0.
\end{equation}
Taking into account that both $k$ and $u$ vanish at the vertex O, Pontryagin's Hamiltonian function at $t=t_0$ is $\Pi = \psi_0 + \psi_1(t_0) + \psi_2(t_0) + \psi_3(t_0) = 0$, which entails that
\begin{equation}\label{eq:psi0-inequality-EL}
    \psi_0 = -\psi_1(t_0) - \psi_2(t_0) - \psi_3(t_0) 
    < \frac{z_3(t_0)}{2z_1(t_0)}\psi_3(t_0) + \frac{z_3(t_0)}{2z_2(t_0)}\psi_3(t_0) - \psi_3(t_0) = \psi_3(t_0) \left(\frac{z_3(t_0)}{2z_1(t_0)} + \frac{z_3(t_0)}{2z_2(t_0)} - 1 \right),
\end{equation}
where we have substituted the inequalities from Eq.~\eqref{eq:inequalities-1}. Now we have two possibilities: (i) either $t_0=0$, implying that we start the optimal protocol at O, or (ii) $t_0 \ne 0$, corresponding to the scenario for which we had a previous Euler-Lagrange window from which we switched towards O. In the former, we substitute the initial state into Eq.~\eqref{eq:psi0-inequality-EL}, giving
\begin{equation}
    \psi_0 < \psi_3(t_0)\left(\frac{1+u_i}{2} + \frac{1-u_i}{2} - 1 \right) = 0 \ \Rightarrow \ \psi_0 < 0.
\end{equation}
In the latter, we take into account that $z_1(t_0)=z_2(t_0)=z_3(t_0)$ in the initial Euler-Lagrange window, so Eq.~\eqref{eq:psi0-inequality-EL} now implies
\begin{equation}
    \psi_0 < \psi_3(t_0)\left(\frac{1}{2} + \frac{1}{2} - 1 \right) = 0 \ \Rightarrow \ \psi_0 < 0.
\end{equation}
In both cases,  $\psi_0 < 0$:  this is inconsistent with the fact that, during an Euler-Lagrange window, $\psi_0$ must be zero. Since $\psi_0$ is constant and cannot change during the entire protocol, this rules out the possibility of having mixed Euler-Lagrange protocols with O time windows. 

We proceed in a similar manner for the possible switching between P/N and Euler-Lagrange windows. Let us now consider a P window, starting again at some time $t_0$. Then, the switching functions must verify $\phi_{\text{OP}}(t_0)>0$ and $\phi_{\text{NP}}(t_0)>0$, which, using  again Eq.~\eqref{eq:switching-functions}, imply the conditions
\begin{equation}\label{eq:inequalities-2}
    2\psi_1(t_0)z_1(t_0) + \psi_3(t_0)z_3(t_0) <0, \quad \psi_1(t_0)z_1(t_0) - \psi_2(t_0)z_2(t_0) < 0.
\end{equation}
Now, Pontryagin's Hamiltonian at time $t_0$ is $\Pi= \psi_0 + \psi_1(t_0) + \psi_2(t_0) + \psi_3(t_0) - 2k_{\max}(2\psi_1(t_0)z_1(t_0) + \psi_3(t_0)z_3(t_0)) = 0$, which entails
\begin{align}
    \psi_0 &= -\psi_1(t_0) - \psi_2(t_0) - \psi_3(t_0) + 2k_{\max}(2\psi_1(t_0)z_1(t_0) + \psi_3(t_0)z_3(t_0))\nonumber
    \\ 
    &< -\psi_1(t_0)\left( 1+\frac{z_1(t_0)}{z_2(t_0)}\right) - \psi_3(t_0)+ 2k_{\max}(2\psi_1(t_0)z_1(t_0) + \psi_3(t_0)z_3(t_0)),
\end{align}
where we have substituted the inequalities from Eq.~\eqref{eq:inequalities-2}. On the one hand, had we initially started at vertex P, i.e. $t_0 = 0$, then we would have
\begin{equation}
    \psi_0 < 2(k_{\max}-1)\left(2\psi_1(t_0)z_1(t_0) + \psi_3(t_0)z_3(t_0) \right) < 0 \ \Rightarrow \ \psi_0 < 0,
\end{equation}
{since $k_i=1<k_{\max}$.} On the other hand, had we previously applied an Euler-Lagrange window, then
\begin{equation}
    \psi_0 < {(2k_{\max}z_1(t_0)-1)}(2\psi_1(t_0) + \psi_3(t_0)) < 0 \ \Rightarrow \ \psi_0 < 0,
\end{equation}
{ since $2k_{\max}z_1(t_0)-1 >0$---due to the evolution equation for $u=0$.} Once again, as in both cases we have that $\psi_0 < 0$, then we may disregard mixed protocols involving Euler-Lagrange and P windows. Let us remark that, due to the symmetry between the P and N points, the above proof extends to N time windows as well, and thus, we may conclude that there cannot be mixed protocols involving Euler-Lagrange and bang-bang windows.

\section{\label{app:singular-protocols} Characterisation of singular protocols}

As mentioned in the main text, singular protocols are those for which at least one of the switching functions $\phi_{\text{OP}}$, $\phi_{\text{ON}}$ or $\phi_{\text{NP}}$ identically vanishes during a time window within {$(t_1,t_2)\in(0,t_f)$  and, in addition, the maximum of Pontryagin's Hamiltonian is attained over the corresponding edge of the triangular control set. In this section we show that singular protocols, if they exist, extend to the whole time interval $(0,t_f)$, i.e. switchings between regular and singular protocols is not possible.}

Let us start by considering the edge $\overline{\text{OP}}$, where $k=u$, and its corresponding switching function $\phi_{\text{OP}}$, which is given by Eq.~\eqref{eq:switching-functions-OP}.  Since this function is identically zero in $(t_1,t_2)$, its time derivatives must also cancel:
\begin{align}
    \dot{\phi}_{\text{OP}} &= -2(2\psi_1 + \psi_3) = 0,
    &
    \ddot{\phi}_{\text{OP}} &= -4k(4\psi_1 + \psi_3) = 0.
\end{align}
{Since $k(t)$ does not identically vanish over the edge $\overline{\text{OP}}$}, the only available solution is $\psi_1(t) = \psi_3(t) = 0$, $\forall t\in(t_1,t_2)$. However, recalling Hamilton's canonical equations \eqref{eq:canonical}, we have that { $\dot{\psi}_1=4k\psi_1$ and $\dot{\psi}_3=2k\psi_3$.} Therefore, if $\psi_{1,3}$ vanishes at some specific time, $\psi_{1,3}(t)$ must be zero $\forall t$. Hence, $\psi_1(t) = \psi_3(t) = 0$, $\forall t\in(0,t_f)$, and  $\phi_{\text{OP}}(t)$ also vanishes over the whole time interval: the singular protocol over the edge $\overline{\text{OP}}$ extends to all times in $(0,t_f)$. Now, invoking the third condition of PMP, we have that
\begin{equation}
    0 = \Pi = \psi_0 + \cancelto{0}{\psi_1}(-4kz_1+1) + \psi_2 + \cancelto{0}{\psi_3}(-2kz_3+1) \ \Rightarrow \ \psi_2 = -\psi_0 > 0, \; \forall t\in (0,t_f).
\end{equation}
That is, $\psi_2(t)$ should be constant for singular protocols, and moreover, it cannot cancel. { This is consistent with the evolution equations: over the $\overline{\text{OP}}$ edge, the evolution equation for $\psi_2$ reduces to $\dot{\psi}_2=0$.}

It is straightforward to extend the above analysis to the $\overline{\text{ON}}$ edge and the $\phi_{\text{ON}}$ switching function, due to the symmetry between the P and N points. In such a case, the same results apply, upon exchanging the behaviour found above for the $\psi_1$ and $\psi_2$ momenta. { Wrapping things up, we have shown that singular protocols over the $\overline{\text{OP}}$ and $\overline{\text{ON}}$ edges may appear, but they have to extend to the whole time interval $(0,t_f)$.}

Let us now consider the last edge $\overline{\text{NP}}$ and the corresponding switching function, given by Eq.~\eqref{eq:switching-functions-NP}. Again, since $\phi_{\text{NP}}$ identically vanishes in $(t_1,t_2)$, taking time derivatives we get 
\begin{align}\label{subeq:PN-branch}
    \dot{\phi}_{\text{NP}} &= -2(\psi_1 - \psi_2) = 0, 
    &
    \ddot{\phi}_{\text{NP}} &= -4u(\psi_1 + \psi_2) = 0.
\end{align}
{ Similarly, since $u(t)$ does not identically vanish over the edge $\overline{\text{NP}}$}, we have that $\psi_1(t)=\psi_2(t)=0$ for all $t\in(t_1,t_2)$. Again, from the evolution equations for the momenta, we conclude that $\psi_1(t)=\psi_2(t)=0$ for all $t\in(0,t_f)$, the singular time window extends to the whole interval. Invoking again that Pontryagin's Hamiltonian vanishes over the time optimal protocol,
\begin{align}
        0 = \Pi &= \psi_0 + \cancelto{0}{\psi_1}(-2(k_{\max}+u)z_1 + 1) + \cancelto{0}{\psi_2}(-2(k_{\max}-u)z_1 + 1) + \psi_3(-2k_{\max}z_3+1) ,
\end{align}
we have
\begin{align}\label{eq:psi-3}
 \psi_3 = \frac{\psi_0}{2k_{\max}z_3-1}<0,
\end{align}
{where we have taken into account the inequality $2k_{\max}z_3(t)-1 \geq 0$, which stems from the evolution equation for $z_3$. Again, Eq.~\eqref{eq:psi-3} is consistent with the evolution equations:
\begin{equation}
   \dot{\psi}_3=-\frac{2k_{\max}\psi_0}{(2k_{\max}z_3-1)^2}\dot{z}_3=-\frac{2k_{\max}\psi_0}{(2k_{\max}z_3-1)^2}\left(-2 k_{\max} z_3+1\right)=\frac{2k_{\max}\psi_0}{2k_{\max}z_3-1}=2k_{\max}\psi_3.
\end{equation}
}

Following the preceding discussion, we conclude that the singular protocols corresponding to the $\overline{\text{OP}}$, $\overline{\text{ON}}$, and $\overline{\text{NP}}$ edges share similar features: (i) they correspond to ``pure'' protocols that last for the whole interval $(0,t_f)$, (ii) they involve the vanishing of two conjugate momenta during the entire protocol, and (iii) they are characterised by the action of an unique control variable---either $k(t)$, for the $\overline{\text{OP}}$ and $\overline{\text{ON}}$ edges, or $u(t)$, for the  $\overline{\text{NP}}$ edge. 

An additional interesting feature is that, for each singular protocol, the connection time is always fixed---regardless of the choice of $k(t)$ or $u(t)$ over the considered edge. The physical reason for this behaviour is that the time evolution of one of the variances sets the connection time, independently of the others. More specifically, for the $\overline{\text{OP}}$ edge, we have that $\dot{z}_2 = 1$, thus giving
\begin{equation}
t_f^{({\text{OP}})} = z_2(t_f) - z_2(t_0) = \frac{1}{2}\left(\frac{1}{k_f-u_f} -\frac{1}{1-u_i}\right). 
\end{equation}
Similarly, for the $\overline{\text{OP}}$ branch, we have $\dot{z_1} = 1$, implying that
\begin{equation}
t_f^{({\text{ON}})} = z_1(t_f) - z_1(t_0) = \frac{1}{2}\left(\frac{1}{k_f+u_f} -\frac{1}{1+u_i}\right). 
\end{equation}
And finally, for the $\overline{\text{NP}}$, we have $\dot{z}_3 = -2k_{\max}z_3+1$, which gives
\begin{equation}
t_f^{({\text{NP}})} = \frac{1}{2k_{\max}} \ln \left(k_f \frac{k_{\max}-1}{k_{\max}-k_f} \right).
\end{equation}
Each of the singular branches has still two dynamical variables whose behaviour is determined via one control parameter $k(t)$ or $u(t)$. In principle, since there are no further restrictions on the values of $k$ or $u$, there is an infinite number of possible optimal protocols achieving the desired connection in the given connection time $t_f$. One could take such degeneracy as an opportunity for optimising an additional figure of merit, such as the irreversible work or the information cost.


\section{Explicit equations for optimal bang-bang protocols}\label{app:explicit-bang-bang}

Here we will explicitly write the systems of algebraic equations for each of the two-bang and three-bang protocols, in the $k_{\max}\rightarrow +\infty$ limit, that are presented in Sec.~\ref{sec:brachistochrone}. These sets of equations provide the solutions that are depicted in Figs.~\ref{fig:two-bang-protocols} - \ref{fig:time-slices}. Throughout this section, in order to simplify the presentayion, we directly make the substitutions $z_j(0) = (2\omega_j(0))^{-1}$ and $z_j(t_f) = (2\omega_j(t_f))^{-1}$.

\subsection{Two-bang protocols}

There are six possible permutations of the points (O, P, N) giving rise to two-bang protocols. The symmetry between the P and N points translate into the PN and NP protocols providing the same algebraic system of equations.
\begin{itemize}
    \item \textbf{OP}: Variables $(t_f, \xi)$.
    \begin{subequations}
        \begin{align}
            \frac{1}{2(k_f + u_f)} &= \left(\tilde{\mathcal{E}}_{\xi^2}\circ \mathcal{E}_0^{(t_f)}\right)\left(\frac{1}{2(1+u_i)}\right)= \left(\frac{1}{2(1+u_i)} + t_f\right)\xi^2,
            \\
            \frac{1}{2(k_f - u_f)} &= \left(\tilde{\mathcal{E}}_{1}\circ \mathcal{E}_0^{(t_f)}\right)\left(\frac{1}{2(1-u_i)}\right) = \frac{1}{2(1-u_i)} + t_f,
            \\
            \frac{1}{2k_f} &= \left(\tilde{\mathcal{E}}_{\xi}\circ \mathcal{E}_0^{(t_f)}\right)\left(\frac{1}{2}\right)= \left(\frac{1}{2} + t_f\right)\xi.
        \end{align}
    \end{subequations}
    \item \textbf{ON}: Variables $(t_f, \xi)$.
    \begin{subequations}
        \begin{align}
            \frac{1}{2(k_f + u_f)} &= \left(\tilde{\mathcal{E}}_{1}\circ \mathcal{E}_0^{(t_f)}\right)\left(\frac{1}{2(1+u_i)}\right) = \frac{1}{2(1+u_i)} + t_f,
            \\
            \frac{1}{2(k_f - u_f)} &= \left(\tilde{\mathcal{E}}_{\xi^2}\circ \mathcal{E}_0^{(t_f)}\right)\left(\frac{1}{2(1-u_i)}\right)= \left(\frac{1}{2(1-u_i)} + t_f\right)\xi^2,
            \\
            \frac{1}{2k_f} &= \left(\tilde{\mathcal{E}}_{\xi}\circ \mathcal{E}_0^{(t_f)}\right)\left(\frac{1}{2}\right)= \left(\frac{1}{2} + t_f\right)\xi.
        \end{align}
    \end{subequations}
    \item \textbf{PO}: Variables $(t_f, \xi)$.
    \begin{subequations}
        \begin{align}
            \frac{1}{2(k_f+u_f)} &= \left(\mathcal{E}_0^{(t_f)}\circ \tilde{\mathcal{E}}_{\xi^2}\right)\left(\frac{1}{2(1+u_i)}\right)= \frac{\xi^2}{2(1+u_i)} + t_f,
            \\
            \frac{1}{2(k_f-u_f)} &= \left(\mathcal{E}_0^{(t_f)}\circ \tilde{\mathcal{E}}_{1}\right)\left(\frac{1}{2(1-u_i)}\right)= \frac{1}{2(1-u_i)} + t_f,
            \\
            \frac{1}{2k_f} &=\left(\mathcal{E}_0^{(t_f)}\circ \tilde{\mathcal{E}}_{\xi}\right)\left(\frac{1}{2}\right)= \frac{\xi}{2} + t_f.
        \end{align}
    \end{subequations}
    \item \textbf{NO}: Variables $(t_f, \xi)$.
    \begin{subequations}
        \begin{align}
            \frac{1}{2(k_f+u_f)} &= \left(\mathcal{E}_0^{(t_f)}\circ \tilde{\mathcal{E}}_{1}\right)\left(\frac{1}{2(1+u_i)}\right)= \frac{1}{2(1+u_i)} + t_f,
            \\
            \frac{1}{2(k_f-u_f)} &= \left(\mathcal{E}_0^{(t_f)}\circ \tilde{\mathcal{E}}_{\xi^2}\right)\left(\frac{1}{2(1-u_i)}\right)= \frac{\xi^2}{2(1-u_i)} + t_f,
            \\
            \frac{1}{2k_f} &=\left(\mathcal{E}_0^{(t_f)}\circ \tilde{\mathcal{E}}_{\xi}\right)\left(\frac{1}{2}\right)= \frac{\xi}{2} + t_f.
        \end{align}
    \end{subequations}
    \item \textbf{PN and NP}: Variables $(\xi_1, \xi_2)$, and thus a null connection time.
    \begin{subequations}\label{subeq:PN-protocol}
        \begin{align}
            \frac{1}{2(k_f+u_f)} &= \left(\tilde{\mathcal{E}}_{1}\circ \tilde{\mathcal{E}}_{\xi_1^2}\right)\left(\frac{1}{2(1+u_i)}\right) = \frac{\xi_1^2}{2(1+u_i)}, 
            \\
            \frac{1}{2(k_f-u_f)} &= \left( \tilde{\mathcal{E}}_{\xi_2^2}\circ\tilde{\mathcal{E}}_{1}\right)\left(\frac{1}{2(1-u_i)}\right) = \frac{\xi_2^2}{2(1-u_i)},
            \\
            \frac{1}{2k_f} &= \left(\tilde{\mathcal{E}}_{\xi_2}\circ \tilde{\mathcal{E}}_{\xi_1}\right)\left(\frac{1}{2}\right) = \frac{\xi_1\xi_2}{2},
        \end{align}
    \end{subequations}
\end{itemize}


\subsection{Three-bang protocols}

In principle, there are twelve possible permutations of the vertices (O, P, N). Although the protocols POP and NON are actually \textit{singular} protocols---as mentioned in the main text and thoroughly studied in Appendix~\ref{app:singular-bang-bang}, we  consider them here because they have the three-bang structure as well. Similarly to what occurred for the two-bang protocols, the symmetry between the P and N points reduces our possible combinations from twelve to nine, as the pairs of protocols (PNO, NPO), (PNP, NPN) and (OPN, ONP) give the same algebraic equations. 
\begin{itemize}
    \item \textbf{OPO}: Variables $(\tau_1, \tau_2, \xi)$, with $\tau_1 + \tau_2 = t_f$.
    \begin{subequations}
        \begin{align}
            \frac{1}{2(k_f + u_f)} &= \left(\mathcal{E}_0^{(\tau_2)} \circ \tilde{\mathcal{E}}_{\xi^2}\circ \mathcal{E}_0^{(\tau_1)}\right)\left(\frac{1}{2(1+u_i)}\right)= \left(\frac{1}{2(1+u_i)} + \tau_1\right)\xi^2 +\tau_2,
            \\
            \frac{1}{2(k_f - u_f)} &= \left(\mathcal{E}_0^{(\tau_2)} \circ\tilde{\mathcal{E}}_{1}\circ \mathcal{E}_0^{(\tau_1)}\right)\left(\frac{1}{2(1-u_i)}\right) = \frac{1}{2(1-u_i)} + \tau_1 + \tau_2,
            \\
            \frac{1}{2k_f} &= \left(\mathcal{E}_0^{(\tau_2)} \circ\tilde{\mathcal{E}}_{\xi}\circ \mathcal{E}_0^{(\tau_1)}\right)\left(\frac{1}{2}\right)= \left(\frac{1}{2} + \tau_1\right)\xi + \tau_2.
        \end{align}
    \end{subequations}
    \item \textbf{ONO}: Variables $(\tau_1, \tau_2, \xi)$, with $\tau_1 + \tau_2 = t_f$.
    \begin{subequations}
        \begin{align}
            \frac{1}{2(k_f + u_f)} &= \left(\mathcal{E}_0^{(\tau_2)} \circ\tilde{\mathcal{E}}_{1}\circ \mathcal{E}_0^{(\tau_1)}\right)\left(\frac{1}{2(1+u_i)}\right) = \frac{1}{2(1+u_i)} + \tau_1 + \tau_2,
            \\
            \frac{1}{2(k_f - u_f)} &= \left(\mathcal{E}_0^{(\tau_2)} \circ \tilde{\mathcal{E}}_{\xi^2}\circ \mathcal{E}_0^{(\tau_1)}\right)\left(\frac{1}{2(1-u_i)}\right)= \left(\frac{1}{2(1-u_i)} + \tau_1\right)\xi^2 +\tau_2,
            \\
            \frac{1}{2k_f} &= \left(\mathcal{E}_0^{(\tau_2)} \circ\tilde{\mathcal{E}}_{\xi}\circ \mathcal{E}_0^{(\tau_1)}\right)\left(\frac{1}{2}\right)= \left(\frac{1}{2} + \tau_1\right)\xi + \tau_2.
        \end{align}
    \end{subequations}
    \item \textbf{POP} (Singular): Variables $(t_f, \xi_1, \xi_2)$.
    \begin{subequations}
        \begin{align}
            \frac{1}{2(k_f+u_f)} &= \left(\tilde{\mathcal{E}}_{\xi_2^2} \circ \mathcal{E}_0^{(t_f)}\circ \tilde{\mathcal{E}}_{\xi_1^2}\right)\left(\frac{1}{2(1+u_i)}\right)= \left(\frac{\xi_1^2}{2(1+u_i)} + t_f\right) \xi_2^2,
            \\
            \frac{1}{2(k_f-u_f)} &= \left(\tilde{\mathcal{E}}_{1} \circ \mathcal{E}_0^{(t_f)}\circ \tilde{\mathcal{E}}_{1}\right)\left(\frac{1}{2(1-u_i)}\right)= \frac{1}{2(1-u_i)} + t_f,
            \\
            \frac{1}{2k_f} &=\left(\tilde{\mathcal{E}}_{\xi_2}\circ \mathcal{E}_0^{(t_f)}\circ \tilde{\mathcal{E}}_{\xi_1}\right)\left(\frac{1}{2}\right)= \left(\frac{\xi_1}{2} + t_f\right) \xi_2.
        \end{align}
    \end{subequations}
    \item \textbf{NON} (Singular): Variables $(t_f, \xi_1, \xi_2)$.
    \begin{subequations}
        \begin{align}
            \frac{1}{2(k_f+u_f)} &= \left(\tilde{\mathcal{E}}_{1} \circ \mathcal{E}_0^{(t_f)}\circ \tilde{\mathcal{E}}_{1}\right)\left(\frac{1}{2(1+u_i)}\right)= \frac{1}{2(1+u_i)} + t_f,
            \\
            \frac{1}{2(k_f-u_f)} &= \left(\tilde{\mathcal{E}}_{\xi_2^2} \circ \mathcal{E}_0^{(t_f)}\circ \tilde{\mathcal{E}}_{\xi_1^2}\right)\left(\frac{1}{2(1-u_i)}\right)= \left(\frac{\xi_1^2}{2(1-u_i)} + t_f\right) \xi_2^2,
            \\
            \frac{1}{2k_f} &=\left(\tilde{\mathcal{E}}_{\xi_2}\circ \mathcal{E}_0^{(t_f)}\circ \tilde{\mathcal{E}}_{\xi_1}\right)\left(\frac{1}{2}\right)= \left(\frac{\xi_1}{2} + t_f\right) \xi_2.
        \end{align}
    \end{subequations}
    \item \textbf{OPN and ONP}: Variables $(t_f, \xi_1, \xi_2)$.
    \begin{subequations}
        \begin{align}
            \frac{1}{2(k_f + u_f)} &= \left(\tilde{\mathcal{E}}_{1}\circ \tilde{\mathcal{E}}_{\xi_1^2}\circ \mathcal{E}_0^{(t_f)}\right)\left(\frac{1}{2(1+u_i)}\right)= \left(\frac{1}{2(1+u_i)} + t_f\right)\xi_1^2,
            \\
            \frac{1}{2(k_f - u_f)} &= \left(\tilde{\mathcal{E}}_{\xi_2^2}\circ \tilde{\mathcal{E}}_{1}\circ \mathcal{E}_0^{(t_f)}\right)\left(\frac{1}{2(1-u_i)}\right) = \left(\frac{1}{2(1-u_i)} + t_f\right)\xi_2^2,
            \\
            \frac{1}{2k_f} &= \left(\tilde{\mathcal{E}}_{\xi_2}\circ \tilde{\mathcal{E}}_{\xi_1}\circ \mathcal{E}_0^{(t_f)}\right)\left(\frac{1}{2}\right)= \left(\frac{1}{2} + t_f\right)\xi_1\xi_2.
        \end{align}
    \end{subequations}
    \item \textbf{PON}: Variables $(t_f, \xi_1, \xi_2)$.
    \begin{subequations}\label{eq:PON-infinite-kmax}
        \begin{align}
            \frac{1}{2(k_f+u_f)} &= \left(\tilde{\mathcal{E}}_{1}\circ \mathcal{E}_0^{(t_f)}\circ \tilde{\mathcal{E}}_{\xi_1^2}\right)\left(\frac{1}{2(1+u_i)}\right)= \frac{\xi_1^2}{2(1+u_i)} + t_f,
            \\
            \frac{1}{2(k_f-u_f)} &= \left(\tilde{\mathcal{E}}_{\xi_2^2}\circ\mathcal{E}_0^{(t_f)}\circ \tilde{\mathcal{E}}_{1}\right)\left(\frac{1}{2(1-u_i)}\right)= \left(\frac{1}{2(1-u_i)} + t_f\right)\xi_2^2,
            \\
            \frac{1}{2k_f} &=\left(\tilde{\mathcal{E}}_{\xi_2}\circ \mathcal{E}_0^{(t_f)}\circ \tilde{\mathcal{E}}_{\xi_1}\right)\left(\frac{1}{2}\right)= \left(\frac{\xi_1}{2} + t_f\right)\xi_2.
        \end{align}
    \end{subequations}
    \item \textbf{NOP}: Variables $(t_f, \xi_1, \xi_2)$.
    \begin{subequations}
        \begin{align}
            \frac{1}{2(k_f+u_f)} &= \left(\tilde{\mathcal{E}}_{\xi_2^2}\circ\mathcal{E}_0^{(t_f)}\circ \tilde{\mathcal{E}}_{1}\right)\left(\frac{1}{2(1+u_i)}\right)= \left(\frac{1}{2(1+u_i)} + t_f\right)\xi_2^2,
            \\
            \frac{1}{2(k_f-u_f)} &= \left(\tilde{\mathcal{E}}_{1}\circ \mathcal{E}_0^{(t_f)}\circ \tilde{\mathcal{E}}_{\xi_1^2}\right)\left(\frac{1}{2(1-u_i)}\right)= \frac{\xi_1^2}{2(1-u_i)} + t_f,
            \\
            \frac{1}{2k_f} &=\left(\tilde{\mathcal{E}}_{\xi_2}\circ \mathcal{E}_0^{(t_f)}\circ \tilde{\mathcal{E}}_{\xi_1}\right)\left(\frac{1}{2}\right)= \left(\frac{\xi_1}{2} + t_f\right)\xi_2.
        \end{align}
    \end{subequations}
    \item \textbf{PNO and NPO}: Variables $(t_f, \xi_1, \xi_2)$.
    \begin{subequations}
        \begin{align}
            \frac{1}{2(k_f+u_f)} &= \left(\mathcal{E}_0^{(t_f)}\circ\tilde{\mathcal{E}}_{1}\circ \tilde{\mathcal{E}}_{\xi_1^2}\right)\left(\frac{1}{2(1+u_i)}\right) = \frac{\xi_1^2}{2(1+u_i)} + t_f, 
            \\
            \frac{1}{2(k_f-u_f)} &= \left(\mathcal{E}_0^{(t_f)}\circ \tilde{\mathcal{E}}_{\xi_2^2}\circ\tilde{\mathcal{E}}_{1}\right)\left(\frac{1}{2(1-u_i)}\right) = \frac{\xi_2^2}{2(1-u_i)} + t_f,
            \\
            \frac{1}{2k_f} &= \left(\mathcal{E}_0^{(t_f)}\circ\tilde{\mathcal{E}}_{\xi_2}\circ \tilde{\mathcal{E}}_{\xi_1}\right)\left(\frac{1}{2}\right) = \frac{\xi_1\xi_2}{2} + t_f.
        \end{align}
    \end{subequations}
    \item \textbf{PNP and NPN}: Variables $(\xi_1, \xi_2, \xi_3)$, and thus a null connection time.
    \begin{subequations}
        \begin{align}
            \frac{1}{2(k_f+u_f)} &= \left(\tilde{\mathcal{E}}_{\xi_3^2} \circ\tilde{\mathcal{E}}_{1}\circ \tilde{\mathcal{E}}_{\xi_1^2}\right)\left(\frac{1}{2(1+u_i)}\right) = \frac{\xi_1^2\xi_3^2}{2(1+u_i)}, 
            \\
            \frac{1}{2(k_f-u_f)} &= \left(\tilde{\mathcal{E}}_{1}\circ \tilde{\mathcal{E}}_{\xi_2^2}\circ\tilde{\mathcal{E}}_{1}\right)\left(\frac{1}{2(1-u_i)}\right) = \frac{\xi_2^2}{2(1-u_i)},
            \\
            \frac{1}{2k_f} &= \left(\tilde{\mathcal{E}}_{\xi_3} \circ\tilde{\mathcal{E}}_{\xi_2}\circ \tilde{\mathcal{E}}_{\xi_1}\right)\left(\frac{1}{2}\right) = \frac{\xi_1\xi_2\xi_3}{2},
        \end{align}
    \end{subequations}
\end{itemize}
We highlight that the PNP and NPN protocols present an undetermined system of equations, as the variables $\xi_1$ and $\xi_3$ always appear together as $\xi_1 \xi_3$. Therefore, we could just refer to that product as a new quenching variable $\tilde{\xi}_1 \equiv \xi_1 \xi_3$. Thus, the resulting system is equivalent to that for the PN and NP protocols in Eq.~\eqref{subeq:PN-protocol} with $\xi_1\to\tilde{\xi}_1$.


\section{Singular bang-bang protocols}\label{app:singular-bang-bang}

Let us start by considering the three-bang protocol POP---due to the symmetry between the P and N points, the results obtained here are also valid for the protocol NON. We do not consider the limit $k_{\max}\rightarrow +\infty$ in this section, since it becomes a singular limit when studying the derivatives of Pontryagin's Hamiltonian function. It suffices to proceed with the forthcoming analysis for $k_{\max}$ finite, and then extend our results to the $k_{\max}\rightarrow +\infty$ regime, which is the one we employ throughout the majority of this work. 

Let us recall that the switching function $\phi_{\text{OP}}(t)$ is given by Eq.~\eqref{eq:switching-functions-OP}, it accounts for the derivative of Pontryagin's Hamiltonian function along the $\overline{\text{OP}}$ edge of the triangular control set. Since the POP protocol involves two switchings---one switch from P to O at some time $t_1 \in (0,t_f)$ followed by another one from O to P at $t_2 \in (t_1,t_f)$, $\phi_{\text{OP}}(t)$ must have at least two zeros on the entire interval $(0,t_f)$. The protocols starts from vertex P, so we have 
\begin{equation}\label{eq:bound-initial}
    \phi_{\text{OP}}(0) = -2\left[2\psi_1(0) z_1(0) + \psi_3(0) z_3(0)\right] = -2\left(\frac{\psi_{1,0}}{1+u_i} + \frac{\psi_{3,0}}{2} \right) > 0.
\end{equation}
Now, by integrating Hamilton's canonical equations \eqref{eq:canonical} at vertex P, we  write our switching function explicitly as
\begin{equation}
    \phi_{\text{OP}}(t) = \phi_{\text{OP}}(0) - \frac{\psi_{1,0}}{k_{\max}}\left(e^{4k_{\max}t}-1\right) -\frac{\psi_{3,0}}{k_{\max}}\left(e^{2k_{\max}t}-1\right), \quad 0\leq t \leq t_1.
\end{equation}
This function constitutes a second-degree polynomial for the variable $\text{exp}(2k_{\max}t)$, and thus, it presents two roots at most---i.e. up to two candidates for the switching time $t_1$, for which $\phi_{\text{OP}}(t_1)=0$. The fact that $t_1$ must be positive sets bounds on the sign of the ratio $\psi_{1,0}/\psi_{3,0}$, which must be consistent with that in Eq.~\eqref{eq:bound-initial}. Once we switch to the point O, by integrating once again the canonical equations at vertex O, we get
\begin{equation}
    \phi_{\text{OP}}(t) = \cancelto{0}{\phi_{\text{OP}}(t_1)} -2\left(2\psi_{1,0}e^{4k_{\max}t_1} + \psi_{3,0}e^{2k_{\max}t_1}\right)(t-t_1), \quad t \geq t_1.
\end{equation}
However, this corresponds to a linear function starting from zero, which does not have any roots apart from the trivial one at $t = t_1$, and thus, the switching back to the last time window at point P is forbidden. { Therefore, a regular three-bang POP protocol is not possible. Still, a three-bang PON protocol would be indeed feasible, since the switching to vertex N is ruled by the switching function $\phi_{\text{ON}}(t)$, for which we do not have any restriction.}

Another possibility to overcome the found issue could be to consider $\phi_{\text{OP}}(t) = 0$ for $t\geq t_1$, which would set the value of the ratio $\psi_{1,0}/\psi_{3,0}$ {from the condition $2\psi_{1,0}e^{4k_{\max}t_1} + \psi_{3,0}e^{2k_{\max}t_1}=0$}. In this case, the time window at point O would become a singular time interval. However, by virtue of our analysis of singular protocols in Appendix~\ref{app:singular-protocols}, if $\phi_{\text{OP}}(t) = 0$ for some time window, then it must be zero fon the whole interval $(0,t_f)$. Henceforth, the protocols POP and NON---and any other higher order protocol involving the POP and/or NON combinations---belong in the class of singular protocols depicted in Fig.~\ref{fig:switching-funcs-singular}.


\section{Brachistochrones for uncoupled oscillators}\label{app:uncoupled-oscillators}

Let us consider the $u(t)=0$ case for the Brownian Gyrator, which corresponds to connecting arbitrary stationary states for two uncorrelated overdamped oscillators coupled to two different thermal baths. Such states are thus characterised uniquely by the final value of the control parameter $k(t)$, i.e. $k_f$. As $u_i = u_f = 0$, all the dynamical variables of our original system degenerate towards a unique one $z_j(t) = z(t)$, $j=1,2,3$, whose dynamical evolution is given by
\begin{equation}
\label{equilibrium-eq}
    \dot{z} = -2k(t)z + 1,
\end{equation}
with $k(t) \in [0,k_{\max}]$ once again. Due to the simplicity of this situation---since we are dealing only with one dynamical variable---we derive the minimum connection times by other means. 

Equation~\eqref{equilibrium-eq} can be integrated analytically, thus providing an integral form for the final connection time
\begin{equation}
\label{connection-time-integral}
    t_f = \int_{z_i}^{z_f} \frac{dz}{1-2 k(z)z},
\end{equation}
with $z_i \equiv z(t=0) = 1/2$---recall, once more, that $k_i = 1$ in our scaled units--- and $z_f = z(t=t_f) = 1/(2k_f)$. Now, $t_f$ does not diverge as long as the denominator $1-2k(z)z \neq 0 \ \forall z \in [z_i,z_f]$. This implies that the evolution of $z$ must be monotonic for the optimal time connection: given an initial $z_i$, if $z_f > z_i$ ($z_f < z_i$), then we need $\dot{z} > 0$ ($\dot{z} < 0$). Thus, we have two possibilities:
\begin{itemize}
    \item $2k(z)z > 1$: In this case, the integrand becomes negative during the entire integration interval. The connection time will only be positive thus if $z_f < z_i$, implying that $k_f > 1$. The integral from Eq.~\eqref{connection-time-integral} is thus bounded from below by
    \begin{equation}
        t_f = \int_{z_i}^{z_f} \frac{dz}{1-2 k(z)z} = \int_{z_f}^{z_i} \frac{dz}{2 k(z)z-1} \geq \int_{z_f}^{z_i} \frac{dz}{2 k_{\max}z-1} = \frac{1}{2k_{\max}} \ln \left(k_f \frac{k_{\max}-1}{k_{\max}-k_f} \right),
    \end{equation}
    which corresponds to the final connection time for a bang-bang protocol with $k(t) = k_{\max}$.
    \item $2k(z)z < 1$: Here, the integrand is positive, thus implying that $z_f > z_i$ and $k_f < 1$. In this case, the integral from Eq.~\eqref{connection-time-integral} is bounded from below by
    \begin{equation}
        t_f = \int_{z_i}^{z_f} \frac{dz}{1-2 k(z)z} \geq \int_{z_i}^{z_f}dz = z_f - z_i = \frac{1}{2}\left(\frac{1}{k_f} - 1 \right),
    \end{equation}
    which corresponds to the final connection time for a bang-bang protocol with $k(t) = 0$.
\end{itemize}

Given the above, we have proved that optimal protocols achieving the fastest connection between stationary states for uncorrelated oscillators are also the bang-bang type, specifically corresponding to one-bang protocols of the type $k(t) = 0$, for $k_f<1$, and $k(t) = k_{\max}$, for $k_f>1$, for all $t \in (0,t_f)$. 

{

\section{Connection times as functions of $k_{\max}$}\label{app:finite-kmax}

Let us analyse the convergence to the $k_{\max} \to \infty$ limit considered in the main text. To be concrete, we consider here the PON three-bang protocol to illustrate the results---although the discussion may also be applied to the rest of bang-bang protocols. As a three-bang protocol, its time duration $t_f$ is divided into three windows of lengths $t_{\text{P}}$, $t_{\text{O}}$ and $t_{\text{N}}$, corresponding to the P, O, and N vertices, respectively. Employing the $\mathcal{E}^{(\tau)}_{\omega_{j,\text{V}}}(z_{j,0})$ evolution operators introduced in Sec.~\ref{sec:constructing-bang-bang} gives the following system of non-linear equations:
\begin{subequations}\label{eq:PON-finite-kmax}
        \begin{align}
            \frac{1}{2(k_f+u_f)} &= \left(\mathcal{E}_{0}^{(t_{\text{N}})}\circ \mathcal{E}_0^{(t_{\text{O}})}\circ \mathcal{E}_{2k_{\max}}^{(t_{\text{P}})}\right)\left(\frac{1}{2(1+u_i)}\right)= \frac{1}{4k_{\max}} + \left(\frac{1}{2(1+u_i)}-\frac{1}{4k_{\max}}\right)e^{-4k_{\max}t_{\text{P}}}+t_{\text{O}}+t_{\text{N}},
            \\
            \frac{1}{2(k_f-u_f)} &= \left(\mathcal{E}_{2k_{\max}}^{(t_{\text{N}})}\circ\mathcal{E}_0^{(t_{\text{O}})}\circ \mathcal{E}_{0}^{(t_{\text{P}})}\right)\left(\frac{1}{2(1-u_i)}\right)= \frac{1}{4k_{\max}} + \left(\frac{1}{2(1-u_i)}+t_{\text{P}}+t_{\text{O}}-\frac{1}{4k_{\max}}\right)e^{-4k_{\max}t_{\text{N}}},
            \\
            \frac{1}{2k_f} &=\left(\mathcal{E}_{k_{\max}}^{(t_{\text{N}})}\circ \mathcal{E}_0^{(t_{\text{O}})}\circ \mathcal{E}_{k_{\max}}^{(t_{\text{P}})}\right)\left(\frac{1}{2}\right)= \frac{1}{2k_{\max}} + \left[\left(\frac{1}{2}-\frac{1}{2k_{\max}} \right)e^{-2k_{\max}t_{\text{P}}}+t_{\text{O}} \right]e^{-2k_{\max}t_{\text{N}}}.
        \end{align}
    \end{subequations}
At variance with the algebraic system of equations \eqref{eq:PON-infinite-kmax} for $(t_f,\xi_1,\xi_2)$ in the limit $k_{\max}\rightarrow \infty$, the system \eqref{eq:PON-finite-kmax} is no longer algebraic for the times $(t_{\text{P}},t_{\text{O}},t_{\text{N}})$.  

Since one cannot derive closed expressions for $(t_{\text{P}},t_{\text{O}},t_{\text{N}})$ from the system of equations~\eqref{eq:PON-finite-kmax} for finite  $k_{\max}$, we develop in the following a perturbative theory in powers of $k_{\max}^{-1}$ to gather some insight on these times as a function of $k_{\max}$. Later, we will compare the so obtained perturbative solution with the numerical solution of Eq.~\eqref{eq:PON-finite-kmax}. Specifically, we introduce the following expansions for $(t_{\text{P}},t_{\text{O}},t_{\text{N}})$:
\begin{equation}\label{eq:pert-expansion}
    t_{\text{P,N}} = \frac{t_{\text{P,N}}^{(0)}}{k_{\max}} + \frac{t_{\text{P,N}}^{(1)}}{k_{\max}^2} + \cdots=\frac{1}{k_{\max}} \left( t_{\text{P,N}}^{(0)} + \frac{t_{\text{P,N}}^{(1)}}{k_{\max}}+ \cdots\right), \qquad t_{\text{O}} = t_{\text{O}}^{(0)} + \frac{t_{\text{O}}^{(1)}}{k_{\max}} + \cdots.
\end{equation}
Note that the leading term for both $t_{\text{P}}$ and $t_{\text{N}}$ is of the order $O(k_{\max}^{-1})$, whereas the leading term for $t_{\text{O}}$ is $O(1)$. This is consistent with our discussion in Sec.~\ref{sec:constructing-bang-bang}, where it was argued that the times spent at the vertices P and N vanish as $k_{\max}^{-1}$ for large $k_{\max}$. The total time for the connection is
\begin{equation}
    t_f=t_{\text{P}}+t_{\text{O}}+t_{\text{N}}=t_f^{(0)}+\frac{t_f^{(1)}}{k_{\max}}+\cdots,
\end{equation}
where
\begin{equation}
    t_f^{(0)}=t_{\text{O}}^{(0)}, \qquad t_f^{(1)}=t_{\text{O}}^{(1)}+t_{\text{P}}^{(0)}+t_{\text{N}}^{(0)}.
\end{equation}

Inserting the expansions~\eqref{eq:pert-expansion} into the system~\eqref{eq:PON-finite-kmax} provides us with a hierarchy of equations for the coefficients in the expansion, which can be solved recursively. To the lowest order, we get
\begin{subequations}\label{eq:PON-O1}
        \begin{align}
            \frac{1}{2(k_f+u_f)} &= \frac{e^{-4t_{\text{P}}^{(0)}}}{2(1+u_i)} + t_{\text{O}}^{(0)},
            \\
            \frac{1}{2(k_f-u_f)} &=  \left(\frac{1}{2(1-u_i)} + t_{\text{O}}^{(0)}\right)e^{-4t_{\text{N}}^{(0)}},
            \\
            \frac{1}{2k_f} &= \left(\frac{e^{-2t_{\text{P}}^{(0)}}}{2} + t_{\text{O}}^{(0)}\right)e^{-2t_{\text{N}}^{(0)}}.
        \end{align}
    \end{subequations}
With the identifications, consistently with Eq.~\eqref{eq:quenching-factor},
\begin{equation}
    t_{\text{P}}^{(0)} = -\frac{1}{2}\ln \xi_1, \quad t_{\text{N}}^{(0)} = -\frac{1}{2}\ln \xi_2.
\end{equation}
we conclude that the system~\eqref{eq:PON-O1} is, logically, equivalent to that in Eq.~\eqref{eq:PON-infinite-kmax} of Appendix~\ref{app:explicit-bang-bang}, derived there in the limit $k_{\max}\to\infty$---with the role of $t_f$ played here by $t_{\text{O}}^{(0)}=t_f^{(0)}$. The following order gives
\begin{subequations}\label{eq:PON-O2}
        \begin{align}
            0 &= \frac{1}{4}-\xi_1^2\left(\frac{1}{4}+\frac{2t_{\text{P}}^{(1)}}{1+u_i}\right) -\frac{1}{2}\ln \xi_2 + t_{\text{O}}^{(1)},
            \\
            0 &=  \frac{1}{4} + \xi_2^2\left[-2t_{\text{N}}^{(1)}\left(\frac{1}{1-u_i}+2t_f^{(0)}\right)-\frac{1}{2}\ln \xi_1 + t_{\text{O}}^{(1)}-\frac{1}{4}\right],
            \\
            0 &= \frac{1}{2} + \xi_2 \left[-t_{\text{N}}^{(1)}\left(\xi_1+2t_f^{(0)}\right)+t_{\text{O}}^{(1)}-\xi_1\left(\frac{1}{2} + t_{\text{P}}^{(1)}\right)\right],
        \end{align}
\end{subequations}
which makes it possible to obtain the first order corrections to the $k_{\max}\to\infty$ result. The system~\eqref{eq:PON-O2} is linear in the variables $(t_{\text{P}}^{(1)},t_{\text{O}}^{(1)},t_{\text{N}}^{(1)})$, thus allowing for explicit analytical solutions thereof as functions of $(t_f^{(0)}, \xi_1, \xi_2)$. Nevertheless, we do not display such analytical expressions here, since they involve rather cumbersome formulae and do not provide any useful insight. 

\begin{figure}
  \centering\includegraphics[width=0.65\textwidth]{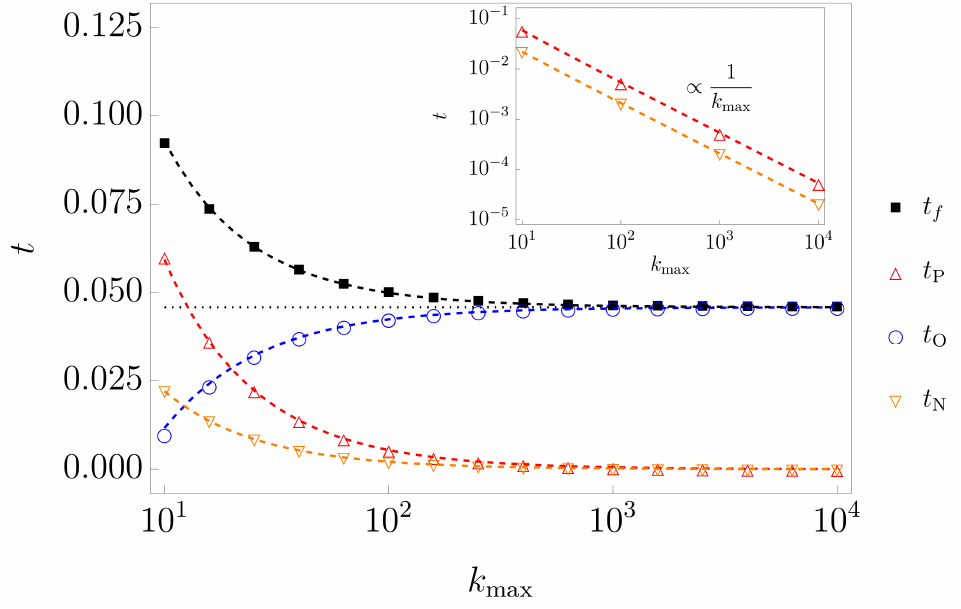}
  \caption{{Time spans $(t_{\text{P}},t_{\text{O}},t_{\text{N}})$ and total connection time $t_f = t_{\text{P}} + t_{\text{O}} + t_{\text{N}}$ as a function of $k_{\max}$, for the regular three-bang PON protocol. Specifically, the plotted data correspond to the connection between the initial NESS with $(k_i,u_i) = (1,0.5)$ and the target NESS with $(k_f,u_f) = (3.5,2.4)$. Symbols give the numerical solution of Eq.~\eqref{eq:PON-finite-kmax} for different values of $k_{\max}$ for each of the time spans: $t_f$ (filled black squares), $t_{\text{P}}$ (empty red up triangles), $t_{\text{O}}$ (empty blue circles), and $t_{\text{N}}$ (empty orange down triangles), whereas the dashed curves account for the perturbative expansions from Eq.~\eqref{eq:pert-expansion}, keeping their first two non-vanishing terms. The black dotted line corresponds to the asymptotic value $t_f^{(0)}$, which is approached by both $t_f$ and $t_{\text{O}}$ in the limit $k_{\max}\rightarrow \infty$. The inset shows the behaviour of both $t_{\text{P}}$ and $t_{\text{N}}$ in logarithmic scale, in order to appreciate its approximate $1/k_{\max}$ behaviour.}
  }
  \label{fig:tf-finite-km}
\end{figure}
In Fig.~\ref{fig:tf-finite-km}, we show the behaviour of the spans $(t_{\text{P}},t_{\text{O}},t_{\text{N}})$, corresponding to each of the time windows of the PON three-bang protocol, as functions of $k_{\max}$. We have chosen to study the connection between the states $(k_i,u_i) = (1,0.5)$ and $(k_f,u_f) = (3.5,2.4)$, such that we ensure that the final state $(k_f,u_f)$ lies deep within the PON yellow region for $k_{\max}\to\infty$ shown in Fig.~\ref{fig:three-bangs-uipos}. As can be neatly observed, the agreement between the numerical solution of the system of equations~\eqref{eq:PON-finite-kmax} (symbols) and the perturbative expansions up two the first two non-vanishing contributions from Eq.~\eqref{eq:pert-expansion} (dashed curves) is excellent, even for the smallest value of $k_{\max}$ displayed, $k_{\max} = 10$. As expected on physical grounds, the connection time $t_f$ decreases with $k_{\max}$: a higher capacity for compression and thus a larger control set leads to a faster brachistochrone. This monotonic tendency is also shared by both $t_{\text{P}}$ and $t_{\text{N}}$, for which our perturbative approach predicts that they decay as $k_{\max}^{-1}$---as shown in the inset. However, $t_{\text{O}}$ follows a different trend, increasing monotonically with $k_{\max}$ towards its asymptotic value $t_f^{(0)}$. In fact, had we kept decreasing the value for $k_{\max}$, it would have hit zero around $k_{\max}\simeq 7.99$: at that point, the PON protocol to reach the final state $(k_f,u_f) = (3.5,2.4)$ coalesces with the PN protocol. Such limiting value of $k_{\max}$ reflects the fact that the reachable states for the PN two-bang protocol also vary with $k_{\max}$---the orange curves plotted in Fig.~\ref{fig:two-bang-protocols} correspond to the limit $k_{\max}\rightarrow \infty$. In particular, the point $(k_f,u_f) = (3.5,2.4)$ can be shown to belong to the PN line for $k_{\max}\simeq 7.99$.
}

\section{Bound on the connection from the speed limit inequality~\eqref{eq:speed-limit}}\label{app:speed-limit-bound}

In this appendix, we evaluate the contribution to the irreversible work $\expval{W_{\irr}}$ defined in Eq.~\eqref{eq:irr-work} from the bangs at the different vertices O, P, and N. Throughout the appendix, we consider that the bang starts at a certain time $t_1$ and ends at a certain time $t_2\ge t_1$, so that the duration of the bang is $\tau=t_2-t_1$. It is useful to write $\expval{W_{\irr}}$ as
\begin{equation}
    \expval{W_{\irr}}=\expval{W_{\irr}^{(1)}}+\expval{W_{\irr}^{(2)}}, \quad \expval{W_{\irr}^{(j)}}=\gamma \int_0^{t_f}dt\, \left(\frac{d\sigma_j}{dt}\right)^2. 
\end{equation}

Consistently with the non-dimensionalisation introduced in Eq.~\eqref{eq:dimensionless-variables}, we introduce dimensionless standard deviations $\sigma_j^*$ as:
\begin{equation}
    \sigma_j=\sqrt{\frac{k_B(T_x+T_y)}{k_i}}\sigma_j^* \iff \sigma_j^*=\sqrt{z_j}.
\end{equation}
Therefore, we obtain
\begin{equation}
   \expval{W_{\irr}^{(j)}}=\cancel{\gamma} \frac{\cancel{k_i}}{\cancel{\gamma}} \frac{k_B(T_x+T_y)}{\cancel{k_i}}\int_0^{t_f^*} dt^*\, \left(\frac{d\sigma_j^*}{dt^*}\right)^2 ,
\end{equation}
which suggest the following definition for the non-dimensionless work:
\begin{equation}
    \expval{W_{\irr}^{(j)}}=k_B(T_x+T_y)\expval{W_{\irr}^{(j)}}^* \implies \expval{W_{\irr}^{(j)}}^*=\int_0^{t_f^*} dt^*\, \left(\frac{d\sigma_j^*}{dt^*}\right)^2 .
\end{equation}
As in the remainder of the paper, we drop the asterisks in the following to simplify our notation.

\subsection{Vertex O}

At the vertex O, we have that $k=u=0$, so the evolution equations for the variances $z_j=\sigma_j^2$ are $\dot{z}_j=1$ and thus
\begin{equation}
    z_j(t)=z_j(t_1)+(t-t_1) \implies \sigma_j(t)=\sqrt{\sigma_j^2(t_1)+(t-t_1)}, \quad t_1\le t\le t_2.
\end{equation}
Therefore, we have that
\begin{equation}
    \expval{W_{\irr}^{(j)}}= \int_{t_1}^{t_2}dt\, \left[\sigma_j^2(t_1)+(t-t_1)\right]^{-1}=\ln \left[\frac{\sigma_j^2(t_2)}{\sigma_j^2(t_1)}\right],
\end{equation}
and, finally
\begin{equation}
    \expval{W_{\irr}}=2\ln \left[\frac{\sigma_1(t_2)\sigma_2(t_2)}{\sigma_1(t_1)\sigma_2(t_1)}\right].
\end{equation}

\subsection{Vertex P}

At the vertex P, $k=u=k_{\max}$, so the evolution equation for the variances are
\begin{equation}
    \frac{dz_1}{dt}=-4k_{\max}z_1+1, \quad \frac{dz_2}{dt}=1.
\end{equation}
The equation for $z_2$ is the same as for the vertex O, so we do not need to derive the contribution $\expval{W_{\irr}^{(2)}}$. We thus focus on the evolution of $z_1$---or $\sigma_1$---and its corresponding contribution $\expval{W_{\irr}^{(1)}}$:
\begin{equation}
    \sigma_1(t)=\sqrt{\frac{1}{4k_{\max}}+\left(\sigma_1^2(t_1)-\frac{1}{4k_{\max}}\right)e^{-4k_{\max}t}},
\end{equation}
from which we derive
\begin{equation}
    \expval{W_{\irr}^{(1)}}=k_{\max}\left\{\sigma_1^2(t_1)-\sigma_1^2(t_2)+\frac{1}{4k_{\max}}\ln\left[\frac{\sigma_1^2(t_2)}{\sigma_1^2(t_1)}\right]\right\}.
\end{equation}
In the limit $k_{\max}\to\infty$ we have considered, the bang at vertex P becomes instantaneous, $\tau=t_2-t_1\to 0$, but $\xi=e^{-2k_{\max}t/\gamma}<1$ remains finite---as discussed in Sec.~\ref{sec:constructing-bang-bang}. Then, we get
\begin{equation}
    \expval{W_{\irr}^{(1)}}\sim k_{\max}\,\sigma_1^2(t_1)\left(1-\xi^2\right),
\end{equation}
and this contribution to the irreversible work diverges linearly with $k_{\max}$. An analogous divergent behaviour of the cost of the connection has been found for the thermodynamic geometry cost in other systems~\cite{prados_optimizing_2021,patron_thermal_2022}---although, therein, the parameter that was controlled was the bath temperature instead of the confining potential.

\subsection{Vertex N}

At the vertex N, $k=-u=k_{\max}$, so the evolution equation for the variances are
\begin{equation}
    \frac{dz_1}{dt}=1, \quad \frac{dz_2}{dt}=-4k_{\max}z_2+1.
\end{equation}
As repeatedly commented along the paper, there is a symmetry between the P and N vertices. For the calculation of the contribution to the irreversible work we are considering here, we see that we have only to exchange the labels $1\leftrightarrow 2$. Therefore, for the vertex N, it is $\expval{W_{\irr}^{(1)}}$ that coincides with our calculation for vertex O, whereas
\begin{equation}
    \expval{W_{\irr}^{(2)}}=k_{\max}\left\{\sigma_2^2(t_1)-\sigma_2^2(t_2)+\frac{1}{4k_{\max}}\ln\left[\frac{\sigma_2^2(t_2)}{\sigma_2^2(t_1)}\right]\right\},
\end{equation}
which, in the limit $k_{\max}\to\infty$, reduces to
\begin{equation}
    \expval{W_{\irr}^{(2)}}\sim k_{\max}\,\sigma_2^2(t_1)\left(1-\xi^2\right).
\end{equation}

As a consequence of the above results, the irreversible work for any bang-bang protocol involving at least one P or N vertex---i.e. any bang-bang protocol different from the one-bang at vertex O---diverges linearly with $k_{\max}$.

\bibliography{PPyP24.bib}

\begin{thebibliography}{72}%
\makeatletter
\providecommand \@ifxundefined [1]{%
 \@ifx{#1\undefined}
}%
\providecommand \@ifnum [1]{%
 \ifnum #1\expandafter \@firstoftwo
 \else \expandafter \@secondoftwo
 \fi
}%
\providecommand \@ifx [1]{%
 \ifx #1\expandafter \@firstoftwo
 \else \expandafter \@secondoftwo
 \fi
}%
\providecommand \natexlab [1]{#1}%
\providecommand \enquote  [1]{``#1''}%
\providecommand \bibnamefont  [1]{#1}%
\providecommand \bibfnamefont [1]{#1}%
\providecommand \citenamefont [1]{#1}%
\providecommand \href@noop [0]{\@secondoftwo}%
\providecommand \href [0]{\begingroup \@sanitize@url \@href}%
\providecommand \@href[1]{\@@startlink{#1}\@@href}%
\providecommand \@@href[1]{\endgroup#1\@@endlink}%
\providecommand \@sanitize@url [0]{\catcode `\\12\catcode `\$12\catcode `\&12\catcode `\#12\catcode `\^12\catcode `\_12\catcode `\%12\relax}%
\providecommand \@@startlink[1]{}%
\providecommand \@@endlink[0]{}%
\providecommand \url  [0]{\begingroup\@sanitize@url \@url }%
\providecommand \@url [1]{\endgroup\@href {#1}{\urlprefix }}%
\providecommand \urlprefix  [0]{URL }%
\providecommand \Eprint [0]{\href }%
\providecommand \doibase [0]{https://doi.org/}%
\providecommand \selectlanguage [0]{\@gobble}%
\providecommand \bibinfo  [0]{\@secondoftwo}%
\providecommand \bibfield  [0]{\@secondoftwo}%
\providecommand \translation [1]{[#1]}%
\providecommand \BibitemOpen [0]{}%
\providecommand \bibitemStop [0]{}%
\providecommand \bibitemNoStop [0]{.\EOS\space}%
\providecommand \EOS [0]{\spacefactor3000\relax}%
\providecommand \BibitemShut  [1]{\csname bibitem#1\endcsname}%
\let\auto@bib@innerbib\@empty
\bibitem [{\citenamefont {Guéry-Odelin}\ \emph {et~al.}(2019)\citenamefont {Guéry-Odelin}, \citenamefont {Ruschhaupt}, \citenamefont {Kiely}, \citenamefont {Torrontegui}, \citenamefont {Martínez-Garaot},\ and\ \citenamefont {Muga}}]{guery-odelin_shortcuts_2019}%
  \BibitemOpen
  \bibfield  {author} {\bibinfo {author} {\bibfnamefont {D.}~\bibnamefont {Guéry-Odelin}}, \bibinfo {author} {\bibfnamefont {A.}~\bibnamefont {Ruschhaupt}}, \bibinfo {author} {\bibfnamefont {A.}~\bibnamefont {Kiely}}, \bibinfo {author} {\bibfnamefont {E.}~\bibnamefont {Torrontegui}}, \bibinfo {author} {\bibfnamefont {S.}~\bibnamefont {Martínez-Garaot}},\ and\ \bibinfo {author} {\bibfnamefont {J.~G.}\ \bibnamefont {Muga}},\ }\bibfield  {title} {\bibinfo {title} {Shortcuts to adiabaticity: {Concepts}, methods, and applications},\ }\href {https://doi.org/10.1103/RevModPhys.91.045001} {\bibfield  {journal} {\bibinfo  {journal} {Reviews of Modern Physics}\ }\textbf {\bibinfo {volume} {91}},\ \bibinfo {pages} {045001} (\bibinfo {year} {2019})}\BibitemShut {NoStop}%
\bibitem [{\citenamefont {Chen}\ \emph {et~al.}(2010{\natexlab{a}})\citenamefont {Chen}, \citenamefont {Ruschhaupt}, \citenamefont {Schmidt}, \citenamefont {del Campo}, \citenamefont {Guéry-Odelin},\ and\ \citenamefont {Muga}}]{chen_fast_2010}%
  \BibitemOpen
  \bibfield  {author} {\bibinfo {author} {\bibfnamefont {X.}~\bibnamefont {Chen}}, \bibinfo {author} {\bibfnamefont {A.}~\bibnamefont {Ruschhaupt}}, \bibinfo {author} {\bibfnamefont {S.}~\bibnamefont {Schmidt}}, \bibinfo {author} {\bibfnamefont {A.}~\bibnamefont {del Campo}}, \bibinfo {author} {\bibfnamefont {D.}~\bibnamefont {Guéry-Odelin}},\ and\ \bibinfo {author} {\bibfnamefont {J.~G.}\ \bibnamefont {Muga}},\ }\bibfield  {title} {\bibinfo {title} {Fast {Optimal} {Frictionless} {Atom} {Cooling} in {Harmonic} {Traps}: {Shortcut} to {Adiabaticity}},\ }\href {https://doi.org/10.1103/PhysRevLett.104.063002} {\bibfield  {journal} {\bibinfo  {journal} {Physical Review Letters}\ }\textbf {\bibinfo {volume} {104}},\ \bibinfo {pages} {063002} (\bibinfo {year} {2010}{\natexlab{a}})}\BibitemShut {NoStop}%
\bibitem [{\citenamefont {Chen}\ \emph {et~al.}(2010{\natexlab{b}})\citenamefont {Chen}, \citenamefont {Lizuain}, \citenamefont {Ruschhaupt}, \citenamefont {Guéry-Odelin},\ and\ \citenamefont {Muga}}]{chen_shortcut_2010}%
  \BibitemOpen
  \bibfield  {author} {\bibinfo {author} {\bibfnamefont {X.}~\bibnamefont {Chen}}, \bibinfo {author} {\bibfnamefont {I.}~\bibnamefont {Lizuain}}, \bibinfo {author} {\bibfnamefont {A.}~\bibnamefont {Ruschhaupt}}, \bibinfo {author} {\bibfnamefont {D.}~\bibnamefont {Guéry-Odelin}},\ and\ \bibinfo {author} {\bibfnamefont {J.~G.}\ \bibnamefont {Muga}},\ }\bibfield  {title} {\bibinfo {title} {Shortcut to {Adiabatic} {Passage} in {Two}- and {Three}-{Level} {Atoms}},\ }\href {https://doi.org/10.1103/PhysRevLett.105.123003} {\bibfield  {journal} {\bibinfo  {journal} {Physical Review Letters}\ }\textbf {\bibinfo {volume} {105}},\ \bibinfo {pages} {123003} (\bibinfo {year} {2010}{\natexlab{b}})}\BibitemShut {NoStop}%
\bibitem [{\citenamefont {Patra}\ and\ \citenamefont {Jarzynski}(2017)}]{patra_shortcuts_2017}%
  \BibitemOpen
  \bibfield  {author} {\bibinfo {author} {\bibfnamefont {A.}~\bibnamefont {Patra}}\ and\ \bibinfo {author} {\bibfnamefont {C.}~\bibnamefont {Jarzynski}},\ }\bibfield  {title} {\bibinfo {title} {Shortcuts to adiabaticity using flow fields},\ }\href {https://doi.org/10.1088/1367-2630/aa924c} {\bibfield  {journal} {\bibinfo  {journal} {New Journal of Physics}\ }\textbf {\bibinfo {volume} {19}},\ \bibinfo {pages} {125009} (\bibinfo {year} {2017})}\BibitemShut {NoStop}%
\bibitem [{\citenamefont {Guéry-Odelin}\ \emph {et~al.}(2014)\citenamefont {Guéry-Odelin}, \citenamefont {Muga}, \citenamefont {Ruiz-Montero},\ and\ \citenamefont {Trizac}}]{guery-odelin_nonequilibrium_2014}%
  \BibitemOpen
  \bibfield  {author} {\bibinfo {author} {\bibfnamefont {D.}~\bibnamefont {Guéry-Odelin}}, \bibinfo {author} {\bibfnamefont {J.~G.}\ \bibnamefont {Muga}}, \bibinfo {author} {\bibfnamefont {M.~J.}\ \bibnamefont {Ruiz-Montero}},\ and\ \bibinfo {author} {\bibfnamefont {E.}~\bibnamefont {Trizac}},\ }\bibfield  {title} {\bibinfo {title} {Nonequilibrium {Solutions} of the {Boltzmann} {Equation} under the {Action} of an {External} {Force}},\ }\href {https://doi.org/10.1103/PhysRevLett.112.180602} {\bibfield  {journal} {\bibinfo  {journal} {Physical Review Letters}\ }\textbf {\bibinfo {volume} {112}},\ \bibinfo {pages} {180602} (\bibinfo {year} {2014})}\BibitemShut {NoStop}%
\bibitem [{\citenamefont {Martínez}\ \emph {et~al.}(2016{\natexlab{a}})\citenamefont {Martínez}, \citenamefont {Petrosyan}, \citenamefont {Guéry-Odelin}, \citenamefont {Trizac},\ and\ \citenamefont {Ciliberto}}]{martinez_engineered_2016}%
  \BibitemOpen
  \bibfield  {author} {\bibinfo {author} {\bibfnamefont {I.~A.}\ \bibnamefont {Martínez}}, \bibinfo {author} {\bibfnamefont {A.}~\bibnamefont {Petrosyan}}, \bibinfo {author} {\bibfnamefont {D.}~\bibnamefont {Guéry-Odelin}}, \bibinfo {author} {\bibfnamefont {E.}~\bibnamefont {Trizac}},\ and\ \bibinfo {author} {\bibfnamefont {S.}~\bibnamefont {Ciliberto}},\ }\bibfield  {title} {\bibinfo {title} {Engineered swift equilibration of a {Brownian} particle},\ }\href {https://doi.org/10.1038/nphys3758} {\bibfield  {journal} {\bibinfo  {journal} {Nature Physics}\ }\textbf {\bibinfo {volume} {12}},\ \bibinfo {pages} {843} (\bibinfo {year} {2016}{\natexlab{a}})}\BibitemShut {NoStop}%
\bibitem [{\citenamefont {Li}\ \emph {et~al.}(2017)\citenamefont {Li}, \citenamefont {Quan},\ and\ \citenamefont {Tu}}]{li_shortcuts_2017}%
  \BibitemOpen
  \bibfield  {author} {\bibinfo {author} {\bibfnamefont {G.}~\bibnamefont {Li}}, \bibinfo {author} {\bibfnamefont {H.~T.}\ \bibnamefont {Quan}},\ and\ \bibinfo {author} {\bibfnamefont {Z.~C.}\ \bibnamefont {Tu}},\ }\bibfield  {title} {\bibinfo {title} {Shortcuts to isothermality and nonequilibrium work relations},\ }\href {https://doi.org/10.1103/PhysRevE.96.012144} {\bibfield  {journal} {\bibinfo  {journal} {Physical Review E}\ }\textbf {\bibinfo {volume} {96}},\ \bibinfo {pages} {012144} (\bibinfo {year} {2017})}\BibitemShut {NoStop}%
\bibitem [{\citenamefont {Funo}\ \emph {et~al.}(2020)\citenamefont {Funo}, \citenamefont {Lambert}, \citenamefont {Nori},\ and\ \citenamefont {Flindt}}]{funo_shortcuts_2020}%
  \BibitemOpen
  \bibfield  {author} {\bibinfo {author} {\bibfnamefont {K.}~\bibnamefont {Funo}}, \bibinfo {author} {\bibfnamefont {N.}~\bibnamefont {Lambert}}, \bibinfo {author} {\bibfnamefont {F.}~\bibnamefont {Nori}},\ and\ \bibinfo {author} {\bibfnamefont {C.}~\bibnamefont {Flindt}},\ }\bibfield  {title} {\bibinfo {title} {Shortcuts to {Adiabatic} {Pumping} in {Classical} {Stochastic} {Systems}},\ }\href {https://doi.org/10.1103/PhysRevLett.124.150603} {\bibfield  {journal} {\bibinfo  {journal} {Physical Review Letters}\ }\textbf {\bibinfo {volume} {124}},\ \bibinfo {pages} {150603} (\bibinfo {year} {2020})}\BibitemShut {NoStop}%
\bibitem [{\citenamefont {Guéry-Odelin}\ \emph {et~al.}(2023)\citenamefont {Guéry-Odelin}, \citenamefont {Jarzynski}, \citenamefont {Plata}, \citenamefont {Prados},\ and\ \citenamefont {Trizac}}]{guery-odelin_driving_2023}%
  \BibitemOpen
  \bibfield  {author} {\bibinfo {author} {\bibfnamefont {D.}~\bibnamefont {Guéry-Odelin}}, \bibinfo {author} {\bibfnamefont {C.}~\bibnamefont {Jarzynski}}, \bibinfo {author} {\bibfnamefont {C.~A.}\ \bibnamefont {Plata}}, \bibinfo {author} {\bibfnamefont {A.}~\bibnamefont {Prados}},\ and\ \bibinfo {author} {\bibfnamefont {E.}~\bibnamefont {Trizac}},\ }\bibfield  {title} {\bibinfo {title} {Driving rapidly while remaining in control: classical shortcuts from {Hamiltonian} to stochastic dynamics},\ }\href {https://doi.org/10.1088/1361-6633/acacad} {\bibfield  {journal} {\bibinfo  {journal} {Reports on Progress in Physics}\ }\textbf {\bibinfo {volume} {86}},\ \bibinfo {pages} {035902} (\bibinfo {year} {2023})}\BibitemShut {NoStop}%
\bibitem [{\citenamefont {Schmiedl}\ and\ \citenamefont {Seifert}(2007)}]{schmiedl_optimal_2007}%
  \BibitemOpen
  \bibfield  {author} {\bibinfo {author} {\bibfnamefont {T.}~\bibnamefont {Schmiedl}}\ and\ \bibinfo {author} {\bibfnamefont {U.}~\bibnamefont {Seifert}},\ }\bibfield  {title} {\bibinfo {title} {Optimal {Finite}-{Time} {Processes} {In} {Stochastic} {Thermodynamics}},\ }\href {https://doi.org/10.1103/PhysRevLett.98.108301} {\bibfield  {journal} {\bibinfo  {journal} {Physical Review Letters}\ }\textbf {\bibinfo {volume} {98}},\ \bibinfo {pages} {108301} (\bibinfo {year} {2007})}\BibitemShut {NoStop}%
\bibitem [{\citenamefont {Schmiedl}\ and\ \citenamefont {Seifert}(2008)}]{schmiedl_efficiency_2008}%
  \BibitemOpen
  \bibfield  {author} {\bibinfo {author} {\bibfnamefont {T.}~\bibnamefont {Schmiedl}}\ and\ \bibinfo {author} {\bibfnamefont {U.}~\bibnamefont {Seifert}},\ }\bibfield  {title} {\bibinfo {title} {Efficiency at maximum power: {An} analytically solvable model for stochastic heat engines},\ }\href {https://doi.org/10.1209/0295-5075/81/20003} {\bibfield  {journal} {\bibinfo  {journal} {EPL (Europhysics Letters)}\ }\textbf {\bibinfo {volume} {81}},\ \bibinfo {pages} {20003} (\bibinfo {year} {2008})}\BibitemShut {NoStop}%
\bibitem [{\citenamefont {Aurell}\ \emph {et~al.}(2011)\citenamefont {Aurell}, \citenamefont {Mejía-Monasterio},\ and\ \citenamefont {Muratore-Ginanneschi}}]{aurell_optimal_2011}%
  \BibitemOpen
  \bibfield  {author} {\bibinfo {author} {\bibfnamefont {E.}~\bibnamefont {Aurell}}, \bibinfo {author} {\bibfnamefont {C.}~\bibnamefont {Mejía-Monasterio}},\ and\ \bibinfo {author} {\bibfnamefont {P.}~\bibnamefont {Muratore-Ginanneschi}},\ }\bibfield  {title} {\bibinfo {title} {Optimal {Protocols} and {Optimal} {Transport} in {Stochastic} {Thermodynamics}},\ }\href {https://doi.org/10.1103/PhysRevLett.106.250601} {\bibfield  {journal} {\bibinfo  {journal} {Physical Review Letters}\ }\textbf {\bibinfo {volume} {106}},\ \bibinfo {pages} {250601} (\bibinfo {year} {2011})}\BibitemShut {NoStop}%
\bibitem [{\citenamefont {Aurell}\ \emph {et~al.}(2012{\natexlab{a}})\citenamefont {Aurell}, \citenamefont {Mejía-Monasterio},\ and\ \citenamefont {Muratore-Ginanneschi}}]{aurell_boundary_2012}%
  \BibitemOpen
  \bibfield  {author} {\bibinfo {author} {\bibfnamefont {E.}~\bibnamefont {Aurell}}, \bibinfo {author} {\bibfnamefont {C.}~\bibnamefont {Mejía-Monasterio}},\ and\ \bibinfo {author} {\bibfnamefont {P.}~\bibnamefont {Muratore-Ginanneschi}},\ }\bibfield  {title} {\bibinfo {title} {Boundary layers in stochastic thermodynamics},\ }\href {https://doi.org/10.1103/PhysRevE.85.020103} {\bibfield  {journal} {\bibinfo  {journal} {Physical Review E}\ }\textbf {\bibinfo {volume} {85}},\ \bibinfo {pages} {020103} (\bibinfo {year} {2012}{\natexlab{a}})}\BibitemShut {NoStop}%
\bibitem [{\citenamefont {Plata}\ \emph {et~al.}(2019)\citenamefont {Plata}, \citenamefont {Guéry-Odelin}, \citenamefont {Trizac},\ and\ \citenamefont {Prados}}]{plata_optimal_2019}%
  \BibitemOpen
  \bibfield  {author} {\bibinfo {author} {\bibfnamefont {C.~A.}\ \bibnamefont {Plata}}, \bibinfo {author} {\bibfnamefont {D.}~\bibnamefont {Guéry-Odelin}}, \bibinfo {author} {\bibfnamefont {E.}~\bibnamefont {Trizac}},\ and\ \bibinfo {author} {\bibfnamefont {A.}~\bibnamefont {Prados}},\ }\bibfield  {title} {\bibinfo {title} {Optimal work in a harmonic trap with bounded stiffness},\ }\href {https://doi.org/10.1103/PhysRevE.99.012140} {\bibfield  {journal} {\bibinfo  {journal} {Physical Review E}\ }\textbf {\bibinfo {volume} {99}},\ \bibinfo {pages} {012140} (\bibinfo {year} {2019})}\BibitemShut {NoStop}%
\bibitem [{\citenamefont {Muratore-Ginanneschi}(2014)}]{muratore-ginanneschi_extremals_2014}%
  \BibitemOpen
  \bibfield  {author} {\bibinfo {author} {\bibfnamefont {P.}~\bibnamefont {Muratore-Ginanneschi}},\ }\bibfield  {title} {\bibinfo {title} {On extremals of the entropy production by ‘{Langevin}–{Kramers}’ dynamics},\ }\href {https://doi.org/10.1088/1742-5468/2014/05/P05013} {\bibfield  {journal} {\bibinfo  {journal} {Journal of Statistical Mechanics: Theory and Experiment}\ }\textbf {\bibinfo {volume} {2014}},\ \bibinfo {pages} {P05013} (\bibinfo {year} {2014})}\BibitemShut {NoStop}%
\bibitem [{\citenamefont {Muratore-Ginanneschi}\ and\ \citenamefont {Schwieger}(2017)}]{muratore-ginanneschi_application_2017}%
  \BibitemOpen
  \bibfield  {author} {\bibinfo {author} {\bibfnamefont {P.}~\bibnamefont {Muratore-Ginanneschi}}\ and\ \bibinfo {author} {\bibfnamefont {K.}~\bibnamefont {Schwieger}},\ }\bibfield  {title} {\bibinfo {title} {An {Application} of {Pontryagin}’s {Principle} to {Brownian} {Particle} {Engineered} {Equilibration}},\ }\href {https://doi.org/10.3390/e19070379} {\bibfield  {journal} {\bibinfo  {journal} {Entropy}\ }\textbf {\bibinfo {volume} {19}},\ \bibinfo {pages} {379} (\bibinfo {year} {2017})}\BibitemShut {NoStop}%
\bibitem [{\citenamefont {Zhang}(2020{\natexlab{a}})}]{zhang_work_2020}%
  \BibitemOpen
  \bibfield  {author} {\bibinfo {author} {\bibfnamefont {Y.}~\bibnamefont {Zhang}},\ }\bibfield  {title} {\bibinfo {title} {Work needed to drive a thermodynamic system between two distributions},\ }\href {https://doi.org/10.1209/0295-5075/128/30002} {\bibfield  {journal} {\bibinfo  {journal} {EPL (Europhysics Letters)}\ }\textbf {\bibinfo {volume} {128}},\ \bibinfo {pages} {30002} (\bibinfo {year} {2020}{\natexlab{a}})}\BibitemShut {NoStop}%
\bibitem [{\citenamefont {Zhang}(2020{\natexlab{b}})}]{zhang_optimization_2020}%
  \BibitemOpen
  \bibfield  {author} {\bibinfo {author} {\bibfnamefont {Y.}~\bibnamefont {Zhang}},\ }\bibfield  {title} {\bibinfo {title} {Optimization of {Stochastic} {Thermodynamic} {Machines}},\ }\href {https://doi.org/10.1007/s10955-020-02508-0} {\bibfield  {journal} {\bibinfo  {journal} {Journal of Statistical Physics}\ }\textbf {\bibinfo {volume} {178}},\ \bibinfo {pages} {1336} (\bibinfo {year} {2020}{\natexlab{b}})}\BibitemShut {NoStop}%
\bibitem [{\citenamefont {Plata}\ \emph {et~al.}(2021)\citenamefont {Plata}, \citenamefont {Prados}, \citenamefont {Trizac},\ and\ \citenamefont {Gu{\'e}ry-Odelin}}]{plata_taming_2021}%
  \BibitemOpen
  \bibfield  {author} {\bibinfo {author} {\bibfnamefont {C.~A.}\ \bibnamefont {Plata}}, \bibinfo {author} {\bibfnamefont {A.}~\bibnamefont {Prados}}, \bibinfo {author} {\bibfnamefont {E.}~\bibnamefont {Trizac}},\ and\ \bibinfo {author} {\bibfnamefont {D.}~\bibnamefont {Gu{\'e}ry-Odelin}},\ }\bibfield  {title} {\bibinfo {title} {Taming the {Time} {Evolution} in {Overdamped} {Systems}: {Shortcuts} {Elaborated} from {Fast}-{Forward} and {Time}-{Reversed} {Protocols}},\ }\href {https://doi.org/10.1103/PhysRevLett.127.190605} {\bibfield  {journal} {\bibinfo  {journal} {Physical Review Letters}\ }\textbf {\bibinfo {volume} {127}},\ \bibinfo {pages} {190605} (\bibinfo {year} {2021})}\BibitemShut {NoStop}%
\bibitem [{\citenamefont {Mart\'inez}\ \emph {et~al.}(2015)\citenamefont {Mart\'inez}, \citenamefont {Rold\'an}, \citenamefont {Dinis}, \citenamefont {Petrov},\ and\ \citenamefont {Rica}}]{martinez_adiabatic_2015}%
  \BibitemOpen
  \bibfield  {author} {\bibinfo {author} {\bibfnamefont {I.~A.}\ \bibnamefont {Mart\'inez}}, \bibinfo {author} {\bibfnamefont {E.}~\bibnamefont {Rold\'an}}, \bibinfo {author} {\bibfnamefont {L.}~\bibnamefont {Dinis}}, \bibinfo {author} {\bibfnamefont {D.}~\bibnamefont {Petrov}},\ and\ \bibinfo {author} {\bibfnamefont {R.~A.}\ \bibnamefont {Rica}},\ }\bibfield  {title} {\bibinfo {title} {Adiabatic {Processes} {Realized} with a {Trapped} {Brownian} {Particle}},\ }\href {https://doi.org/10.1103/PhysRevLett.114.120601} {\bibfield  {journal} {\bibinfo  {journal} {Physical Review Letters}\ }\textbf {\bibinfo {volume} {114}},\ \bibinfo {pages} {120601} (\bibinfo {year} {2015})}\BibitemShut {NoStop}%
\bibitem [{\citenamefont {Martínez}\ \emph {et~al.}(2017)\citenamefont {Martínez}, \citenamefont {Roldán}, \citenamefont {Dinis},\ and\ \citenamefont {Rica}}]{martinez_colloidal_2017}%
  \BibitemOpen
  \bibfield  {author} {\bibinfo {author} {\bibfnamefont {I.~A.}\ \bibnamefont {Martínez}}, \bibinfo {author} {\bibfnamefont {E.}~\bibnamefont {Roldán}}, \bibinfo {author} {\bibfnamefont {L.}~\bibnamefont {Dinis}},\ and\ \bibinfo {author} {\bibfnamefont {R.~A.}\ \bibnamefont {Rica}},\ }\bibfield  {title} {\bibinfo {title} {Colloidal heat engines: a review},\ }\href {https://doi.org/10.1039/C6SM00923A} {\bibfield  {journal} {\bibinfo  {journal} {Soft Matter}\ }\textbf {\bibinfo {volume} {13}},\ \bibinfo {pages} {22} (\bibinfo {year} {2017})}\BibitemShut {NoStop}%
\bibitem [{\citenamefont {Chupeau}\ \emph {et~al.}(2018)\citenamefont {Chupeau}, \citenamefont {Besga}, \citenamefont {Guéry-Odelin}, \citenamefont {Trizac}, \citenamefont {Petrosyan},\ and\ \citenamefont {Ciliberto}}]{chupeau_thermal_2018}%
  \BibitemOpen
  \bibfield  {author} {\bibinfo {author} {\bibfnamefont {M.}~\bibnamefont {Chupeau}}, \bibinfo {author} {\bibfnamefont {B.}~\bibnamefont {Besga}}, \bibinfo {author} {\bibfnamefont {D.}~\bibnamefont {Guéry-Odelin}}, \bibinfo {author} {\bibfnamefont {E.}~\bibnamefont {Trizac}}, \bibinfo {author} {\bibfnamefont {A.}~\bibnamefont {Petrosyan}},\ and\ \bibinfo {author} {\bibfnamefont {S.}~\bibnamefont {Ciliberto}},\ }\bibfield  {title} {\bibinfo {title} {Thermal bath engineering for swift equilibration},\ }\href {https://doi.org/10.1103/PhysRevE.98.010104} {\bibfield  {journal} {\bibinfo  {journal} {Physical Review E}\ }\textbf {\bibinfo {volume} {98}},\ \bibinfo {pages} {010104} (\bibinfo {year} {2018})}\BibitemShut {NoStop}%
\bibitem [{\citenamefont {Plata}\ \emph {et~al.}(2020{\natexlab{a}})\citenamefont {Plata}, \citenamefont {Guéry-Odelin}, \citenamefont {Trizac},\ and\ \citenamefont {Prados}}]{plata_finite-time_2020}%
  \BibitemOpen
  \bibfield  {author} {\bibinfo {author} {\bibfnamefont {C.~A.}\ \bibnamefont {Plata}}, \bibinfo {author} {\bibfnamefont {D.}~\bibnamefont {Guéry-Odelin}}, \bibinfo {author} {\bibfnamefont {E.}~\bibnamefont {Trizac}},\ and\ \bibinfo {author} {\bibfnamefont {A.}~\bibnamefont {Prados}},\ }\bibfield  {title} {\bibinfo {title} {Finite-time adiabatic processes: {Derivation} and speed limit},\ }\href {https://doi.org/10.1103/PhysRevE.101.032129} {\bibfield  {journal} {\bibinfo  {journal} {Physical Review E}\ }\textbf {\bibinfo {volume} {101}},\ \bibinfo {pages} {032129} (\bibinfo {year} {2020}{\natexlab{a}})}\BibitemShut {NoStop}%
\bibitem [{\citenamefont {Plata}\ \emph {et~al.}(2020{\natexlab{b}})\citenamefont {Plata}, \citenamefont {Guéry-Odelin}, \citenamefont {Trizac},\ and\ \citenamefont {Prados}}]{plata_building_2020}%
  \BibitemOpen
  \bibfield  {author} {\bibinfo {author} {\bibfnamefont {C.~A.}\ \bibnamefont {Plata}}, \bibinfo {author} {\bibfnamefont {D.}~\bibnamefont {Guéry-Odelin}}, \bibinfo {author} {\bibfnamefont {E.}~\bibnamefont {Trizac}},\ and\ \bibinfo {author} {\bibfnamefont {A.}~\bibnamefont {Prados}},\ }\bibfield  {title} {\bibinfo {title} {Building an irreversible {Carnot}-like heat engine with an overdamped harmonic oscillator},\ }\href {https://doi.org/10.1088/1742-5468/abb0e1} {\bibfield  {journal} {\bibinfo  {journal} {Journal of Statistical Mechanics: Theory and Experiment}\ }\textbf {\bibinfo {volume} {2020}},\ \bibinfo {pages} {093207} (\bibinfo {year} {2020}{\natexlab{b}})}\BibitemShut {NoStop}%
\bibitem [{\citenamefont {Prados}(2021)}]{prados_optimizing_2021}%
  \BibitemOpen
  \bibfield  {author} {\bibinfo {author} {\bibfnamefont {A.}~\bibnamefont {Prados}},\ }\bibfield  {title} {\bibinfo {title} {Optimizing the relaxation route with optimal control},\ }\href {https://doi.org/10.1103/PhysRevResearch.3.023128} {\bibfield  {journal} {\bibinfo  {journal} {Physical Review Research}\ }\textbf {\bibinfo {volume} {3}},\ \bibinfo {pages} {023128} (\bibinfo {year} {2021})}\BibitemShut {NoStop}%
\bibitem [{\citenamefont {Patrón}\ \emph {et~al.}(2022)\citenamefont {Patrón}, \citenamefont {Prados},\ and\ \citenamefont {Plata}}]{patron_thermal_2022}%
  \BibitemOpen
  \bibfield  {author} {\bibinfo {author} {\bibfnamefont {A.}~\bibnamefont {Patrón}}, \bibinfo {author} {\bibfnamefont {A.}~\bibnamefont {Prados}},\ and\ \bibinfo {author} {\bibfnamefont {C.~A.}\ \bibnamefont {Plata}},\ }\bibfield  {title} {\bibinfo {title} {Thermal brachistochrone for harmonically confined {Brownian} particles},\ }\href {https://doi.org/10.1140/epjp/s13360-022-03150-3} {\bibfield  {journal} {\bibinfo  {journal} {The European Physical Journal Plus}\ }\textbf {\bibinfo {volume} {137}},\ \bibinfo {pages} {1011} (\bibinfo {year} {2022})}\BibitemShut {NoStop}%
\bibitem [{\citenamefont {Ruiz-Pino}\ and\ \citenamefont {Prados}(2022)}]{ruiz-pino_optimal_2022}%
  \BibitemOpen
  \bibfield  {author} {\bibinfo {author} {\bibfnamefont {N.}~\bibnamefont {Ruiz-Pino}}\ and\ \bibinfo {author} {\bibfnamefont {A.}~\bibnamefont {Prados}},\ }\bibfield  {title} {\bibinfo {title} {Optimal {Control} of {Uniformly} {Heated} {Granular} {Fluids} in {Linear} {Response}},\ }\href {https://doi.org/10.3390/e24010131} {\bibfield  {journal} {\bibinfo  {journal} {Entropy}\ }\textbf {\bibinfo {volume} {24}},\ \bibinfo {pages} {131} (\bibinfo {year} {2022})}\BibitemShut {NoStop}%
\bibitem [{\citenamefont {Blaber}\ and\ \citenamefont {Sivak}(2023)}]{blaber_optimal_2023}%
  \BibitemOpen
  \bibfield  {author} {\bibinfo {author} {\bibfnamefont {S.}~\bibnamefont {Blaber}}\ and\ \bibinfo {author} {\bibfnamefont {D.~A.}\ \bibnamefont {Sivak}},\ }\bibfield  {title} {\bibinfo {title} {Optimal control in stochastic thermodynamics},\ }\href {https://doi.org/10.1088/2399-6528/acbf04} {\bibfield  {journal} {\bibinfo  {journal} {Journal of Physics Communications}\ }\textbf {\bibinfo {volume} {7}},\ \bibinfo {pages} {033001} (\bibinfo {year} {2023})}\BibitemShut {NoStop}%
\bibitem [{\citenamefont {Pontryagin}(1987)}]{pontryagin_mathematical_1987}%
  \BibitemOpen
  \bibfield  {author} {\bibinfo {author} {\bibfnamefont {L.~S.}\ \bibnamefont {Pontryagin}},\ }\href@noop {} {\emph {\bibinfo {title} {Mathematical {Theory} of {Optimal} {Processes}}}}\ (\bibinfo  {publisher} {CRC Press},\ \bibinfo {year} {1987})\BibitemShut {NoStop}%
\bibitem [{\citenamefont {Liberzon}(2012)}]{liberzon_calculus_2012}%
  \BibitemOpen
  \bibfield  {author} {\bibinfo {author} {\bibfnamefont {D.}~\bibnamefont {Liberzon}},\ }\href@noop {} {\emph {\bibinfo {title} {Calculus of {Variations} and {Optimal} {Control} {Theory}: {A} {Concise} {Introduction}}}}\ (\bibinfo  {publisher} {Princeton University Press},\ \bibinfo {year} {2012})\BibitemShut {NoStop}%
\bibitem [{\citenamefont {Pires}\ \emph {et~al.}(2023)\citenamefont {Pires}, \citenamefont {Goerlich}, \citenamefont {da~Fonseca}, \citenamefont {Debiossac}, \citenamefont {Hervieux}, \citenamefont {Manfredi},\ and\ \citenamefont {Genet}}]{pires_optimal_2023}%
  \BibitemOpen
  \bibfield  {author} {\bibinfo {author} {\bibfnamefont {L.~B.}\ \bibnamefont {Pires}}, \bibinfo {author} {\bibfnamefont {R.}~\bibnamefont {Goerlich}}, \bibinfo {author} {\bibfnamefont {A.~L.}\ \bibnamefont {da~Fonseca}}, \bibinfo {author} {\bibfnamefont {M.}~\bibnamefont {Debiossac}}, \bibinfo {author} {\bibfnamefont {P.-A.}\ \bibnamefont {Hervieux}}, \bibinfo {author} {\bibfnamefont {G.}~\bibnamefont {Manfredi}},\ and\ \bibinfo {author} {\bibfnamefont {C.}~\bibnamefont {Genet}},\ }\bibfield  {title} {\bibinfo {title} {Optimal {Time}-{Entropy} {Bounds} and {Speed} {Limits} for {Brownian} {Thermal} {Shortcuts}},\ }\href {https://doi.org/10.1103/PhysRevLett.131.097101} {\bibfield  {journal} {\bibinfo  {journal} {Phys. Rev. Lett.}\ }\textbf {\bibinfo {volume} {131}},\ \bibinfo {pages} {097101} (\bibinfo {year} {2023})},\ \bibinfo {note} {publisher: American Physical Society}\BibitemShut {NoStop}%
\bibitem [{\citenamefont {Filliger}\ and\ \citenamefont {Reimann}(2007)}]{filliger_brownian_2007}%
  \BibitemOpen
  \bibfield  {author} {\bibinfo {author} {\bibfnamefont {R.}~\bibnamefont {Filliger}}\ and\ \bibinfo {author} {\bibfnamefont {P.}~\bibnamefont {Reimann}},\ }\bibfield  {title} {\bibinfo {title} {Brownian {Gyrator}: {A} {Minimal} {Heat} {Engine} on the {Nanoscale}},\ }\href {https://doi.org/10.1103/PhysRevLett.99.230602} {\bibfield  {journal} {\bibinfo  {journal} {Physical Review Letters}\ }\textbf {\bibinfo {volume} {99}},\ \bibinfo {pages} {230602} (\bibinfo {year} {2007})}\BibitemShut {NoStop}%
\bibitem [{\citenamefont {Dotsenko}\ \emph {et~al.}(2013)\citenamefont {Dotsenko}, \citenamefont {Maciołek}, \citenamefont {Vasilyev},\ and\ \citenamefont {Oshanin}}]{dotsenko_two-temperature_2013}%
  \BibitemOpen
  \bibfield  {author} {\bibinfo {author} {\bibfnamefont {V.}~\bibnamefont {Dotsenko}}, \bibinfo {author} {\bibfnamefont {A.}~\bibnamefont {Maciołek}}, \bibinfo {author} {\bibfnamefont {O.}~\bibnamefont {Vasilyev}},\ and\ \bibinfo {author} {\bibfnamefont {G.}~\bibnamefont {Oshanin}},\ }\bibfield  {title} {\bibinfo {title} {Two-temperature {Langevin} dynamics in a parabolic potential},\ }\href {https://doi.org/10.1103/PhysRevE.87.062130} {\bibfield  {journal} {\bibinfo  {journal} {Physical Review E}\ }\textbf {\bibinfo {volume} {87}},\ \bibinfo {pages} {062130} (\bibinfo {year} {2013})}\BibitemShut {NoStop}%
\bibitem [{\citenamefont {Cerasoli}\ \emph {et~al.}(2018)\citenamefont {Cerasoli}, \citenamefont {Dotsenko}, \citenamefont {Oshanin},\ and\ \citenamefont {Rondoni}}]{cerasoli_asymmetry_2018}%
  \BibitemOpen
  \bibfield  {author} {\bibinfo {author} {\bibfnamefont {S.}~\bibnamefont {Cerasoli}}, \bibinfo {author} {\bibfnamefont {V.}~\bibnamefont {Dotsenko}}, \bibinfo {author} {\bibfnamefont {G.}~\bibnamefont {Oshanin}},\ and\ \bibinfo {author} {\bibfnamefont {L.}~\bibnamefont {Rondoni}},\ }\bibfield  {title} {\bibinfo {title} {Asymmetry relations and effective temperatures for biased {Brownian} gyrators},\ }\href {https://doi.org/10.1103/PhysRevE.98.042149} {\bibfield  {journal} {\bibinfo  {journal} {Physical Review E}\ }\textbf {\bibinfo {volume} {98}},\ \bibinfo {pages} {042149} (\bibinfo {year} {2018})}\BibitemShut {NoStop}%
\bibitem [{\citenamefont {Baldassarri}\ \emph {et~al.}(2020)\citenamefont {Baldassarri}, \citenamefont {Puglisi},\ and\ \citenamefont {Sesta}}]{baldassarri_engineered_2020}%
  \BibitemOpen
  \bibfield  {author} {\bibinfo {author} {\bibfnamefont {A.}~\bibnamefont {Baldassarri}}, \bibinfo {author} {\bibfnamefont {A.}~\bibnamefont {Puglisi}},\ and\ \bibinfo {author} {\bibfnamefont {L.}~\bibnamefont {Sesta}},\ }\bibfield  {title} {\bibinfo {title} {Engineered swift equilibration of a {Brownian} gyrator},\ }\href {https://doi.org/10.1103/PhysRevE.102.030105} {\bibfield  {journal} {\bibinfo  {journal} {Physical Review E}\ }\textbf {\bibinfo {volume} {102}},\ \bibinfo {pages} {030105} (\bibinfo {year} {2020})}\BibitemShut {NoStop}%
\bibitem [{\citenamefont {Miangolarra}\ \emph {et~al.}(2022)\citenamefont {Miangolarra}, \citenamefont {Taghvaei}, \citenamefont {Chen},\ and\ \citenamefont {Georgiou}}]{miangolarra_thermodynamic_2022}%
  \BibitemOpen
  \bibfield  {author} {\bibinfo {author} {\bibfnamefont {O.~M.}\ \bibnamefont {Miangolarra}}, \bibinfo {author} {\bibfnamefont {A.}~\bibnamefont {Taghvaei}}, \bibinfo {author} {\bibfnamefont {Y.}~\bibnamefont {Chen}},\ and\ \bibinfo {author} {\bibfnamefont {T.~T.}\ \bibnamefont {Georgiou}},\ }\bibfield  {title} {\bibinfo {title} {Thermodynamic engine powered by anisotropic fluctuations},\ }\href {https://doi.org/10.1103/PhysRevResearch.4.023218} {\bibfield  {journal} {\bibinfo  {journal} {Physical Review Research}\ }\textbf {\bibinfo {volume} {4}},\ \bibinfo {pages} {023218} (\bibinfo {year} {2022})}\BibitemShut {NoStop}%
\bibitem [{\citenamefont {Ciliberto}\ \emph {et~al.}(2013)\citenamefont {Ciliberto}, \citenamefont {Imparato}, \citenamefont {Naert},\ and\ \citenamefont {Tanase}}]{ciliberto_heat_2013}%
  \BibitemOpen
  \bibfield  {author} {\bibinfo {author} {\bibfnamefont {S.}~\bibnamefont {Ciliberto}}, \bibinfo {author} {\bibfnamefont {A.}~\bibnamefont {Imparato}}, \bibinfo {author} {\bibfnamefont {A.}~\bibnamefont {Naert}},\ and\ \bibinfo {author} {\bibfnamefont {M.}~\bibnamefont {Tanase}},\ }\bibfield  {title} {\bibinfo {title} {Heat {Flux} and {Entropy} {Produced} by {Thermal} {Fluctuations}},\ }\href {https://doi.org/10.1103/PhysRevLett.110.180601} {\bibfield  {journal} {\bibinfo  {journal} {Physical Review Letters}\ }\textbf {\bibinfo {volume} {110}},\ \bibinfo {pages} {180601} (\bibinfo {year} {2013})}\BibitemShut {NoStop}%
\bibitem [{\citenamefont {Chiang}\ \emph {et~al.}(2017)\citenamefont {Chiang}, \citenamefont {Lee}, \citenamefont {Lai},\ and\ \citenamefont {Chen}}]{chiang_electrical_2017}%
  \BibitemOpen
  \bibfield  {author} {\bibinfo {author} {\bibfnamefont {K.-H.}\ \bibnamefont {Chiang}}, \bibinfo {author} {\bibfnamefont {C.-L.}\ \bibnamefont {Lee}}, \bibinfo {author} {\bibfnamefont {P.-Y.}\ \bibnamefont {Lai}},\ and\ \bibinfo {author} {\bibfnamefont {Y.-F.}\ \bibnamefont {Chen}},\ }\bibfield  {title} {\bibinfo {title} {Electrical autonomous {Brownian} gyrator},\ }\href {https://doi.org/10.1103/PhysRevE.96.032123} {\bibfield  {journal} {\bibinfo  {journal} {Physical Review E}\ }\textbf {\bibinfo {volume} {96}},\ \bibinfo {pages} {032123} (\bibinfo {year} {2017})}\BibitemShut {NoStop}%
\bibitem [{\citenamefont {Argun}\ \emph {et~al.}(2017)\citenamefont {Argun}, \citenamefont {Soni}, \citenamefont {Dabelow}, \citenamefont {Bo}, \citenamefont {Pesce}, \citenamefont {Eichhorn},\ and\ \citenamefont {Volpe}}]{argun_experimental_2017}%
  \BibitemOpen
  \bibfield  {author} {\bibinfo {author} {\bibfnamefont {A.}~\bibnamefont {Argun}}, \bibinfo {author} {\bibfnamefont {J.}~\bibnamefont {Soni}}, \bibinfo {author} {\bibfnamefont {L.}~\bibnamefont {Dabelow}}, \bibinfo {author} {\bibfnamefont {S.}~\bibnamefont {Bo}}, \bibinfo {author} {\bibfnamefont {G.}~\bibnamefont {Pesce}}, \bibinfo {author} {\bibfnamefont {R.}~\bibnamefont {Eichhorn}},\ and\ \bibinfo {author} {\bibfnamefont {G.}~\bibnamefont {Volpe}},\ }\bibfield  {title} {\bibinfo {title} {Experimental realization of a minimal microscopic heat engine},\ }\href {https://doi.org/10.1103/PhysRevE.96.052106} {\bibfield  {journal} {\bibinfo  {journal} {Physical Review E}\ }\textbf {\bibinfo {volume} {96}},\ \bibinfo {pages} {052106} (\bibinfo {year} {2017})}\BibitemShut {NoStop}%
\bibitem [{\citenamefont {Cerasoli}\ \emph {et~al.}(2022)\citenamefont {Cerasoli}, \citenamefont {Ciliberto}, \citenamefont {Marinari}, \citenamefont {Oshanin}, \citenamefont {Peliti},\ and\ \citenamefont {Rondoni}}]{cerasoli_spectral_2022}%
  \BibitemOpen
  \bibfield  {author} {\bibinfo {author} {\bibfnamefont {S.}~\bibnamefont {Cerasoli}}, \bibinfo {author} {\bibfnamefont {S.}~\bibnamefont {Ciliberto}}, \bibinfo {author} {\bibfnamefont {E.}~\bibnamefont {Marinari}}, \bibinfo {author} {\bibfnamefont {G.}~\bibnamefont {Oshanin}}, \bibinfo {author} {\bibfnamefont {L.}~\bibnamefont {Peliti}},\ and\ \bibinfo {author} {\bibfnamefont {L.}~\bibnamefont {Rondoni}},\ }\bibfield  {title} {\bibinfo {title} {Spectral fingerprints of nonequilibrium dynamics: {The} case of a {Brownian} gyrator},\ }\href {https://doi.org/10.1103/PhysRevE.106.014137} {\bibfield  {journal} {\bibinfo  {journal} {Physical Review E}\ }\textbf {\bibinfo {volume} {106}},\ \bibinfo {pages} {014137} (\bibinfo {year} {2022})}\BibitemShut {NoStop}%
\bibitem [{\citenamefont {Deffner}\ \emph {et~al.}(2014)\citenamefont {Deffner}, \citenamefont {Jarzynski},\ and\ \citenamefont {Del~Campo}}]{deffner_classical_2014}%
  \BibitemOpen
  \bibfield  {author} {\bibinfo {author} {\bibfnamefont {S.}~\bibnamefont {Deffner}}, \bibinfo {author} {\bibfnamefont {C.}~\bibnamefont {Jarzynski}},\ and\ \bibinfo {author} {\bibfnamefont {A.}~\bibnamefont {Del~Campo}},\ }\bibfield  {title} {\bibinfo {title} {Classical and {Quantum} {Shortcuts} to {Adiabaticity} for {Scale}-{Invariant} {Driving}},\ }\href {https://doi.org/10.1103/PhysRevX.4.021013} {\bibfield  {journal} {\bibinfo  {journal} {Physical Review X}\ }\textbf {\bibinfo {volume} {4}},\ \bibinfo {pages} {021013} (\bibinfo {year} {2014})}\BibitemShut {NoStop}%
\bibitem [{\citenamefont {Tu}(2014)}]{tu_stochastic_2014}%
  \BibitemOpen
  \bibfield  {author} {\bibinfo {author} {\bibfnamefont {Z.~C.}\ \bibnamefont {Tu}},\ }\bibfield  {title} {\bibinfo {title} {Stochastic heat engine with the consideration of inertial effects and shortcuts to adiabaticity},\ }\href {https://doi.org/10.1103/PhysRevE.89.052148} {\bibfield  {journal} {\bibinfo  {journal} {Physical Review E}\ }\textbf {\bibinfo {volume} {89}},\ \bibinfo {pages} {052148} (\bibinfo {year} {2014})}\BibitemShut {NoStop}%
\bibitem [{\citenamefont {Li}\ \emph {et~al.}(2022)\citenamefont {Li}, \citenamefont {Chen}, \citenamefont {Sun},\ and\ \citenamefont {Dong}}]{li_geodesic_2022}%
  \BibitemOpen
  \bibfield  {author} {\bibinfo {author} {\bibfnamefont {G.}~\bibnamefont {Li}}, \bibinfo {author} {\bibfnamefont {J.-F.}\ \bibnamefont {Chen}}, \bibinfo {author} {\bibfnamefont {C.}~\bibnamefont {Sun}},\ and\ \bibinfo {author} {\bibfnamefont {H.}~\bibnamefont {Dong}},\ }\bibfield  {title} {\bibinfo {title} {Geodesic {Path} for the {Minimal} {Energy} {Cost} in {Shortcuts} to {Isothermality}},\ }\href {https://doi.org/10.1103/PhysRevLett.128.230603} {\bibfield  {journal} {\bibinfo  {journal} {Physical Review Letters}\ }\textbf {\bibinfo {volume} {128}},\ \bibinfo {pages} {230603} (\bibinfo {year} {2022})}\BibitemShut {NoStop}%
\bibitem [{\citenamefont {Sivak}\ and\ \citenamefont {Crooks}(2012)}]{sivak_thermodynamic_2012}%
  \BibitemOpen
  \bibfield  {author} {\bibinfo {author} {\bibfnamefont {D.~A.}\ \bibnamefont {Sivak}}\ and\ \bibinfo {author} {\bibfnamefont {G.~E.}\ \bibnamefont {Crooks}},\ }\bibfield  {title} {\bibinfo {title} {Thermodynamic {Metrics} and {Optimal} {Paths}},\ }\href {https://doi.org/10.1103/PhysRevLett.108.190602} {\bibfield  {journal} {\bibinfo  {journal} {Physical Review Letters}\ }\textbf {\bibinfo {volume} {108}},\ \bibinfo {pages} {190602} (\bibinfo {year} {2012})}\BibitemShut {NoStop}%
\bibitem [{\citenamefont {Aurell}\ \emph {et~al.}(2012{\natexlab{b}})\citenamefont {Aurell}, \citenamefont {Gawedzki}, \citenamefont {Mejia-Monasterio}, \citenamefont {Mohayaee},\ and\ \citenamefont {Muratore-Ginanneschi}}]{aurell_refined_2012}%
  \BibitemOpen
  \bibfield  {author} {\bibinfo {author} {\bibfnamefont {E.}~\bibnamefont {Aurell}}, \bibinfo {author} {\bibfnamefont {K.}~\bibnamefont {Gawedzki}}, \bibinfo {author} {\bibfnamefont {C.}~\bibnamefont {Mejia-Monasterio}}, \bibinfo {author} {\bibfnamefont {R.}~\bibnamefont {Mohayaee}},\ and\ \bibinfo {author} {\bibfnamefont {P.}~\bibnamefont {Muratore-Ginanneschi}},\ }\bibfield  {title} {\bibinfo {title} {Refined {Second} {Law} of {Thermodynamics} for {Fast} {Random} {Processes}},\ }\href {https://doi.org/10.1007/s10955-012-0478-x} {\bibfield  {journal} {\bibinfo  {journal} {Journal of Statistical Physics}\ }\textbf {\bibinfo {volume} {147}},\ \bibinfo {pages} {487} (\bibinfo {year} {2012}{\natexlab{b}})}\BibitemShut {NoStop}%
\bibitem [{\citenamefont {Dechant}(2022)}]{dechant_minimum_2022}%
  \BibitemOpen
  \bibfield  {author} {\bibinfo {author} {\bibfnamefont {A.}~\bibnamefont {Dechant}},\ }\bibfield  {title} {\bibinfo {title} {Minimum entropy production, detailed balance and {Wasserstein} distance for continuous-time {Markov} processes},\ }\href {https://doi.org/10.1088/1751-8121/ac4ac0} {\bibfield  {journal} {\bibinfo  {journal} {Journal of Physics A: Mathematical and Theoretical}\ }\textbf {\bibinfo {volume} {55}},\ \bibinfo {pages} {094001} (\bibinfo {year} {2022})}\BibitemShut {NoStop}%
\bibitem [{\citenamefont {Ding}\ \emph {et~al.}(2020)\citenamefont {Ding}, \citenamefont {Huang}, \citenamefont {Paul}, \citenamefont {Hao},\ and\ \citenamefont {Chen}}]{ding_smooth_2020}%
  \BibitemOpen
  \bibfield  {author} {\bibinfo {author} {\bibfnamefont {Y.}~\bibnamefont {Ding}}, \bibinfo {author} {\bibfnamefont {T.-Y.}\ \bibnamefont {Huang}}, \bibinfo {author} {\bibfnamefont {K.}~\bibnamefont {Paul}}, \bibinfo {author} {\bibfnamefont {M.}~\bibnamefont {Hao}},\ and\ \bibinfo {author} {\bibfnamefont {X.}~\bibnamefont {Chen}},\ }\bibfield  {title} {\bibinfo {title} {Smooth bang-bang shortcuts to adiabaticity for atomic transport in a moving harmonic trap},\ }\href {https://doi.org/10.1103/PhysRevA.101.063410} {\bibfield  {journal} {\bibinfo  {journal} {Physical Review A}\ }\textbf {\bibinfo {volume} {101}},\ \bibinfo {pages} {063410} (\bibinfo {year} {2020})}\BibitemShut {NoStop}%
\bibitem [{\citenamefont {Sekimoto}(2010)}]{sekimoto_stochastic_2010}%
  \BibitemOpen
  \bibfield  {author} {\bibinfo {author} {\bibfnamefont {K.}~\bibnamefont {Sekimoto}},\ }\href@noop {} {\emph {\bibinfo {title} {Stochastic {Energetics}}}}\ (\bibinfo  {publisher} {Springer},\ \bibinfo {year} {2010})\BibitemShut {NoStop}%
\bibitem [{\citenamefont {Peliti}\ and\ \citenamefont {Pigolotti}(2021)}]{peliti_stochastic_2021}%
  \BibitemOpen
  \bibfield  {author} {\bibinfo {author} {\bibfnamefont {L.}~\bibnamefont {Peliti}}\ and\ \bibinfo {author} {\bibfnamefont {S.}~\bibnamefont {Pigolotti}},\ }\href@noop {} {\emph {\bibinfo {title} {Stochastic Thermodynamics: an Introduction}}},\ \bibinfo {edition} {1st}\ ed.\ (\bibinfo  {publisher} {Princeton University Press},\ \bibinfo {year} {2021})\BibitemShut {NoStop}%
\bibitem [{\citenamefont {Shiraishi}\ \emph {et~al.}(2018)\citenamefont {Shiraishi}, \citenamefont {Funo},\ and\ \citenamefont {Saito}}]{shiraishi_speed_2018}%
  \BibitemOpen
  \bibfield  {author} {\bibinfo {author} {\bibfnamefont {N.}~\bibnamefont {Shiraishi}}, \bibinfo {author} {\bibfnamefont {K.}~\bibnamefont {Funo}},\ and\ \bibinfo {author} {\bibfnamefont {K.}~\bibnamefont {Saito}},\ }\bibfield  {title} {\bibinfo {title} {Speed {Limit} for {Classical} {Stochastic} {Processes}},\ }\href@noop {} {\bibfield  {journal} {\bibinfo  {journal} {Physical Review Letters}\ }\textbf {\bibinfo {volume} {121}},\ \bibinfo {pages} {070601} (\bibinfo {year} {2018})}\BibitemShut {NoStop}%
\bibitem [{\citenamefont {Shanahan}\ \emph {et~al.}(2018)\citenamefont {Shanahan}, \citenamefont {Chenu}, \citenamefont {Margolus},\ and\ \citenamefont {del Campo}}]{shanahan_quantum_2018}%
  \BibitemOpen
  \bibfield  {author} {\bibinfo {author} {\bibfnamefont {B.}~\bibnamefont {Shanahan}}, \bibinfo {author} {\bibfnamefont {A.}~\bibnamefont {Chenu}}, \bibinfo {author} {\bibfnamefont {N.}~\bibnamefont {Margolus}},\ and\ \bibinfo {author} {\bibfnamefont {A.}~\bibnamefont {del Campo}},\ }\bibfield  {title} {\bibinfo {title} {Quantum {Speed} {Limits} across the {Quantum}-to-{Classical} {Transition}},\ }\href {https://doi.org/10.1103/PhysRevLett.120.070401} {\bibfield  {journal} {\bibinfo  {journal} {Physical Review Letters}\ }\textbf {\bibinfo {volume} {120}},\ \bibinfo {pages} {070401} (\bibinfo {year} {2018})}\BibitemShut {NoStop}%
\bibitem [{\citenamefont {Funo}\ \emph {et~al.}(2019)\citenamefont {Funo}, \citenamefont {Shiraishi},\ and\ \citenamefont {Saito}}]{funo_speed_2019}%
  \BibitemOpen
  \bibfield  {author} {\bibinfo {author} {\bibfnamefont {K.}~\bibnamefont {Funo}}, \bibinfo {author} {\bibfnamefont {N.}~\bibnamefont {Shiraishi}},\ and\ \bibinfo {author} {\bibfnamefont {K.}~\bibnamefont {Saito}},\ }\bibfield  {title} {\bibinfo {title} {Speed limit for open quantum systems},\ }\href {https://doi.org/10.1088/1367-2630/aaf9f5} {\bibfield  {journal} {\bibinfo  {journal} {New Journal of Physics}\ }\textbf {\bibinfo {volume} {21}},\ \bibinfo {pages} {013006} (\bibinfo {year} {2019})}\BibitemShut {NoStop}%
\bibitem [{\citenamefont {Shiraishi}\ and\ \citenamefont {Saito}(2021)}]{shiraishi_speed_2020}%
  \BibitemOpen
  \bibfield  {author} {\bibinfo {author} {\bibfnamefont {N.}~\bibnamefont {Shiraishi}}\ and\ \bibinfo {author} {\bibfnamefont {K.}~\bibnamefont {Saito}},\ }\bibfield  {title} {\bibinfo {title} {Speed limit for open systems coupled to general environments},\ }\href {https://doi.org/10.1103/PhysRevResearch.3.023074} {\bibfield  {journal} {\bibinfo  {journal} {Phys. Rev. Research}\ }\textbf {\bibinfo {volume} {3}},\ \bibinfo {pages} {023074} (\bibinfo {year} {2021})}\BibitemShut {NoStop}%
\bibitem [{\citenamefont {Ito}\ and\ \citenamefont {Dechant}(2020)}]{ito_stochastic_2020}%
  \BibitemOpen
  \bibfield  {author} {\bibinfo {author} {\bibfnamefont {S.}~\bibnamefont {Ito}}\ and\ \bibinfo {author} {\bibfnamefont {A.}~\bibnamefont {Dechant}},\ }\bibfield  {title} {\bibinfo {title} {Stochastic time-evolution, information geometry and the {Cramer}-{Rao} {Bound}},\ }\href {https://doi.org/10.1103/PhysRevX.10.021056} {\bibfield  {journal} {\bibinfo  {journal} {Physical Review X}\ }\textbf {\bibinfo {volume} {10}},\ \bibinfo {pages} {021056} (\bibinfo {year} {2020})}\BibitemShut {NoStop}%
\bibitem [{\citenamefont {Van~Vu}\ and\ \citenamefont {Hasegawa}(2021)}]{van_vu_geometrical_2021}%
  \BibitemOpen
  \bibfield  {author} {\bibinfo {author} {\bibfnamefont {T.}~\bibnamefont {Van~Vu}}\ and\ \bibinfo {author} {\bibfnamefont {Y.}~\bibnamefont {Hasegawa}},\ }\bibfield  {title} {\bibinfo {title} {Geometrical {Bounds} of the {Irreversibility} in {Markovian} {Systems}},\ }\href {https://doi.org/10.1103/PhysRevLett.126.010601} {\bibfield  {journal} {\bibinfo  {journal} {Physical Review Letters}\ }\textbf {\bibinfo {volume} {126}},\ \bibinfo {pages} {010601} (\bibinfo {year} {2021})}\BibitemShut {NoStop}%
\bibitem [{\citenamefont {Lee}\ \emph {et~al.}(2022)\citenamefont {Lee}, \citenamefont {Lee}, \citenamefont {Kwon},\ and\ \citenamefont {Park}}]{lee_speed_2022}%
  \BibitemOpen
  \bibfield  {author} {\bibinfo {author} {\bibfnamefont {J.~S.}\ \bibnamefont {Lee}}, \bibinfo {author} {\bibfnamefont {S.}~\bibnamefont {Lee}}, \bibinfo {author} {\bibfnamefont {H.}~\bibnamefont {Kwon}},\ and\ \bibinfo {author} {\bibfnamefont {H.}~\bibnamefont {Park}},\ }\bibfield  {title} {\bibinfo {title} {Speed {Limit} for a {Highly} {Irreversible} {Process} and {Tight} {Finite}-{Time} {Landauer}’s {Bound}},\ }\href {https://doi.org/10.1103/PhysRevLett.129.120603} {\bibfield  {journal} {\bibinfo  {journal} {Physical Review Letters}\ }\textbf {\bibinfo {volume} {129}},\ \bibinfo {pages} {120603} (\bibinfo {year} {2022})}\BibitemShut {NoStop}%
\bibitem [{\citenamefont {Landi}\ and\ \citenamefont {Paternostro}(2021)}]{landi_irreversible_2021}%
  \BibitemOpen
  \bibfield  {author} {\bibinfo {author} {\bibfnamefont {G.~T.}\ \bibnamefont {Landi}}\ and\ \bibinfo {author} {\bibfnamefont {M.}~\bibnamefont {Paternostro}},\ }\bibfield  {title} {\bibinfo {title} {Irreversible entropy production: {From} classical to quantum},\ }\href {https://doi.org/10.1103/RevModPhys.93.035008} {\bibfield  {journal} {\bibinfo  {journal} {Reviews of Modern Physics}\ }\textbf {\bibinfo {volume} {93}},\ \bibinfo {pages} {035008} (\bibinfo {year} {2021})}\BibitemShut {NoStop}%
\bibitem [{\citenamefont {Crooks}(2007)}]{crooks_measuring_2007}%
  \BibitemOpen
  \bibfield  {author} {\bibinfo {author} {\bibfnamefont {G.~E.}\ \bibnamefont {Crooks}},\ }\bibfield  {title} {\bibinfo {title} {Measuring {Thermodynamic} {Length}},\ }\href {https://doi.org/10.1103/PhysRevLett.99.100602} {\bibfield  {journal} {\bibinfo  {journal} {Physical Review Letters}\ }\textbf {\bibinfo {volume} {99}},\ \bibinfo {pages} {100602} (\bibinfo {year} {2007})}\BibitemShut {NoStop}%
\bibitem [{\citenamefont {Ito}(2018)}]{ito_stochastic_2018}%
  \BibitemOpen
  \bibfield  {author} {\bibinfo {author} {\bibfnamefont {S.}~\bibnamefont {Ito}},\ }\bibfield  {title} {\bibinfo {title} {Stochastic {Thermodynamic} {Interpretation} of {Information} {Geometry}},\ }\href {https://doi.org/10.1103/PhysRevLett.121.030605} {\bibfield  {journal} {\bibinfo  {journal} {Physical Review Letters}\ }\textbf {\bibinfo {volume} {121}},\ \bibinfo {pages} {030605} (\bibinfo {year} {2018})}\BibitemShut {NoStop}%
\bibitem [{\citenamefont {Esposito}\ \emph {et~al.}(2010)\citenamefont {Esposito}, \citenamefont {Kawai}, \citenamefont {Lindenberg},\ and\ \citenamefont {Van~den Broeck}}]{esposito_efficiency_2010}%
  \BibitemOpen
  \bibfield  {author} {\bibinfo {author} {\bibfnamefont {M.}~\bibnamefont {Esposito}}, \bibinfo {author} {\bibfnamefont {R.}~\bibnamefont {Kawai}}, \bibinfo {author} {\bibfnamefont {K.}~\bibnamefont {Lindenberg}},\ and\ \bibinfo {author} {\bibfnamefont {C.}~\bibnamefont {Van~den Broeck}},\ }\bibfield  {title} {\bibinfo {title} {Efficiency at {Maximum} {Power} of {Low}-{Dissipation} {Carnot} {Engines}},\ }\href {https://doi.org/10.1103/PhysRevLett.105.150603} {\bibfield  {journal} {\bibinfo  {journal} {Physical Review Letters}\ }\textbf {\bibinfo {volume} {105}},\ \bibinfo {pages} {150603} (\bibinfo {year} {2010})}\BibitemShut {NoStop}%
\bibitem [{\citenamefont {Blickle}\ and\ \citenamefont {Bechinger}(2012)}]{blickle_realization_2012}%
  \BibitemOpen
  \bibfield  {author} {\bibinfo {author} {\bibfnamefont {V.}~\bibnamefont {Blickle}}\ and\ \bibinfo {author} {\bibfnamefont {C.}~\bibnamefont {Bechinger}},\ }\bibfield  {title} {\bibinfo {title} {Realization of a micrometre-sized stochastic heat engine},\ }\href {https://doi.org/10.1038/nphys2163} {\bibfield  {journal} {\bibinfo  {journal} {Nature Physics}\ }\textbf {\bibinfo {volume} {8}},\ \bibinfo {pages} {143} (\bibinfo {year} {2012})}\BibitemShut {NoStop}%
\bibitem [{\citenamefont {Deng}\ \emph {et~al.}(2013)\citenamefont {Deng}, \citenamefont {Wang}, \citenamefont {Liu}, \citenamefont {H{\"a}nggi},\ and\ \citenamefont {Gong}}]{deng_boosting_2013}%
  \BibitemOpen
  \bibfield  {author} {\bibinfo {author} {\bibfnamefont {J.}~\bibnamefont {Deng}}, \bibinfo {author} {\bibfnamefont {Q.-h.}\ \bibnamefont {Wang}}, \bibinfo {author} {\bibfnamefont {Z.}~\bibnamefont {Liu}}, \bibinfo {author} {\bibfnamefont {P.}~\bibnamefont {H{\"a}nggi}},\ and\ \bibinfo {author} {\bibfnamefont {J.}~\bibnamefont {Gong}},\ }\bibfield  {title} {\bibinfo {title} {Boosting work characteristics and overall heat-engine performance via shortcuts to adiabaticity: {Quantum} and classical systems},\ }\href {https://doi.org/10.1103/PhysRevE.88.062122} {\bibfield  {journal} {\bibinfo  {journal} {Physical Review E}\ }\textbf {\bibinfo {volume} {88}},\ \bibinfo {pages} {062122} (\bibinfo {year} {2013})}\BibitemShut {NoStop}%
\bibitem [{\citenamefont {Muratore-Ginanneschi}\ and\ \citenamefont {Schwieger}(2015)}]{muratore-ginanneschi_efficient_2015}%
  \BibitemOpen
  \bibfield  {author} {\bibinfo {author} {\bibfnamefont {P.}~\bibnamefont {Muratore-Ginanneschi}}\ and\ \bibinfo {author} {\bibfnamefont {K.}~\bibnamefont {Schwieger}},\ }\bibfield  {title} {\bibinfo {title} {Efficient protocols for {Stirling} heat engines at the micro-scale},\ }\href {https://doi.org/10.1209/0295-5075/112/20002} {\bibfield  {journal} {\bibinfo  {journal} {EPL (Europhysics Letters)}\ }\textbf {\bibinfo {volume} {112}},\ \bibinfo {pages} {20002} (\bibinfo {year} {2015})}\BibitemShut {NoStop}%
\bibitem [{\citenamefont {Martínez}\ \emph {et~al.}(2016{\natexlab{b}})\citenamefont {Martínez}, \citenamefont {Roldán}, \citenamefont {Dinis}, \citenamefont {Petrov}, \citenamefont {Parrondo},\ and\ \citenamefont {Rica}}]{martinez_brownian_2016}%
  \BibitemOpen
  \bibfield  {author} {\bibinfo {author} {\bibfnamefont {I.~A.}\ \bibnamefont {Martínez}}, \bibinfo {author} {\bibfnamefont {E.}~\bibnamefont {Roldán}}, \bibinfo {author} {\bibfnamefont {L.}~\bibnamefont {Dinis}}, \bibinfo {author} {\bibfnamefont {D.}~\bibnamefont {Petrov}}, \bibinfo {author} {\bibfnamefont {J.~M.~R.}\ \bibnamefont {Parrondo}},\ and\ \bibinfo {author} {\bibfnamefont {R.~A.}\ \bibnamefont {Rica}},\ }\bibfield  {title} {\bibinfo {title} {Brownian {Carnot} engine},\ }\href {https://doi.org/10.1038/nphys3518} {\bibfield  {journal} {\bibinfo  {journal} {Nature Physics}\ }\textbf {\bibinfo {volume} {12}},\ \bibinfo {pages} {67} (\bibinfo {year} {2016}{\natexlab{b}})}\BibitemShut {NoStop}%
\bibitem [{\citenamefont {Abah}\ and\ \citenamefont {Paternostro}(2019)}]{abah_shortcut--adiabaticity_2019}%
  \BibitemOpen
  \bibfield  {author} {\bibinfo {author} {\bibfnamefont {O.}~\bibnamefont {Abah}}\ and\ \bibinfo {author} {\bibfnamefont {M.}~\bibnamefont {Paternostro}},\ }\bibfield  {title} {\bibinfo {title} {Shortcut-to-adiabaticity {Otto} engine: {A} twist to finite-time thermodynamics},\ }\href {https://doi.org/10.1103/PhysRevE.99.022110} {\bibfield  {journal} {\bibinfo  {journal} {Physical Review E}\ }\textbf {\bibinfo {volume} {99}},\ \bibinfo {pages} {022110} (\bibinfo {year} {2019})}\BibitemShut {NoStop}%
\bibitem [{\citenamefont {Albay}\ \emph {et~al.}(2019)\citenamefont {Albay}, \citenamefont {Wulaningrum}, \citenamefont {Kwon}, \citenamefont {Lai},\ and\ \citenamefont {Jun}}]{albay_thermodynamic_2019}%
  \BibitemOpen
  \bibfield  {author} {\bibinfo {author} {\bibfnamefont {J.~A.~C.}\ \bibnamefont {Albay}}, \bibinfo {author} {\bibfnamefont {S.~R.}\ \bibnamefont {Wulaningrum}}, \bibinfo {author} {\bibfnamefont {C.}~\bibnamefont {Kwon}}, \bibinfo {author} {\bibfnamefont {P.-Y.}\ \bibnamefont {Lai}},\ and\ \bibinfo {author} {\bibfnamefont {Y.}~\bibnamefont {Jun}},\ }\bibfield  {title} {\bibinfo {title} {Thermodynamic cost of a shortcuts-to-isothermal transport of a {Brownian} particle},\ }\href {https://doi.org/10.1103/PhysRevResearch.1.033122} {\bibfield  {journal} {\bibinfo  {journal} {Physical Review Research}\ }\textbf {\bibinfo {volume} {1}},\ \bibinfo {pages} {033122} (\bibinfo {year} {2019})}\BibitemShut {NoStop}%
\bibitem [{\citenamefont {Albay}\ \emph {et~al.}(2020)\citenamefont {Albay}, \citenamefont {Lai},\ and\ \citenamefont {Jun}}]{albay_realization_2020}%
  \BibitemOpen
  \bibfield  {author} {\bibinfo {author} {\bibfnamefont {J.~A.~C.}\ \bibnamefont {Albay}}, \bibinfo {author} {\bibfnamefont {P.-Y.}\ \bibnamefont {Lai}},\ and\ \bibinfo {author} {\bibfnamefont {Y.}~\bibnamefont {Jun}},\ }\bibfield  {title} {\bibinfo {title} {Realization of finite-rate isothermal compression and expansion using optical feedback trap},\ }\href {https://doi.org/10.1063/1.5143602} {\bibfield  {journal} {\bibinfo  {journal} {Applied Physics Letters}\ }\textbf {\bibinfo {volume} {116}},\ \bibinfo {pages} {103706} (\bibinfo {year} {2020})}\BibitemShut {NoStop}%
\bibitem [{\citenamefont {Nakamura}\ \emph {et~al.}(2020)\citenamefont {Nakamura}, \citenamefont {Matrasulov},\ and\ \citenamefont {Izumida}}]{nakamura_fast_2020}%
  \BibitemOpen
  \bibfield  {author} {\bibinfo {author} {\bibfnamefont {K.}~\bibnamefont {Nakamura}}, \bibinfo {author} {\bibfnamefont {J.}~\bibnamefont {Matrasulov}},\ and\ \bibinfo {author} {\bibfnamefont {Y.}~\bibnamefont {Izumida}},\ }\bibfield  {title} {\bibinfo {title} {Fast-forward approach to stochastic heat engine},\ }\href {https://doi.org/10.1103/PhysRevE.102.012129} {\bibfield  {journal} {\bibinfo  {journal} {Physical Review E}\ }\textbf {\bibinfo {volume} {102}},\ \bibinfo {pages} {012129} (\bibinfo {year} {2020})}\BibitemShut {NoStop}%
\bibitem [{\citenamefont {Tu}(2021)}]{tu_abstract_2021}%
  \BibitemOpen
  \bibfield  {author} {\bibinfo {author} {\bibfnamefont {Z.~C.}\ \bibnamefont {Tu}},\ }\bibfield  {title} {\bibinfo {title} {Abstract models for heat engines},\ }\href {https://doi.org/10.1007/s11467-020-1029-6} {\bibfield  {journal} {\bibinfo  {journal} {Frontiers of Physics}\ }\textbf {\bibinfo {volume} {16}},\ \bibinfo {pages} {33202} (\bibinfo {year} {2021})}\BibitemShut {NoStop}%
\bibitem [{\citenamefont {Frim}\ and\ \citenamefont {DeWeese}(2022)}]{frim_optimal_2022}%
  \BibitemOpen
  \bibfield  {author} {\bibinfo {author} {\bibfnamefont {A.~G.}\ \bibnamefont {Frim}}\ and\ \bibinfo {author} {\bibfnamefont {M.~R.}\ \bibnamefont {DeWeese}},\ }\bibfield  {title} {\bibinfo {title} {Optimal finite-time {Brownian} {Carnot} engine},\ }\href {https://doi.org/10.1103/PhysRevE.105.L052103} {\bibfield  {journal} {\bibinfo  {journal} {Physical Review E}\ }\textbf {\bibinfo {volume} {105}},\ \bibinfo {pages} {L052103} (\bibinfo {year} {2022})}\BibitemShut {NoStop}%
\bibitem [{\citenamefont {Pedram}\ \emph {et~al.}(2023)\citenamefont {Pedram}, \citenamefont {Kad{\i}o{\u{g}}lu}, \citenamefont {Kabakç{\i}o\u{g}lu},\ and\ \citenamefont {M\"ustecapl{\i}o{\u{g}}lu}}]{pedram_quantum_2023}%
  \BibitemOpen
  \bibfield  {author} {\bibinfo {author} {\bibfnamefont {A.}~\bibnamefont {Pedram}}, \bibinfo {author} {\bibfnamefont {S.~C.}\ \bibnamefont {Kad{\i}o{\u{g}}lu}}, \bibinfo {author} {\bibfnamefont {A.}~\bibnamefont {Kabakç{\i}o\u{g}lu}},\ and\ \bibinfo {author} {\bibfnamefont {{\"O}.~E.}\ \bibnamefont {M\"ustecapl{\i}o{\u{g}}lu}},\ }\bibfield  {title} {\bibinfo {title} {A quantum {Otto} engine with shortcuts to thermalization and adiabaticity},\ }\href {https://doi.org/10.1088/1367-2630/ad0857} {\bibfield  {journal} {\bibinfo  {journal} {New Journal of Physics}\ }\textbf {\bibinfo {volume} {25}},\ \bibinfo {pages} {113014} (\bibinfo {year} {2023})}\BibitemShut {NoStop}%
\bibitem [{\citenamefont {Deng}\ \emph {et~al.}(2024)\citenamefont {Deng}, \citenamefont {Ai}, \citenamefont {Wang}, \citenamefont {Shao}, \citenamefont {Liu},\ and\ \citenamefont {Cui}}]{deng_exploring_2024}%
  \BibitemOpen
  \bibfield  {author} {\bibinfo {author} {\bibfnamefont {G.-x.}\ \bibnamefont {Deng}}, \bibinfo {author} {\bibfnamefont {H.}~\bibnamefont {Ai}}, \bibinfo {author} {\bibfnamefont {B.}~\bibnamefont {Wang}}, \bibinfo {author} {\bibfnamefont {W.}~\bibnamefont {Shao}}, \bibinfo {author} {\bibfnamefont {Y.}~\bibnamefont {Liu}},\ and\ \bibinfo {author} {\bibfnamefont {Z.}~\bibnamefont {Cui}},\ }\bibfield  {title} {\bibinfo {title} {Exploring the optimal cycle for a quantum heat engine using reinforcement learning},\ }\href {https://doi.org/10.1103/PhysRevA.109.022246} {\bibfield  {journal} {\bibinfo  {journal} {Physical Review A}\ }\textbf {\bibinfo {volume} {109}},\ \bibinfo {pages} {022246} (\bibinfo {year} {2024})}\BibitemShut {NoStop}%
\end{thebibliography}%

\end{document}